\documentclass[aps,twocolumn,superscriptaddress,preprintnumbers,showpacs,article,prl]{revtex4-1}

\usepackage[latin9]{inputenc}
\usepackage{color}
\usepackage{verbatim}
\usepackage{amsmath}
\usepackage{amssymb}
\usepackage{graphicx}
\usepackage{esint}
\usepackage{esvect}
\usepackage{braket}
\usepackage{amsmath,bm}
\usepackage{graphicx}
\usepackage{subfigure}
\usepackage{url}
\usepackage{lipsum}
\usepackage{lipsum}
\usepackage[dvipsnames]{xcolor}
\usepackage{hyperref}
\usepackage[resetlabels]{multibib}

\makeatletter
\@ifundefined{textcolor}{}
{%
 \definecolor{BLACK}{gray}{0}
 \definecolor{WHITE}{gray}{1}
 \definecolor{RED}{rgb}{1,0,0}
 \definecolor{GREEN}{rgb}{0,1,0}
 \definecolor{BLUE}{rgb}{0,0,1}
 \definecolor{CYAN}{cmyk}{1,0,0,0}
 \definecolor{MAGENTA}{cmyk}{0,1,0,0}
 \definecolor{YELLOW}{cmyk}{0,0,1,0}
}

\usepackage{mathrsfs}

\draft
\makeatother
\begin{document}

\title{Structure of domain walls in chiral spin liquids}
\author{Yan-Qi Wang}
\affiliation{Department of Physics, University of California,  Berkeley, California 94720, USA}
\affiliation{Materials Sciences Division, Lawrence Berkeley National Laboratory, Berkeley, California 94720, USA}
\author{Chunxiao Liu}
\affiliation{Department of Physics, University of California,  Berkeley, California 94720, USA}
\author{Joel E. Moore}
\affiliation{Department of Physics, University of California,  Berkeley, California 94720, USA}
\affiliation{Materials Sciences Division, Lawrence Berkeley National Laboratory, Berkeley, California 94720, USA}

\begin{abstract}
The chiral spin liquid is one of the canonical examples of a topological state of quantum spins coexisting with symmetry-breaking chiral order; its experimental realization has been actively discussed in the past few years. Here, motivated by the interplay between topology and symmetry breaking, we examine the physics of the interface between two chiral spin liquid domains with opposite chiralities. We show that a self-consistent mean-field description for the spinons exists that describes both the change of chirality at the domain wall and the gapless edge modes living on it.  A Ginzburg--Landau theory for the domain wall is formulated based on the mean-field picture, from which we obtain the non-universal properties of the domain wall such as the wall width and tension. We show that the velocity of the topologically protected domain wall edge states can be accessed through the Jackiw-Rebbi mechanism. We further argue that the gapless modes at the edge contribute an extra, non-analytic $|\phi^3|$ term to the domain wall theory, and find numerical evidence for this non-analyticity.

\end{abstract}

\maketitle

 {\it Introduction--} The two most prevalent classes of ordered states in quantum materials are those arising from spontaneous symmetry breaking and from topological order.  The nature of domains and domain walls or boundaries in these two kinds of ordered states is generally quite different: most, but not all, topological states have protected gapless excitations at their boundaries, while symmetry-breaking states need not.  The focus of the present work is on the nature of domain walls in perhaps the most studied example of a state with simultaneous symmetry breaking and fractional topological order: the chiral spin liquid.

A chiral spin liquid (CSL) is a long-range-entangled spin liquid ground state, which, despite the absence of conventional magnetic order, spontaneously breaks time-reversal symmetry and develops a chiral order \cite{Wenbook,Fradkinbook,Wen1989,Kalmeyer1987,Kalmeyer1989,Barkeshli2013,Bauer2014,Lu2012,Hermele2011,He2014,Gong2014,He2015,Bieri2015,Wietek2015,Gong2015,Hu2015,Hu2016,Nataf2016,Chen2020,Schroeter2007}. 
The prototypical example is the Kalmeyer--Laughlin (KL) CSL \cite{Kalmeyer1987,Kalmeyer1989}: as a spin system analog of the $\nu\!=\!1/2$ bosonic fractional quantum Hall state~\cite{Wen1990}, it  hosts both spin-1/2 anyonic quasiparticles (spinons) in the bulk~\cite{Halperin1984} and chiral gapless modes on the edge~\cite{Wen1991}. The universal bulk physics of the KL CSL is captured by the Chern--Simons (CS) effective theory. Many model systems are found to support such a state~\cite{Thomale2009,Nielsen2013,Glasser2015,Greiter2014}; some of them are explicitly time-reversal invariant~\cite{Szasz2020,Zhang2021,Zhu2020a,Cookmeyer2021,Chen2021,Thomale2009,Nielsen2013,Glasser2015,Greiter2014} and host degenerate chiral ground states and possibly domains of either chiralities; variation in domain size between samples could explain the variable thermal conductivities reported in some spin liquid candidates~\cite{Szasz2020}.

While the CS theory contains the full topological data of the bulk, it does not capture the symmetry-breaking properties of a CSL, nor some of the non-universal but experimentally important features of the system. For one thing, the CS theory provides little information about the details of the quasiparticle energetic structure. It is well known that the CS term is not gauge invariant in a system with boundary, and a gapless edge mode must exist at the boundary to restore the gauge invariance~\cite{Callan1985,Chandrasekharan1994,Maeda1996}. The velocity of such an edge mode enters the CS term as an effective parameter~\cite{Wen1995,Lopez1999,Lopez2001,Lee1991,Kao1996}, which, in a quantum Hall fluid, would be determined externally by the electric and magnetic fields at the edge. In a CSL, however, there is no established way to determine the edge current velocity in terms of the fundamental data of the CSL. 

A more interesting, yet less explored, scenario occurs when a CSL contains multiple domains. In this case, an interface of two domains with opposite chiralities hosts gapless mode with a total chiral current carrying spin $S\!=\!2$~\cite{Wen1989,Wenbook,Fradkinbook}. The chiral order parameter across the interface is determined \emph{self-consistently} to minimize the total energy of the multiple domain system, and its spatial profile reveals many system-dependent properties of CSLs. Previously, the self-consistent mean-field theory has been widely applied to the study of homogeneous quantum spin liquid systems~\cite{Wen1989,Wen1991B}. The solution, called a mean-field ansatz, provides a convenient description for the bulk spinons in a quantum spin liquid state. It is thus desirable to develop a microscopic theory to capture both bulk and domain wall physics of CSLs, and the self-consistent mean-field theory, building on its previous successes, can be adapted to provide valuable insight into this problem.

In this work, we study the domain wall structure in CSLs with domains of opposite chiralities. Based on a standard model~\cite{Wen1989}, we obtain a spatially varying, self-consistent mean-field solution for the spinons that describes both the gapless edge modes and the change of chirality at the domain wall. We complement the numerics by an effective field theory analysis for the mean-field bond amplitude fluctuations, and analytically solve the domain wall profile within the conventional Ising domain wall theory.  By modeling the low energy theory for spinons as Dirac fermions with spatially varying mass term, the velocity of the topologically protected domain wall edge states can be accessed through the Jackiw-Rebbi mechanism. We further find that, while the conventional Ising (i.e., $\phi^4$) theory shows approximate agreement with the numerics in terms of the domain wall profile, the gapless modes at the edge should in principle contribute an extra, non-analytic, $|\phi^3|$ term to the theory, and indeed such a term appears in derivatives of the bulk energy.

\begin{figure}[!t]
\centering 
\includegraphics[width=1\columnwidth]{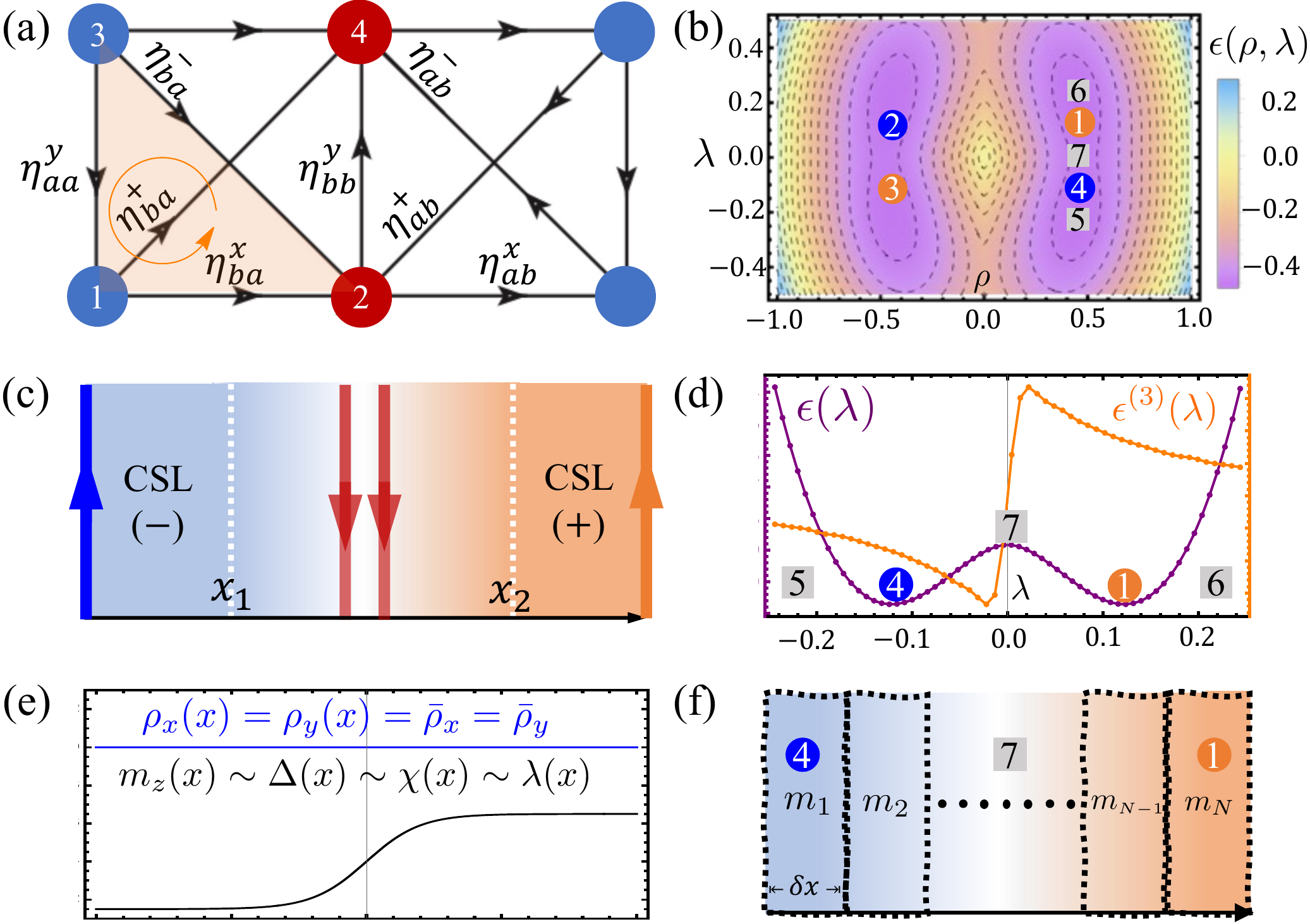}
\caption{\label{Illustration}  (a) Parton mean-field hoppings ($+i$ along the arrow), with blue (red) circle stands for the $a$ ($b$) sublattice. (b) Density-contour plot for the energy \eqref{Energy} in the parameter space, where $\epsilon(\rho,\lambda) = F(\rho,\rho,\lambda)/J_1$. The orange (blue) dots stand for the solution with positive (negative) chirality. (c) The two-domain setup. The system is finite along $x$ direction but periodic along $y$ direction, which can be viewed as a quasi-1D ribbon. The blue (orange) part to the left (right) of $x_1$ ($x_2$) are the chiral spin states with negative (positive) chirality, held fixed during the numerical iteration, while the region between $x_1$ and $x_2$ are subjected to the iteration. (d) The energy $\epsilon(\lambda)$
and its third derivative $\epsilon^{(3)}(\lambda)$ along $5\rightarrow 4 \rightarrow 7 \rightarrow 1\rightarrow 6$ in (b). (e) A converged solution for the domain wall structure. Note that the  local mass $m_z(x)$, gap $\Delta(x)$, and chirality $\chi(x)$ are all proportional to $\lambda(x)$. (f) Partition for the ribbon mentioned in (c).}
\end{figure}

{{\it Spin model and parton mean-field theory}--} Consider a spin-$1/2$ model defined on a square lattice
\begin{equation}\label{SpinModel}
	H =\sum_{{\bm i}{\bm j}} J_{{\bm i} {\bm j}} {\bm S}_{\bm i} \cdot{}{\bm S}_{\bm j}.
\end{equation} 
 with antiferromagnetic nearest-neighbor and next-nearest-neighbor coupling $J_1>J_2>0$~\cite{Wen1989,Wenbook,Fradkinbook}. This is one of the canonical models for CSLs and we review the parton self-consistent mean-field treatment here, although naturally a mean-field theory will not predict the energetics of the ground state reliably. Introduce the fermionic parton operators $f_{{\bm i}\alpha}$, $\alpha = 1,2$,  defined by ${\bm S}_{\bm i} = f^\dagger_{{\bm i} \alpha} {\bm \sigma}_{\alpha \beta} f_{{\bm i} \beta}/2$, with ${\bm \sigma}$ denotes the three-component vector of Pauli matrices. Physically, these operators describe the spin-$1/2$ charge-neutral spinon excitations in the CSL phase. One seeks a spinon mean-field Hamiltonian of the form:
\begin{equation}\label{MeanField}
\begin{aligned}
	H_{\rm mean} =& \sum_{\langle {\bm i} {\bm j} \rangle} - \frac{1}{2} J_{{\bm i} {\bm j}} \bigg{[} (f^\dagger_{{\bm i} \alpha} f_{{\bm j}\alpha} \eta_{{\bm j} {\bm i}} + {\rm h.c.}) - |\eta_{{\bm i} {\bm j}}|^2 \bigg{]} \\
	+& \sum_{\bm i} a_0({\bm i}) (f^\dagger_{{\bm i} \alpha} f_{{\bm i} \alpha} - 1),
\end{aligned}
\end{equation}
constrained by the self-consistency condition for the nearest and next-nearest hoppings, as well as the one-fermion-per-site condition:
\begin{equation}\label{SelfConsistentRequirement}
	 \langle f^\dagger_{{\bm i}\alpha} f_{{\bm j}\alpha} \rangle = \eta_{{\bm i}{\bm j}} , \quad \langle f^\dagger_{{\bm i} \alpha} f_{{\bm i} \alpha}\rangle = 1.
\end{equation} 
Here the $\langle \cdots \rangle$ stands for the expectation value with respect to the ground state. The solutions to Eq.~\eqref{SelfConsistentRequirement}, $\{\bar \eta_{{\bm i}{\bm j}}\}$, give the mean-field ansatz for the CSL phase~\cite{Wen1989,Wenbook,Fradkinbook}. We consider the following ansatz
$		 \eta_{ba}^x({\bm r}_i^{b}, {\bm r}_i^a)\!=\!\eta_{ab}^x({\bm r}_i^b \!+\! \hat e_x , {\bm r}_i^b)\!=\!+ i\rho_x({\bm i}), 
		 \eta_{aa}^y({\bm r}_i^a, {\bm r}_i^a \!+\! \hat e_y)\!=\!- \eta_{bb}^y ({\bm r}_i^b, {\bm r}_i^b \!+\! \hat e_y)\!=\!-i \rho_y ({\bm i}), ~
		 \eta^{+}_{ba}({\bm r}_i^b, {\bm r}_i^b \!+\!\hat e_x \!+\! \hat e_y)\!=\!\eta^-_{ba}({\bm r}_i^b,{\bm r}_i^b \!+ \! \hat e_x \!-\! \hat e_y)\!=\! -\eta_{ab}^{+}({\bm r}_i^a, {\bm r}_i^a \!+\! \hat e_x \!+ \!\hat e_y)\!=\!-\eta_{ab}^{-}({\bm r}_i^a, {\bm r}_i^a \!+ \!\hat e_x \!-\! \hat e_y)\!=\!+ i\lambda({\bm i})$, where ${\bm r}_i^a$ and ${\bm r}_i^b$ stand for the position of two sublattices in the $i$-th unit cell, and $\hat e_x$  ($\hat e_y$) the unit vector along the $+x$ ($+y$) direction, see Fig.~\ref{Illustration}(a) for an illustration. The order parameter for time-reversal symmetry breaking is the spin chirality operator, defined as
\begin{equation}
	\chi({\bm i})\!\equiv\!2{\bm S}({\bm r}_{\bm i}^a) \cdot{} ({\bm S}({\bm r}_{\bm i}^b)\!\times\!{\bm S}({\bm r}_{\bm i}^a + \hat e_y))\!=\!\rho_x({\bm i})   \rho_y({\bm i})  \lambda({\bm i}).
\end{equation}

We first look for a homogeneous solution $\{ \rho_x({\bm i}), \rho_y({\bm i}), \lambda({\bm i}) \} \equiv(\rho_x, \rho_y, \lambda) $ by solving the self-consistency condition~\eqref{SelfConsistentRequirement} in an iterative manner: plugging an initial ansatz for the hopping parameters to the left-hand side of Eq.~\eqref{SelfConsistentRequirement} generates a new ansatz on the right-hand side. We repeat this procedure until the ansatz converges~\cite{Supp,Hermele2011}. When $J_2 > 0.46J_1$, we find eight solutions $(\pm \bar \rho_x, \pm \bar \rho_y, \pm \bar \lambda)$ with $\bar \rho_x\!=\!\bar \rho_y$ and $\bar{\lambda}>0$, ensuring the Hamiltonian is gapped. The solutions can be grouped by their chirality $\chi\!=\!\pm \chi_0$, where $\chi_0\!=\!|\bar \rho_x \bar \rho_y \bar \lambda|$. The corresponding Bloch Hamiltonian reads (we set lattice constant $a_0=1$)
\begin{equation}\label{BlochBand}
\begin{aligned}
	h_0({\bm k}) &\!=\! J_1 |\rho_x|^2 + J_1 |\rho_y|^2 + 2J_2 |\lambda|^2 + {\bm d}({\bm k}) \cdot{} {\bm \sigma},\\
\bm{d}(\bm{k})
&\!=\!
2(J_1\rho_x \sin k_x, 2J_2 \lambda \cos k_x  \cos k_y,-J_1 \rho_y \sin k_y).
	\end{aligned}
\end{equation}
The Chern number of the lower band $C\!=\!-{\rm sgn}(\rho_x \rho_y \lambda)$ is associated with $\chi({\bm i})$ of the original spin model~\cite{Supp,bernevig2013}.

Alternatively, one can use an energy minimization approach to obtain the mean-field ansatz. Define the associated energy $F(\rho_x, \rho_y, \lambda)$ at half-filling:
\begin{equation}\label{Energy}
	F(\rho_x,\rho_y,\lambda) \!=\! J_1 |\rho_x|^2 \!+\! J_1 |\rho_y|^2 \!+\! 2J_1 |\lambda|^2 \!-\! \int_{\rm BZ} [d^2k] |{\bm d}({\bm k})|,
\end{equation}
where $[d^2k]$ stands for the measure of integral in Fourier space, we verified that minimizing the energy $F(\rho_x, \rho_y, \lambda)$ with respect to the hopping parameters gives the same set of results $(\pm \bar \rho_x, \pm \bar \rho_y, \pm \bar \lambda)$, confirming the validity of the iterative method. The energy profile and the solutions are illustrated in Fig.~\ref{Illustration}(b). 

{\it Self-consistent solution for an inhomogeneous CSL with domain wall--} One central approach in this work is a self-consistent mean-field ansatz for a CSL system with both positive and negative chirality domains. For such a spatially inhomogeneous ansatz, the iteration method is particularly efficient compared to the energy minimization approach and hence will be our main tool in the analysis below. For simplicity, we consider a two-domain setup as illustrated in Fig.~\ref{Illustration}(c): the system is finite along $x$ direction and periodic along $y$ direction, with a negative (positive) chirality $-\chi_0$ ($+\chi_0$) CSL domain existing on the left (right). While the chirality is nearly uniform far away from the interface, a profile develops near the interface, leading to a finite domain wall whose width will be determined self-consistently~\cite{Supp}.

\begin{figure}[t]
\centering 
\includegraphics[width=1\columnwidth]{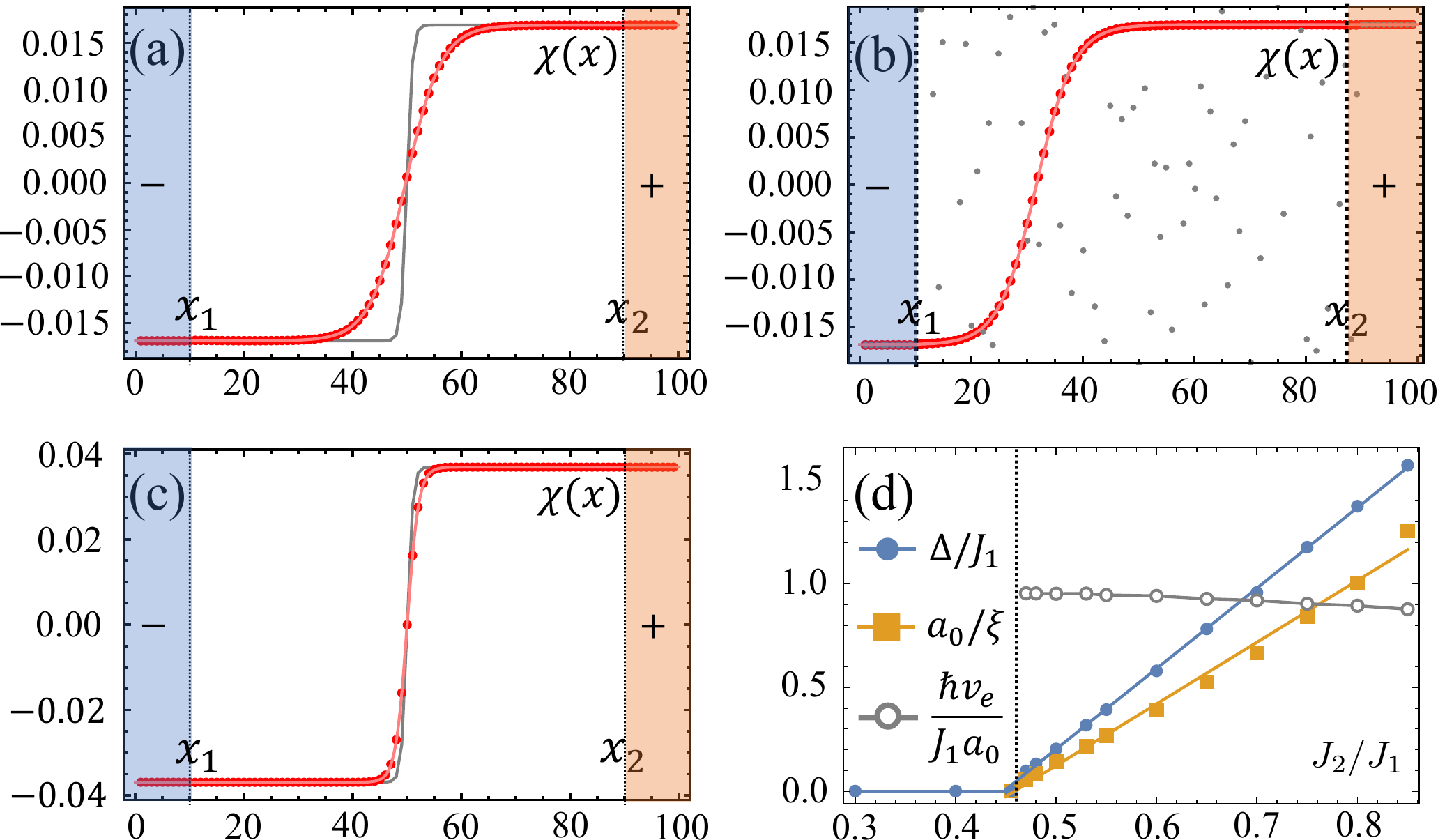}
\caption{\label{DomainWallPattern} (a-c) Domain wall configurations, with the gray line standing for the initially proposed chirality profile and the red dots the final chirality profile of the converged solution. The left blue (right orange) part are the chiral spin states fixed to the ground state value $(\bar \rho_x, \bar \rho_y, -\bar \lambda)$ ($(\bar \rho_x, \bar \rho_y, +\bar \lambda)$) during the iteration. (a) $J_2/J_1 = 0.53$, with the symmetric sharp initial condition for chirality. (b) $J_2/J_1 =0.53$ with the random initial condition for chirality. (c) $J_2/J_1=0.70$, with symmetric sharp initial condition for chirality.  (d) The bulk gap $\Delta$ for homogeneous limit and the inverse correlation length $\xi^{-1}$ with respect to $J_2/J_1$, from which we obtain $J_1 a_0/\xi = 0.77 \Delta$. The values $J_2/J_1$, $\Delta/J_1$, $a_0/\xi$ and $\hbar v_{\rm edge} /J_1a_0$ have a one-to-one correspondence with each other.}
\end{figure}

Given a set of $(J_1,J_2)$, we perform the self-consistent calculation for a system with 50 unit cells ($L\!=\!100$ sites) along the $x$-direction with sufficiently dense $k_y$ points to conduct the numerical $k_y$ integral. For $J_2\geq 0.46 J_1$, we find a converging solution of the following form: $\bar \rho_x$ and $\bar \rho_y$ remain constant in the entire system, while $\lambda(x)$ in the domain wall region extends between $-\bar\lambda$ (the value on the far left) and $\bar\lambda$ (the value on the far right). This solution is shown in Fig.~\ref{DomainWallPattern}. One can fit the $\lambda(x)$ profile and hence the chirality profile very well with a tanh function
\begin{equation}
	\chi(x) = \chi_0 \tanh[(x-x_0)/(\sqrt{2}\xi)],
\end{equation}
here $x_0$ defines the domain wall location. Changing the domain wall location does not affect the shape of the domain wall or the total energy of the system, hence we will manually set $x_0 = L/2$.

It is interesting to study the domain wall width $\xi$ as a function of the exchange parameters $(J_1,J_2)$. We find that $\xi$ is directly proportional to the inverse of the bulk gap (far away from the domain wall), which in turn is a function of $J_2/J_1$.
In the limit where the bulk gap $\Delta =8 J_2\bar \lambda$ vanishes (corresponding to $J_2 \approx 0.46 J_1$), the domain wall width diverges, as shown in Fig.~\ref{DomainWallPattern}(d).

{\it Effective field theory--}  The bulk of a CSL contains both spinon excitations and gauge fluctuations -- both are gapped, due to finite spinon mass and a nonzero spinon Chern number. A third type, the amplitude fluctuations of the spinon hoppings, are often not discussed, as they are irrelevant to the description of a homogeneous CSL and are suppressed in the usual large-$N$ treatment. Nevertheless, the amplitude fluctuations can provide important information for a spatially varying CSL state.

\begin{figure}[t]
\centering 
\includegraphics[width=1\columnwidth]{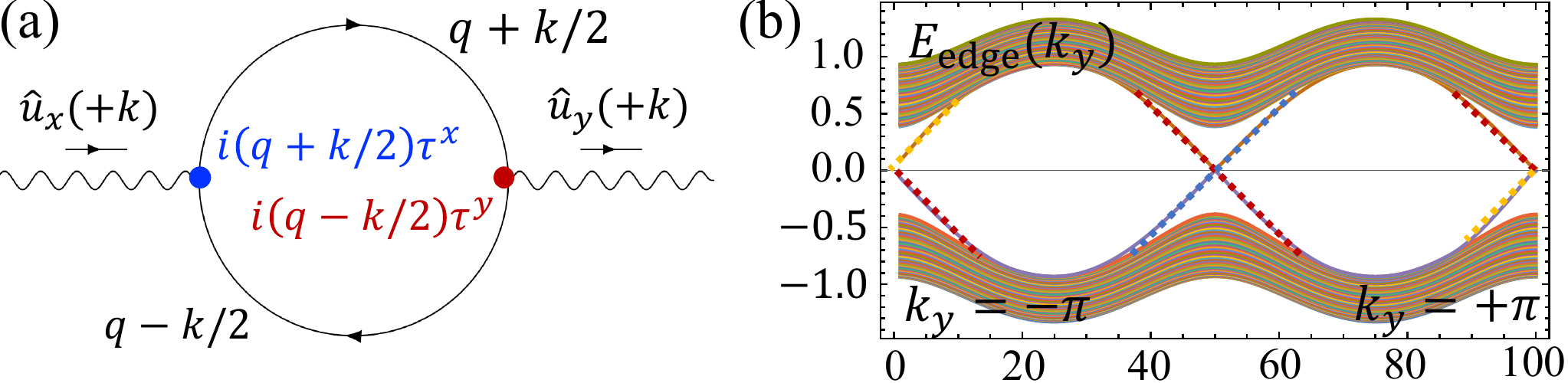}
\caption{\label{EFTEdge} (a) Feynman diagram for spinon-amplitude fluctuation at one-loop level.
(b) Spinon energy spectrum corresponding to Fig.~\ref{DomainWallPattern}(a). Note that there are in total four linear dispersions in the plot. The two linear dispersions marked by red (each has double degeneracy) correspond to the four spinon edge modes right at the domain wall. The blue (orange) dotted lines correspond to the edge modes on the boundary on the far left (far right).
}
\end{figure}

To obtain the theory for the amplitude fluctuations, we first expand the Hamiltonian \eqref{BlochBand} around zero momentum $(k_x,k_y)=(0,0)$. After dropping the constant terms and a proper rotation for Pauli matrices, we arrive at: 
\begin{equation}\label{ContinuousLimit}
	h_0 \sim [(u_x^0 + \hat u_x)  \tilde k_x \tau_x + (u_y^0 + \hat u_y) \tilde  k_y \tau_y + (u_z^0 + \hat m_z) \tau_z],
\end{equation}
where $(\tilde k_x, \tilde k_y)\!=\!(k_x a_0, k_y a_0)$ is the dimensionless momentum. The deviation $(\hat u_x, \hat u_y ,\hat m_z)  $ from the self-consistent ansatz in the homogeneous limit $(u_x^0, u_y^0 , u_z^0) = (2J_1 \bar \rho_x , -2J_1 \bar \rho_y , 4J_2 \bar \lambda)$ captures the amplitude fluctuations around the energy minima.  Next, we integrate out the spinons and keep the effective field theory to quadratic level; this requires evaluating the one-loop diagram as shown in Fig.~\ref{EFTEdge}~\footnote{The main feature of the amplitude fluctuation vertices, compared to the usual gauge vertices, is that they are linear in momentum~\cite{Supp}.}. Treating the deviation as a three-component field $\Psi = (\hat u_x, \hat u_y, \hat m_z )^{\rm T}$, and assuming $\bar \rho_0 = \bar \rho_x = \bar \rho_y$, we arrive at:
\begin{equation}\label{quadratictheory}
		S_{\rm eff} \simeq \frac{\hbar}{2(u_x^0)^2} \int [d^3\tilde k] \Psi^\dagger [{\mathcal M}_0 + {\mathcal T}_{\tilde k} ] \Psi,
\end{equation}
with the mass matrix ${\mathcal M}_0$ and the stiffness matrix ${\mathcal T}_{\tilde k}$. The detailed form of ${\mathcal M}_0$ and ${\mathcal T}_{\tilde k}$ can be found in \cite{Supp}. We define the dimensionless parameter $m\!=\!|u_z^0 /u_x^0| \!=\!|2J_2 \bar \lambda/ J_1 \bar \rho_0|$.
The energy at zero temperature of the field is derived by dropping the frequency dependence. Assuming translation symmetry along the $y$ direction as shown in Fig.~\ref{Illustration}(c), we set $k_y\!=\! 0$ in order to minimize the energy. Transforming back to real space, we obtain the energy of the fluctuations around one local minimum
\begin{equation}\label{FluctuationEnergy}
	\begin{aligned}
		F_T \!\simeq\!& \frac{m  }{u_x^0 a_0}\int dx \bigg{\{} m^2( 16 \hat u_y^2 + 24\hat u_y \hat m_z \!+\! 21 \hat m_z^2) \\
		&+ m^2(16 \hat u_x^2 \!+\! 8\hat u_x \hat u_y \!+\! 24\hat u_x \hat m_z) \\
		&+ a_0^2[2(\partial_x \hat u_y)^2\! +\! 2(\partial_x \hat u_y)(\partial_x \hat m_z) \!+\! \frac{9}{4}(\partial_x \hat m_z)^2]  \bigg{\}}.\quad
	\end{aligned}
\end{equation}
Since ${\mathcal M}_0$ and ${\mathcal T}_{\tilde k}$ are positive definite, any non-zero fluctuations would increase the energy. 

Recall that our inhomogenous ansatz for the two-domain CSL system has constant nearest-neighbor hopping $\bar{\rho}_{x,y}$, and the change of chirality is entirely due to the spatial variation of the next-nearest neighbor $\lambda$. Motivated by this fact, we focus our attention on the fluctuation of $\hat{m}_z$, keeping $\hat u_x\!=\!\hat u_y \!=\!0$ \footnote{We have further justified the legitimacy of this in \cite{Supp}}.
This reduces our effective theory to that of a single scalar field $\hat m_z$.
To restore the energy for a domain wall configuration in the full order parameter space, we plug back $(u_x,u_y,m_z)\!=\!(u_x^0 , u_y^0 , u_z^0 + \hat m_z)$, with each $(u_x,u_y,m_z)$ a parameter point $(\rho_x,\rho_y,\lambda)\!=\!(u_x/2J_1,-u_y/2J_1,m_z/4J_2) $ in Fig.~\ref{Illustration}(b). In the following,  we go beyond the quadratic theory of Eq.~\eqref{quadratictheory} and examine higher order fluctuations of $\hat{m}_z$. We hope that these higher order fluctuations can provide information for not only fluctuations around local minima, but also tunneling between different minima.

{\it Generalized Ginzburg--Landau theory--} Consider the following  Ginzburg--Landau  functional for the spatial varying scalar order parameter $m_z(x) = u_z^0 + \hat m_z$
\begin{equation}\label{GeneralizedGL}
    F[m_z] \simeq 
   \frac{m}{u_x^0 a_0} \int dx [K(\partial_x m_z)^2 + V(m_z)],
\end{equation}
where the potential $V(m_z)$, as illustrated in Fig.~\ref{Illustration}(d), has two degenerate local minima $\pm u_z^0$. The parameter $K$ denotes the stiffness of field $m_z$, and can be read from Eq.~\eqref{FluctuationEnergy} as $K = 9a_0^2/4 $. We use Eq.~\eqref{GeneralizedGL} to model the formation of domain wall in our CSL system: it captures the tunneling from the local minimum with negative chirality (blue points in Fig.~\ref{Illustration}(b)) to the local minimum with positive chirality (red points in Fig.~\ref{Illustration}(b)) in the order parameter space. The spatial profile of $V(m_z)$ can be identified with that of $\epsilon(\lambda)$ given in Fig.~\ref{Illustration}(d) numerically.

Assuming the spatial variation of the chirality domain wall is smooth enough, we can divide the domain wall into multiple stripes, each with width $\delta x$, as shown in Fig.~\ref{Illustration}(f). The set of stripes in Fig.~\ref{Illustration}(f) from the far left to the far right corresponds to a family of locally homogeneous ansatze whose mean-field parameters traverse the phase space (shown in Fig.~\ref{Illustration}(b)) from the $-\chi_0$-chirality minimum to the $+\chi_0$-chirality minimum. If there are multiple paths for changing the chirality, the realistic path will be the one with the lowest energy cost. Naively, the domain wall with minimal energy cost could be the shortest cut in the phase space, as illustrated by the path $4\rightarrow 7 \rightarrow 1$ in Fig.~\ref{Illustration}(b).  We choose to use the standard $\phi^4$ theory $V(m_z) \sim {\mathcal M} m_z^2/2 + U_z m_z^4/4$, where the mass  ${\mathcal M} = -42 m^2$ can be read off from Eq.~\eqref{FluctuationEnergy}. The interaction $U_z$ can be fixed by solving $\frac{dV(m_z)}{dm_z}\big|_{u_z^0} = 0$, i.e., $U_z = 42/(u_z^0)^2$.

Solving the equation of motion under the boundary condition $m_z(0) = -u_z^0$ and $m_z(L) = +u_z^0$ gives the domain wall solution
\begin{equation}\label{GLMeanFieldSolution}
	u_{x,y}(x) \!=\! u_{x,y}^0, ~ m_z(x) \!=\! u_z^0 \tanh[(x-x_0)/\sqrt{2}\xi_z],
\end{equation}
with the domain wall width $\xi_z \!=\! \sqrt{-K/{\mathcal M}} \propto a_0/m \sim u_x^0 a_0/\Delta$, where $\Delta$ is the bulk gap. Eqs.~\eqref{GeneralizedGL} and \eqref{GLMeanFieldSolution} capture the energy in presence of the domain wall, and the domain wall tension ${\mathcal E}$ is the difference between the results above and that in the homogeneous limit. A direct calculation shows that ${\mathcal E}\propto m_0^5 J_1 \ll J_1 \sim J_2$, which is much smaller than the bulk spin gap~\cite{Supp}.

We point out that the gapless edge modes on the domain wall can introduce additional terms to the Ginzburg--Landau free energy. Specifically, in our case a $|m_z^3|$ term appears. To see this, we treat the mass term in Fig.~\ref{Illustration}(f) as constant within each stripe, then the potential $V(m_z)$, according to Eq.~\eqref{Energy}, can be approximated by a massive Dirac fermion $ V(m_z) \sim m_z^2 - \int_{\rm BZ} [d^2k] \sqrt{k^2 + m_z^2} \sim m_z^2 + \frac{1}{3} |m_z|^3 - \frac{1}{3} (k^2_0 + m_z^2)^{3/2}$, while $k_0$ is the cutoff from the BZ. We verify this by computing the energy $\epsilon(\lambda)$ along the path $4\rightarrow 7 \rightarrow 1$ and taking the third derivative with respect to $\lambda$. A discontinuity of $\epsilon^{(3)}(\lambda)$ indeed appears around $\lambda=0$ (orange curve  in Fig.~\ref{Illustration}(d)), indicating the existence of the $|\lambda^3|$ term. 
While it is hard to numerically identify the $|m_z^3|$ term in domain profiles accessible at our system size, this term will have observable effects.

{\it Edge states--} Substituting Eq.~\eqref{GLMeanFieldSolution} back to Eq.~\eqref{ContinuousLimit}, we arrive at the low-energy theory of spinons associated with each Dirac cone:
\begin{equation}\label{EdgeHamiltonian}
	H_D\!\simeq\!\int dx dk_y \{ \psi^\dagger_{k_y} [u_x^0 (-i\partial_x) \tau_x\!+ \!u_y^0 k_y\tau_y\!+\!m_z(x) \tau_z] \psi_{k_y} \}, 
\end{equation}
where $m_z(x)$ change sign at $x_0$ from $x = 0$ to $x= L$, see Fig.~\ref{Illustration}(c,e). Here $\psi_{k_y}$ is a two component spinor. According to the Jackiw-Rebbi mechanism~\cite{Jackiw1976,Qi2011}, there will be edge states localized at $x= x_0$, with the dispersion:
\begin{equation}
	E_{\rm edge}(k_y) = - v_{\rm edge} k_y,  
\end{equation}
with $ v_{\rm edge} = |2J_1 \bar \rho_y a_0/\hbar|$ the edge velocity. Note that the domain wall hosts four $S=1/2$ spinon edge modes, and the full edge theory is described by four copies of Eq.~\eqref{EdgeHamiltonian}, as shown in Fig.~\ref{EFTEdge}(b). By applying the non-double occupancy condition \eqref{SelfConsistentRequirement} one recovers the physical $S\!=\!1$ excitation from the two $S\!=\!1/2$ spinon edge modes from each domain \cite{Supp,Lu2012B,Lai2013}, and the edge excitation has total spin $S=2$.

{\it Conclusion--} We studied the domain wall for $\nu =1/2$ Kalmeyer-Laughlin chrial spin liquid with opposite chiralities. Starting from a spatially varying, self-consistent mean-field ansatz for spinons, we mapped out the spatial profile of the domain wall in terms of the spin chirality. We further proposed an effective Ginzburg--Landau field theory for  mean-field bond amplitude fluctuations which qualitatively produces the change of chirality across the domains, from which we derived the non-topological properties of the multi-domain system, such as the domain wall tension and edge velocity. 
We argue that the gapless modes at the edge contribute an extra, non-analytic, $|\phi^3|$ term to the domain wall theory. As shown in Fig.~\ref{DomainWallPattern}(d), in our toy model, the non-universal features, such as the bulk spinon gap $\Delta$, domain wall width $\xi$ and edge spinon velocity $v_{\rm edge}$, are all determined by the fundamental exchange parameters $J_1$ and $J_2$. While there is no direct way of measuring the exchange strength, other quantities, such as the spinon gap, can be directly read out from e.g. neutron scattering experiments. The experimental determination of any quantities above would provide information about the others.

Note that, although our work focused on a simple $J_1-J_2$ model for CSL on a square lattice, the self-consistent mean-field numerics and Ginzburg--Landau analysis can be carried out in many other contexts, such as other lattices.   It is also possible to extend the Ginzburg--Landau analysis to a larger space of magnetic order parameters, as was recently considered in~\cite{HUANG2021}. Our approach could be applied to the features of other multi-domain systems with coexisting conventional order and topological order, such as Pfaffian\---anti-Pfaffian-domains in the $\nu =5/2$ fractional quantum Hall system~\cite{Zaletel2013,Zhu2020b,Crepel2019b,Crepel2019a} and spontaneous time-reversal symmetry breaking domains in fractional topological insulators~\cite{Neupert_2015,Kourtis2014,Shavit2022}. Another example would be the excitonic condensed phase of quantum  Hall insulators~\cite{Jia2022}, where the order parameter for exciton condensation may have a similar profile as that of the spin chirality in chiral spin liquids.





{\it Acknowledgements--} This work was supported as part of the Center for Novel Pathways to Quantum Coherence in Materials, an Energy Frontier Research Center funded by the U.S. Department of Energy, Office of Science, Basic Energy Sciences (Y.-Q.W. and J.E.M.). C.L. was supported by the EPiQS program of the Gordon and Betty Foundation, and J.E.M. acknowledges support from a Simons Investigatorship.  We thank Dung-Hai Lee, Yuan-Ming Lu, Jason Alicea, Donna Sheng, Michael Zaletel, Leon Balents, Shubhayu Chatterjee, Michal Papaj, Aaron Szasz and Vir Bulchandani for helpful discussion.



\bibliography{apssamp}


\noindent

\onecolumngrid

\renewcommand\theequation{S\arabic{equation}}
\renewcommand\thefigure{S\arabic{figure}}
\renewcommand\bibnumfmt[1]{[S#1]}
\setcounter{equation}{0}
\setcounter{figure}{0}

\section*{Supplementary Material}

\section{More details of the mean-field theory}\label{SecMeanField}
In this section, we first briefly review the mean-field parton construction based on Ref.~\cite{Wenbook,Fradkinbook}. Consider the following spin model:
\begin{equation}\label{SpinHamiltonian}
	H = \sum_{\langle {\bm i} {\bm j} \rangle} J_{{\bm i}{\bm j}}{\bm S}_{\bm i} \cdot {\bm S}_{\bm j}.
\end{equation}
We will use the mean-field approximation to understand its physical properties: let ${\bm S}_{\bm i} = \langle  {\bm S}_{\bm i}\rangle + \delta {\bm S}_{\bm i}$, and we obtain the following mean-field Hamiltonian:
\begin{equation}
	H_{\rm mean} = \sum_{\langle {\bm i} {\bm j} \rangle} J_{{\bm i} {\bm j}} (\langle {\bm S}_i \rangle \cdot{} {\bm S}_j + {\bm S}_i \cdot{} \langle {\bm S}_{\bm j}\rangle - \langle {\bm S}_{\bm i}\rangle \langle {\bm S}_j \rangle).
\end{equation} 
We introduce the spinon operators $f_{{\bm i}\alpha}$, $\alpha = 1,2$, which are spin-1/2 charge-neutral fermionic operators. The spin operator ${\bm S}_{\bm i}$ is represented by:
\begin{equation}
	{\bm S}_{\bm i} = \frac{1}{2} f^\dagger_{{\bm i}\alpha} {\bm \sigma}_{\alpha \beta} f_{{\bm i} \beta}.
\end{equation}
Substitute this into Eq.~\eqref{SpinHamiltonian}:
\begin{equation}\label{SpinonHamiltonian}
	\begin{aligned}
		H &= \sum_{\langle {\bm i} {\bm j} \rangle} J_{{\bm i}{\bm j}} {\bm S}_{\bm i} \cdot{} {\bm S}_{\bm j} = \sum_{\langle {\bm i}{\bm j} \rangle} \frac{J_{{\bm i}{\bm j}}}{4} f^\dagger_{{\bm i}\alpha} {\bm \sigma}_{\alpha \beta} f_{{\bm i} \beta} f_{{\bm j}\alpha^\prime}^\dagger {\bm \sigma}_{\alpha^\prime \beta^\prime} f_{{\bm j} {\beta}^\prime} = \sum_{\langle {\bm i} {\bm j} \rangle} - \frac{J_{{\bm i}{\bm j}}}{2} f^\dagger_{{\bm i}\alpha} f_{{\bm j} \alpha} f^\dagger_{{\bm j} \beta} f_{{\bm i} \beta} + \sum_{\langle {\bm i} {\bm j} \rangle} J_{{\bm i} {\bm j}} \bigg{(} \frac{1}{2} n_{\bm i} - \frac{1}{4} n_{\bm i} n_{\bm j} \bigg{)}.
	\end{aligned}
\end{equation}
Here, we have used ${\bm \sigma}_{\alpha \beta} \cdot{} {\bm \sigma}_{\alpha^\prime \beta^\prime} = 2\delta_{\alpha \beta^\prime} \delta_{\alpha^\prime \beta} - \delta_{\alpha \beta} \delta_{\alpha^\prime \beta^\prime}$ and $n_{\bm i}$ is the number of fermions at site ${\bm i}$. The second term in Eq.~\eqref{SpinonHamiltonian} is a constant and will be dropped in the following discussions. Notice that the Hilbert space of Eq.~\eqref{SpinonHamiltonian} with four states per site is larger than Eq.~\eqref{SpinHamiltonian}, which has two states per site. The equivalence between Eq.~\eqref{SpinonHamiltonian} and Eq.~\eqref{SpinHamiltonian} is valid only in the subspace where there is exactly one fermion per site. Therefore, to use Eq.~\eqref{SpinonHamiltonian} to describe the spin state, we need to impose the constraint:
\begin{equation}\label{Constraint}
	f^\dagger_{{\bm i} \alpha} f_{{\bm i}\alpha} = f^\dagger_{{\bm i} 1} f_{{\bm i}1}  + f^\dagger_{{\bm i} 2} f_{{\bm i}2} = 1, \quad f_{{\bm i} \alpha} f_{{\bm i} \beta} \epsilon_{\alpha \beta} = f_{{\bm i}1}f_{{\bm i}2} + f_{{\bm i}2} f_{{\bm i}1}  =0. 
\end{equation}

\begin{figure}[!h]
\centering 
\includegraphics[width=1\columnwidth]{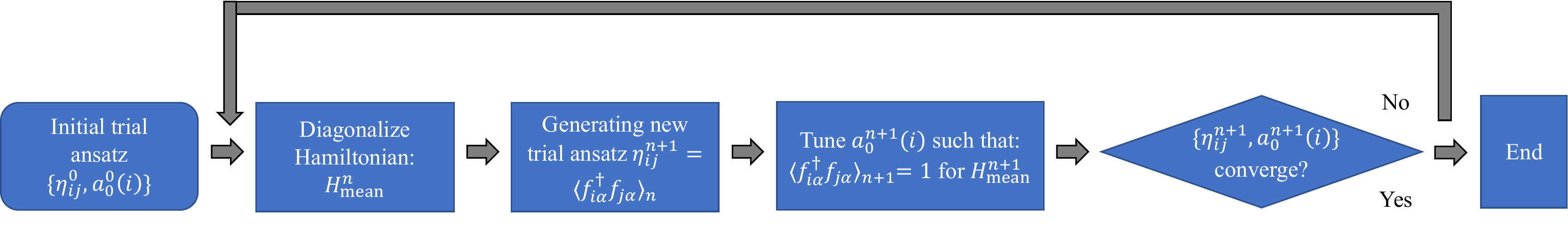}
\caption{\label{SelfConsistentProcedure} The iterative self-consistent procedure. The convergence condition in the last step is set to $\sum_{ij} |\eta_{{\bm i} {\bm j}}^{n+1} - \eta_{{\bm i} {\bm j}}^{n}| < \delta$ and $\sum_i |a_0^{n+1}(i) - a_0^n(i)|< \delta$, while $\delta$ is a positive number which decides the accuracy. }
\end{figure}

A mean-field ground state at zeroth-order is obtained by making the following two approximations, we first replace Eq.~\eqref{Constraint} by its ground-state average:
\begin{equation}\label{GroundStateAverage}
	\langle f^\dagger_{{\bm i} \alpha} f_{{\bm i} \alpha} \rangle = 1.
\end{equation}
such a constraint can be enforced by including a site-dependent and time independent Lagrangian multiplier $a_0({\bm i}) (f^\dagger_{{\bm i} \alpha} f_{{\bm i} \alpha} - 1)$ in the Hamiltonian. Second, we replace the operator $f^\dagger_{{\bm i} \alpha} f_{{\bm j}\alpha}$ by its ground state expectation value $\chi_{{\bm i} {\bm j}}$, again ignoring their fluctuations. In this way, we obtain the zeroth-order mean-field Hamiltonian:
\begin{equation}\label{ZerothOrder}
	H_{\rm mean} = \sum_{\langle {\bm i} {\bm j} \rangle} - \frac{1}{2} J_{{\bm i} {\bm j}} \bigg{[} (f^\dagger_{{\bm i} \alpha} f_{{\bm j}\alpha} \eta_{{\bm j} {\bm i}} + {\rm h.c.}) - |\chi_{{\bm i} {\bm j}}|^2 \bigg{]} + \sum_i a_0({\bm i}) (f^\dagger_{{\bm i} \alpha} f_{{\bm i} \alpha} - 1).
\end{equation}
The $\eta_{{\bm i} {\bm j}}$ in Eq.~\eqref{ZerothOrder} must satisfy the self-consistency condition:
\begin{equation}\label{SelfConsistency}
	\eta_{{\bm i} {\bm j}} =\langle f^\dagger_{{\bm i} \alpha} f_{{\bm j} \alpha} \rangle,
\end{equation}
and the site-dependent chemical potential $a_0({\bm i})$ is chosen such that the Eq.~\eqref{GroundStateAverage} is satisfied by the mean-field ground state. We call the pattern of $\eta_{{\rm i}{\rm j}}$ the mean-field ansatz. Then we conduct the iterative process shown in Fig.~[\ref{SelfConsistentProcedure}].

\section{mean-field ansatz for chiral spin liquid}\label{LatticeModel}
\begin{figure}[!h]
\centering 
\includegraphics[width=0.55\columnwidth]{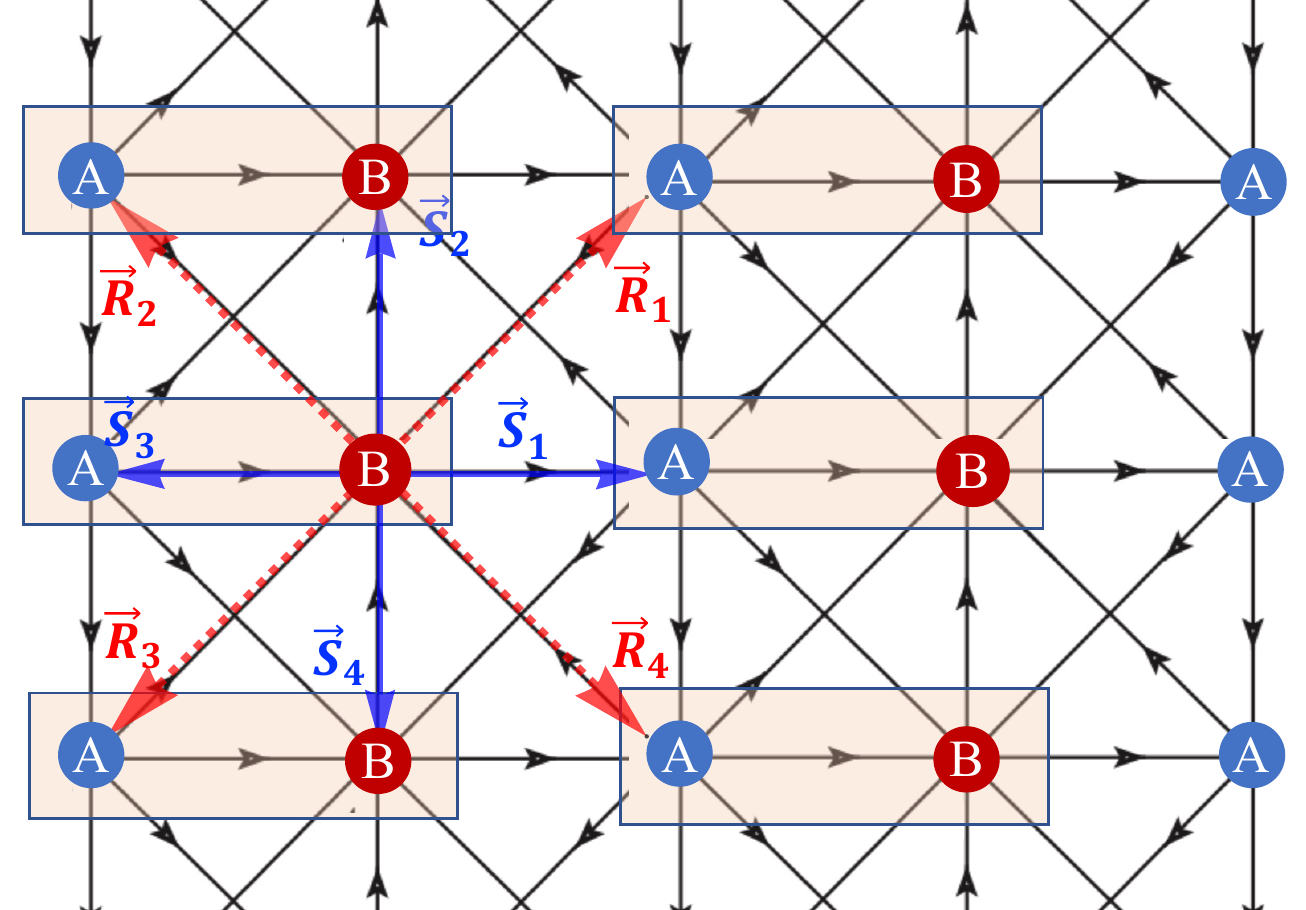}
\caption{\label{ChiralSpinLiquid} Gauge-field conventions for a chiral spin state on a frustrated square lattice based on Fig.~[1.(a)] in the main text~\cite{Wenbook,Fradkinbook}. For each unit cell, there are four sites: $A,B,C,D$. The nearest neighbors are labeled by ${\bm S}_1 = -{\bm S}_3 = (a_0, 0)$, ${\bm S}_2 = - {\bm S}_4 = (0,a_0)$, and the nearest neighbors are labeled by ${\bm R}_1 = -{\bm R}_3 = (a_0 ,a_0)$, ${\bm R}_2 = -{\bm R}_4 = (-a_0, a_0)$, with $a_0$ the distance between two nearest sites. The black arrows on the links represent a phase of $\pi/2$. Since charge for each electron is $-e$ with $e>0$, the Pieles substitution reads: $t_{ij} \rightarrow t_{ij} e^{+i \frac{q}{\hbar} \int_{i}^j {\bm A} \cdot{} d{\bm l}} = t_{ij} e^{-i\frac{e}{\hbar} \int_i^j {\bm A} \cdot{} d{\bm l}}$. Thus along the black arrows we get an additional factor $-i$ compared with the hopping in the absence of flux.}
\end{figure}

Consider the spin-1/2 model defined on the square lattice, with nearest neighbor coupling $J_1$ and next nearest neighbor coupling $J_2$, as the Fig.~[1(a)] in the main text. The geometric detail is shown in Fig.~[\ref{ChiralSpinLiquid}]. The mean-field Hamiltonian based on Eq.~\eqref{ZerothOrder} is given by:
\begin{equation}\label{ModelHamiltonian}
	H = H_1 + H_2,
\end{equation}
with the nearest neighbor terms,
\begin{equation}
	\begin{aligned}
		H_1 \!=\!& -\sum_{i, \sigma} \{ [J_1^x \eta_{ba}^1({\bm r}_i^a) \hat C^\dagger_{a,\sigma}({\bm r}_i^a) \hat C_{b,\sigma}({\bm r}_i^a + {\bm S}_1) + J_1^y \eta_{aa}^2({\bm r}_i^a) \hat C^\dagger_{a,\sigma}({\bm r}_i^a) \hat C_{a,\sigma} ({\bm r}_i^a + {\bm S}_2)+ {\rm H.c.}] + \frac{J_1^x}{2} |\eta^1_{ba}({\bm r}_i^a)|^2 + \frac{J_1^y}{2} |\eta^2_{aa}({\bm r}_i^a)|^2  \}\\
		&- \sum_{i, \sigma} \{ [J_1^x \eta_{ab}^1({\bm r}_i^b) \hat C^\dagger_{b,\sigma}({\bm r}_i^b) \hat C_{a,\sigma}({\bm r}_i^b + {\bm S}_1) + J_1^y \eta_{bb}^2({\bm r}_i^b) \hat C^\dagger_{b,\sigma}({\bm r}_i^b) \hat C_{b,\sigma} ({\bm r}_i^b + {\bm S}_2)+ {\rm H.c.}]  +\frac{J_1^x}{2} |\eta_{ab}^1({\bm r}_i^b)|^2 + \frac{J_1^y}{2} |\eta^2_{bb}({\bm r}_i^b)|^2  \}\\
	\end{aligned}
\end{equation}
and the second nearest neighbor terms,
\begin{equation}
	\begin{aligned}
		H_2 \!=\! & -\sum_{i,\sigma}\{ [J_2^+ \eta^3_{ba}({\bm r}_i^a)\hat C^\dagger_{a,\sigma}({\bm r}_i^a) \hat C_{b,\sigma}({\bm r}_i^a + {\bm R}_1)+J_2^- \eta^4_{ba}({\bm r}_i^a)\hat C^\dagger_{a,\sigma}({\bm r}_i^a) \hat C_{b,\sigma}({\bm r}_i^a + {\bm R}_4) + {\rm H.c.}] + \frac{J_2^+}{2} |\eta^3_{ba}({\bm r}_i^a)|^2 + \frac{J_2^-}{2} |\eta^4_{ba}({\bm r}_i^a)|^2  \}\\
		&-\sum_{i,\sigma} \{ [ J_2^+ \eta^3_{ab} ({\bm r}_i^b)\hat C^\dagger_{b,\sigma}({\bm r}_i^b) \hat C_{a,\sigma}({\bm r}_i^b + {\bm R}_1)+J_2^- \eta^4_{ab}({\bm r}_i^b)\hat C^\dagger_{b,\sigma}({\bm r}_i^b) \hat C_{a,\sigma}({\bm r}_i^b + {\bm R}_4) + {\rm H.c.}] + \frac{J_2^+}{2} |\eta^3_{ab}({\bm r}_i^b)|^2 + \frac{J_2^-}{2} |\eta^4_{ab}({\bm r}_i^b)|^2 \} \\
	\end{aligned}
\end{equation}
Details can be found in Ref.~\cite{Wenbook,Fradkinbook}.
For simplicity, we will drop the constant terms temporarily and only add them back in the free energy Eq.~\eqref{FreeEnergy}. In the homogeneous limit, the following ansatz represents a chiral spin liquid phase: 
\begin{equation}\label{WenAnsatz}
\begin{aligned}
	&\eta^1_{ba}({\bm r}_i^a) \equiv \eta^1_{ba} = + i\bar \rho_x^a, \quad \eta^1_{ab}({\bm r}_i^b) \equiv \eta^1_{ab} = + i \bar \rho^b_x, \quad \eta^2_{aa}({\bm r}_i^a) \equiv \eta^2_{aa} = - i\bar \rho_y^a, \quad \eta^2_{bb}({\bm r}_i^a)\equiv \eta^2_{bb} = + i \bar \rho_y^b, \\
	&\eta^3_{ba}({\bm r}_i^a) \equiv \eta_{ba}^3 =  + i \bar \lambda_+^a, \quad \eta^3_{ab}({\bm r}_i^b) \equiv \eta_{ab}^3 = -i\bar \lambda_+^b, \quad \eta^4_{ba}({\bm r}_i^a) \equiv \eta_{ba}^4 = +i \bar \lambda^a_-, \quad \eta^4_{ab}({\bm r}_i^b) \equiv \eta_{ab}^4 = -i\bar \lambda_-^b.
\end{aligned}
\end{equation}
Introducing the Fourier transformation,
\begin{equation}
	\hat C_a({\bm r}_j) = \frac{1}{\sqrt{N}} \sum_{\bm k} e^{+ i {\bm k} \cdot {\bm r}_j} \hat C_a({\bm k}),  \quad \hat C_{b}({\bm r}_j) = \frac{1}{\sqrt{N}} \sum_{{\bm k}}  e^{+i {\bm k} \cdot{} {\bm r}_j} \hat C_b({\bm k}),
\end{equation}
upon dropping the constant terms, we shall have:
\begin{equation}
	H = \sum_{\sigma, {\bm k}} \begin{pmatrix}
			\hat C^\dagger_{a,\sigma}({\bm k}) & \hat C^\dagger_{b,\sigma}({\bm k})
		\end{pmatrix}  	(h_1 + h_2)
		\begin{pmatrix}
			\hat C_{a,\sigma}({\bm k}) \\
			\hat C_{b,\sigma}({\bm k})
		\end{pmatrix}
\end{equation}
with the kernel:
\begin{equation}
\begin{aligned}
	h_1 &\!=\! -J_1^x (-\bar \rho_x^a + \bar \rho^b_x) \cos ({\bm k} \cdot{} {\bm S}_1)\sigma_y +J_1^x (\bar \rho_x^a + \bar \rho_x^b) \sin ({\bm k} \cdot  {\bm S}_1)\sigma_x -J_1^y(\bar \rho^a_y - \bar \rho_y^b) \sin ({\bm k} \cdot{} {\bm S}_2) \sigma_0 -J_1^y(\bar \rho_y^a + \bar \rho^b_y) \sin ({\bm k}\cdot{} {\bm S}_2)\sigma_z \\
	h_2 &\!=\![ J_2^+ (\bar \lambda_+^a + \bar \lambda^b_+) \cos({\bm k} \cdot{} {\bm R}_1) +  J_2^-(\bar \lambda^a_- + \bar \lambda^b_-) \cos ({\bm k} \cdot{} {\bm R}_4)]\sigma_y +[J_2^+(\bar \lambda_+^a -\bar \lambda_+^b)\sin ({\bm k} \cdot{} {\bm R}_1) + J_2^- (\bar \lambda^a_- - \bar \lambda^b_-) \sin ({\bm k} \cdot{} {\bm R}_4)]\sigma_x.
\end{aligned}
\end{equation}
We can further take the limit for each unit cell the parameters take the same value or $a$ and $b$:
\begin{equation}
	\bar \rho_x^a = \bar \rho_x^b = \bar \rho_x^0, \quad \bar \rho^a_y = \bar \rho_y^b = \bar \rho_y^0 , \quad  J_2^+ = J_2^-\quad \bar \lambda_+^a = \bar \lambda_+^b = \bar  \lambda_+^0, \quad \bar \lambda_-^a = \bar \lambda_-^b = \bar \lambda_-^0.
\end{equation}
The Bloch Hamiltonian can be further simplified as:
\begin{equation}
\begin{aligned}
	h_0 &\!=\! 2J_1^x \bar \rho_x^0 \sin ({\bm k} \cdot{} {\bm S}_1) \sigma_x + [2J_2^+ \bar \lambda^0_+ \cos({\bm k} \cdot{} {\bm R}_1) + 2J_2^- \bar \lambda_-^0 \cos({\bm k} \cdot{} {\bm R}_4) ]\sigma_y - 2J_1^y \bar \rho_y^0 \sin ({\bm k} \cdot{} {\bm S}_2) \sigma_z \\
	&\!=\! 2J_1^x \bar \rho^0_x \sin(k_x a_0) \sigma_x + [2J_2^+ \bar \lambda^0_+ \cos(k_x a_0 + k_y a_0) + 2J_2^- \bar \lambda^0_- \cos(k_x a_0 - k_y a_0)] \sigma_y - 2J_1^y \bar \rho^0_y \sin(k_y a_0) \sigma_z \\
	&\!=\! 2J_1^x \bar \rho_x^0 \sin(k_x a_0) \sigma_x + 2[J_2^+ \bar \lambda^0_+ +J_2^- \bar \lambda^0_-] \cos k_x a_0 \cos k_y a_0 \sigma_y - 2[J_2^+ \bar \lambda^0_+-J_2^- \bar \lambda_-^0]\sin k_x a_0 \sin k_y a_0 \sigma_y -2J_1^y \bar \rho^0_y \sin (k_y a_0) \sigma_z.
\end{aligned}
\end{equation}
In the limit $J_2^+ = J_2^- = J_2$ such that $\lambda^0_+ = \lambda^0_- = \lambda^0$, with the constants added back, we arrive at:
\begin{equation}\label{BlochHamiltonian}
\begin{aligned}
	h_0(k) &= J_1|\bar \rho_x|^2 + J_1 |\bar \rho_y|^2 + 2 J_2 |\bar \lambda|^2  + 2J_1  \rho_x \sin (k_x a_0)\sigma_x + 4J_2  \lambda \cos (k_x a_0) \cos(k_y a_0) \sigma_y - 2J_1 \rho_y \sin (k_y a_0) \sigma_z,
\end{aligned}
\end{equation}
which gives the gap at $(k_x,k_y)=(0,0)$:
\begin{equation}
	\Delta = 2\times 4J_2 \bar \lambda^0 = 8 J_2 \bar \lambda^0.
\end{equation}

\subsection{Self-consistency solution in momentum space}\label{SectionSCSMomentum}
For a two level system (whose energy is normalized to 1):
\begin{equation}
	{\mathcal H}_A = \sin \theta \cos \phi \sigma_x + \sin \theta \sin \phi \sigma_y + \cos \theta \sigma_z,
\end{equation}
we shall have the eigenstates:
\begin{equation}
	\begin{aligned}
		\epsilon = -1, \quad \psi_- =\begin{pmatrix}
			\sin \bigg{(} \frac{\theta}{2} \bigg{)} e^{-i\phi} \\
			-\cos \bigg{(} \frac{\theta}{2} \bigg{)},
		\end{pmatrix}, \quad \epsilon = +1, \quad \psi_+ = \begin{pmatrix}
			\cos \bigg{(} \frac{\theta}{2}\bigg{)} e^{-i \phi}, \\
			\sin \bigg{(} \frac{\theta}{2} \bigg{)}
		\end{pmatrix}.
	\end{aligned}
\end{equation} 
Compared with Eq.~\eqref{BlochHamiltonian}, we can define
\begin{equation}
	\begin{aligned}
	    \hat d_x(k_x,k_y) = \sin \theta \cos \phi &= +\frac{2J_1^x \bar \rho_x^0 \sin(k_x a_0) }{\sqrt{[2J_1^x \bar \rho_x^0 \sin(k_x a_0) ]^2 +[4J_2\lambda^0 \cos(k_x a_0) \cos(k_y a_0)]^2 + [2J_1^y \bar \rho_y^0 \sin(k_y a_0) ]^2 }} \\
	    \hat d_y(k_x,k_y) = \sin \theta \sin \phi &= +\frac{4J_2\lambda^0 \cos(k_x a_0) \cos(k_y a_0)}{\sqrt{[2J_1^x \bar \rho_x^0 \sin(k_x a_0) ]^2 +[4J_2\lambda^0 \cos(k_x a_0) \cos(k_y a_0)]^2 + [2J_1^y \bar \rho_y^0 \sin(k_y a_0) ]^2 }} \\
		\hat d_z(k_x,k_y) = \cos \theta &=  -\frac{2J_1^y \bar \rho_y^0 \sin(k_y a_0) }{\sqrt{[2J_1^x \bar \rho_x^0 \sin(k_x a_0) ]^2 +[4J_2\lambda^0 \cos(k_x a_0) \cos(k_y a_0)]^2 + [2J_1^y \bar \rho_y^0 \sin(k_y a_0) ]^2 }}.\\
	\end{aligned}
\end{equation}
Then we shall have the self consistency equations for the lower subband, 
\begin{equation}\label{SE1}
	\begin{aligned}
		i \bar \rho_x^a = i\bar \rho_x^0\equiv \eta^1_{ba} &= \sum_{\sigma} \int_{-\pi}^{+\pi} \frac{dk_y}{2\pi} \int_{-\pi/2}^{+\pi/2} \frac{dk_x}{\pi} e^{-ik_x a_0} \bigg{[} -\cos \bigg{(} \frac{\theta}{2} \bigg{)}\bigg{]}^* \bigg{[} \sin \bigg{(} \frac{\theta}{2} \bigg{)} e^{- i\phi}\bigg{]} \\
		&= \int_{-\pi}^{+\pi} \frac{dk_y}{2\pi} \int_{-\pi/2}^{+\pi/2} \frac{dk_x}{\pi} \frac{i 2J_1^x \bar \rho_x^0\sin^2 (k_x a_0)}{\sqrt{[2J_1^x \bar \rho_x^0 \sin(k_x a_0) ]^2 +[4J_2\lambda^0 \cos(k_x a_0) \cos(k_y a_0)]^2 + [2J_1^y \bar \rho_y^0 \sin(k_y a_0) ]^2 }},
	\end{aligned}
\end{equation}
and similarly, for the $\eta_{aa}^2$ and $\eta^3_{ba}$
\begin{equation}\label{SE2}
\begin{aligned}
	-i\bar \rho_y^a  &= \int_{-\pi}^{+\pi} \frac{dk_y}{2\pi} \int_{-\pi/2}^{+\pi/2} \frac{dk_x}{\pi} \frac{ - i 2J_1^y \bar \rho_y^0 \sin^2(k_y a_0) }{\sqrt{[2J_1^x \bar \rho_x^0 \sin(k_x a_0) ]^2 +[4J_2\lambda^0 \cos(k_x a_0) \cos(k_y a_0)]^2 + [2J_1^y \bar \rho_y^0 \sin(k_y a_0) ]^2 }} \\
	i\bar \lambda^0 &=\int_{-\pi}^{+\pi} \frac{dk_y}{2\pi} \int_{-\pi/2}^{+\pi/2} \frac{dk_x}{\pi} \frac{i4J_2\lambda^0 \cos^2 (k_xa_0)\cos^2(k_ya_0)}{\sqrt{[2J_1^x \bar \rho_x^0 \sin(k_x a_0) ]^2 +[4J_2\lambda^0 \cos(k_x a_0) \cos(k_y a_0)]^2 + [2J_1^y \bar \rho_y^0 \sin(k_y a_0) ]^2}}.
\end{aligned}
\end{equation}
With above, we have the following recursion relation:
\begin{equation}\label{BlochRecursion}
	\begin{aligned}
		\bar \rho^0_{x}(n+1) &= \int_{-\pi}^{+\pi} \frac{dk_y}{2\pi} \int_{-\pi/2}^{+\pi/2} \frac{dk_x}{\pi} \frac{ 2J_1^x \bar \rho_x^0(n)\sin^2 (k_x a_0)}{\sqrt{[2J_1^x \bar \rho_x^0(n) \sin(k_x a_0) ]^2 +[4J_2\lambda^0(n) \cos(k_x a_0) \cos(k_y a_0)]^2 + [2J_1^y \bar \rho_y^0 (n)\sin(k_y a_0) ]^2 }} \\
		\bar \rho^0_y(n+1) &= \int_{-\pi}^{+\pi} \frac{dk_y}{2\pi} \int_{-\pi/2}^{+\pi/2} \frac{dk_x}{\pi} \frac{  2J_1^y \bar \rho_y^0(n) \sin^2(k_y a_0) }{\sqrt{[2J_1^x \bar \rho_x^0 (n)\sin(k_x a_0) ]^2 +[4J_2\lambda^0(n) \cos(k_x a_0) \cos(k_y a_0)]^2 + [2J_1^y \bar \rho_y^0(n) \sin(k_y a_0) ]^2 }} \\
		\bar \lambda^0(n+1) &=   \int_{-\pi}^{+\pi} \frac{dk_y}{2\pi} \int_{-\pi/2}^{+\pi/2} \frac{dk_x}{\pi} \frac{4J_2\lambda^0(n) \cos^2 (k_xa_0)\cos^2(k_ya_0)}{\sqrt{[2J_1^x \bar \rho_x^0(n) \sin(k_x a_0) ]^2 +[4J_2\lambda^0(n) \cos(k_x a_0) \cos(k_y a_0)]^2 + [2J_1^y \bar \rho_y^0(n) \sin(k_y a_0) ]^2}},
	\end{aligned}
\end{equation}
one can run the iterations for a number of times until reaching convergence as $n \rightarrow \infty$.

\subsection{Self-consistency and minimizing the mean-field energy}\label{SectionFreeEnergy}
The free energy per unit cell per spin reads:
\begin{equation}\label{FreeEnergy}
	\begin{aligned}
		F_\sigma(\rho_x^0,\rho_y^0,\lambda^0) =& J_1^x |\rho^0_x|^2 + J_1^y |\rho^0_y|^2 + 2J_2|\lambda^0|^2 \\
		-& \int_{-\pi}^{+\pi} \frac{dk_y}{2\pi} \int_{-\pi/2}^{+\pi/2} \frac{dk_x}{\pi} \sqrt{[2J_1^x \bar \rho^0_x \sin(k_x a_0)]^2 + [4J_2 \lambda^0 \cos(k_x a_0) \cos(k_y a_0)]^2 + [2J_1^y \bar \rho_y^0 \sin(k_y a_0)]^2}, 
	\end{aligned}
\end{equation}
variational equations:
\begin{equation}
	\frac{\delta F}{\delta \rho_x^0} =0 , \quad \frac{\delta F}{\delta \rho_y^0} =0, \quad \frac{\delta F}{\delta \lambda^0} = 0,
\end{equation}
will give the same relation as given in the self-consistency equations Eq.~\eqref{SE1} and Eq.~\eqref{SE2}.

\subsection{Chern number and chirality}
Consider the following  two level system:
\begin{equation}
	h(k_x,k_y) =  \frac{f({\bm k}) \sigma_x + g({\bm k}) \sigma_y + h({\bm k}) \sigma_z}{\sqrt{f^2({\bm k}) + g^2({\bm k}) + h^2({\bm k})}}  = \hat d_x(k_x,k_y) \sigma_x + \hat d_y(k_x,k_y) \sigma_y + \hat d_z(k_x,k_y) \sigma_z, \quad \hat d_x^2 + \hat d_y^2 + \hat d_z^2 = 1. 
\end{equation}
The Chern number associated with the lower subband of Eq.~(6) in the main text is defined as~\cite{bernevig2013}
\begin{equation}
	C = \frac{1}{2\pi} \int [dk_x dk_y] {\mathcal F}_{xy}, \quad {\mathcal F}_{\mu \nu} = \frac{1}{2} \epsilon^{\alpha \beta \gamma} \hat d_\alpha({\bm k}) \partial_{k_\mu} \hat d_\beta({\bm k}) \partial_{k_\nu} \hat d_\gamma({\bm k}),
\end{equation} 
from which we have
\begin{equation}
	\begin{aligned}
		{\mathcal F}_{xy} 
	&= -\frac{8J_1^2J_2 \lambda \rho_x \rho_y [\cos^2(k_y) \sin^2(k_x) + \cos^2(k_x) \cos^2 (k_y) + \cos^2(k_x)\sin^2(k_y)]}{\{[2J_1^x \bar \rho_x \sin(k_x a_0) ]^2 +[4J_2\lambda \cos(k_x a_0) \cos(k_y a_0)]^2 + [2J_1^y \bar \rho_y \sin(k_y a_0) ]^2\}^{3/2}}.
	\end{aligned}
\end{equation}
We further arrive at the Chern number
\begin{equation}
	C  = -{\rm Sign} ( \rho_x \rho_y \lambda ).
\end{equation}

\section{Self consistent calculations for a 1D ribbon}\label{AppendixSelfConsistency}
If we apply the open boundary condition along $x$ direction but keep infinite boundary condition along $y$ direction for  Eq.~\eqref{ModelHamiltonian}, we shall get a configuration of ribbon, which is the setup for Fig.~1c of the main text. We define the following partial Fourier transformation:
\begin{equation}
	\hat C_{\alpha,\sigma}(x_i^\alpha,y_j^\alpha) = \frac{1}{\sqrt{N_y}} \sum_{k_y} e^{+i k_y y_j^\alpha} \hat C_{\alpha,\sigma}(x_i,k_y).
\end{equation}
Plugging back to Eq.~\eqref{ModelHamiltonian}, we arrive at the partial Fourier transformation for $H_1$
\begin{equation}\label{KyH1}
	\begin{aligned}
		&H_1 = -\sum_{n,k_y,\sigma} [J_1^x \eta^1_{ba}(2n-1) \ket{\Psi_{a,\sigma}(2n-1,k_y)}\bra{\Psi_{b,\sigma}(2n,k_y)} + J_1^x \eta^1_{ab}(2n) \ket{\Psi_{b,\sigma}(2n,k_y)}\bra{\Psi_{a,\sigma}(2n+1,k_y)} + {\rm H.c.}] \\
		&- \sum_{n,k_y,\sigma} [J_1^y \eta^2_{aa}(2n-1) e^{+ik_y a_0} \ket{\Psi_{a,\sigma}(2n-1,k_y)}\bra{\Psi_{a,\sigma}(2n-1,k_y)} + J_1^y \eta^2_{bb}(2n) e^{+ik_ya_0} \ket{\Psi_{b,\sigma}(2n,k_y)} \bra{\Psi_{b,\sigma}(2n,k_y)} + {\rm H.c.}].
	\end{aligned}
\end{equation}
Thus the self-consistency equation for the ribbon reads (the ${\rm occ}$ means the summation runs through all the occupied bands)
\begin{equation}
	\begin{aligned}
		\eta_{ba}^1({\bm r}_i^a)& \!=\! \eta^1_{ba}(x_i^a) \!=\! \eta^1_{ba}(2n-1) \!=\! \langle \hat C^\dagger_{b,\sigma}({\bm r}_i^a + {\bm S}_1) \hat C_{a,\sigma}({\bm r}_i^a)\rangle \!=\! \frac{1}{N_y} \sum_{k_y^1,k_y^2,j,\sigma}  \langle \hat C^\dagger_{b,\sigma}(x_i^a+a_0,k_y^1) e^{-ik_y^1 y_j^a} \hat C_{a,\sigma}(x_i^a, k_y^2) e^{+ik_y^2 y_j^a}\rangle\\
		& \!=\! \frac{1}{N_y} \sum_{k_y} \sum_{\sigma} \langle \hat C^\dagger_{b,\sigma}(x_i^a+a_0,k_y)\hat C_{a,\sigma}(x_i^a,k_y) \rangle \!=\! \frac{1}{N_y} \sum_{\sigma,k_y\in [-\pi,+\pi)} \sum_{\zeta \in {\rm occ}} \Psi^{*,\zeta}_{b,\sigma}(2n,k_y)\Psi^\zeta_{a,\sigma}(2n-1,k_y) \\
		\eta_{ab}^1({\bm r}_i^b) 	&\!=\! \frac{1}{N_y} \sum_{k_y} \sum_\sigma \langle \hat C^\dagger_{a,\sigma}(x_i^b + a_0, k_y) \hat C_{b,\sigma}(x_i^b, k_y)\rangle \!=\! \frac{1}{N_y} \sum_{\sigma,k_y \in [-\pi, +\pi)} \sum_{\zeta \in {\rm occ}}\Psi^{*,\zeta}_{a,\sigma}(2n+1,k_y) \Psi^\zeta_{b,\sigma}(2n,k_y)\\
		\eta_{aa}^2({\bm r}_i^a)	&\!=\! \frac{1}{N_y} \sum_{k_y}\sum_{\sigma} \langle e^{-ik_y a_0}  \hat C^\dagger_{a,\sigma}(x_i^a,k_y) \hat C_{a,\sigma}(x_i^a,k_y) \rangle \!=\! \frac{1}{N_y} \sum_{\sigma,k_y\in[-\pi,+\pi)}\sum_{\zeta \in {\rm occ}} e^{-ik_ya_0} \Psi^{*,\zeta}_{a,\sigma}(2n-1,k_y)\Psi^\zeta_{a,\sigma}(2n-1,k_y) \\
		\eta_{bb}^2({\bm r}_i^b) &\!=\! \frac{1}{N_y} \sum_{k_y} \sum_\sigma \langle  e^{-ik_y a_0}\hat C^\dagger_{b, \sigma}(x_i^b,k_y) \hat C_{b,\sigma}(x_i^b, k_y) \rangle \!=\! \frac{1}{N_y} \sum_{\zeta \in {\rm occ}}\sum_{\sigma, k_y \in [-\pi, +\pi)} e^{-ik_y a_0} \Psi^{*,\zeta}_{b,\sigma}(2n,k_y) \hat \Psi^\zeta_{b,\sigma}(2n,k_y).
 		\end{aligned}
\end{equation}
Similarly, the next-nearest neighbor coupling can also be partially transformed into Fourier space
\begin{equation}\label{KyH2}
	\begin{aligned}
		H_2 	=& -\sum_{i,k_y,\sigma} [J_2^+ \eta^3_{ba}(x_i^a) e^{+ik_ya_0}\ket{\Psi_{a,\sigma}(2n-1,k_y)}\bra{\Psi_{b,\sigma}(2n,k_y)} + {\rm H.c.}  ] \\
		& -\sum_{i,k_y,\sigma} [J_2^+ \eta^3_{ab}(x_i^b) e^{+ik_y a_0} \ket{\Psi_{b,\sigma}(2n,k_y)}\bra{\Psi_{a,\sigma}(2n+1,k_y)} + {\rm H.c.}] \\
		& -\sum_{i,k_y,\sigma} [J_2^- \eta^4_{ba} (x_i^a) e^{-ik_ya_0} \ket{\Psi_{a,\sigma}}(2n-1,k_y) \bra{\Psi_{b,\sigma}(2n,k_y)}  + {\rm H.c.}] \\
		& -\sum_{i,k_y,\sigma} [J_2^- \eta^4_{ab} (x_i^b) e^{-ik_ya_0} \ket{\Psi_{b,\sigma}(2n,k_y)}\bra{\Psi_{a,\sigma}(2n+1,k_y)} + {\rm H.c.}],
	\end{aligned}
\end{equation}
and we have the following self-consistent equations:
\begin{equation}
	\begin{aligned}
		\eta^3_{ba}({\bm r}_i^a) &\!=\! \eta^3_{ba}(x_i^a) \!=\! \eta^3_{ba}(2n-1) \!=\! \langle \hat C^\dagger_{b,\sigma}({\bm r}_i^a + {\bm R}_1) \hat C_{a,\sigma}({\bm r}_i^a) \rangle \!=\! \frac{1}{N_y} \sum_{k_y^1,k_y^2,j,\sigma}\langle \hat C^\dagger_{b,\sigma}(x_i^a + a_0,k_y^1)e^{-ik_y^1(y_j^a+a_0)}\hat C_{a,\sigma}(x_i^a,k_y^2) e^{+ik_y^2 y_j^a}\rangle \\
		&\!=\! \frac{1}{N_y} \sum_{k_y} \sum_\sigma \langle e^{-ik_y a_0}\hat C^\dagger_{b,\sigma}(x_i^a + a_0, k_y)\hat C_{a,\sigma}(x_i^a ,k_y) \rangle \!=\! \frac{1}{N_y} \sum_{\sigma,k_y \in [-\pi,+\pi)} \sum_{\zeta \in {\rm occ}}e^{-ik_y a_0} \Psi^{*,\zeta}_{b,\sigma}(2n,k_y) \Psi^\zeta_{a,\sigma}(2n-1,k_y) \\
		\eta^3_{ab}({\bm r}_i^b)	&\!=\! \frac{1}{N_y} \sum_{k_y} \sum_\sigma \langle e^{-ik_y a_0} \hat C^\dagger_{a,\sigma}(x_i^b + a_0, k_y) \hat C_{b,\sigma}(x_i^b,k_y) \rangle \!=\! \frac{1}{N_y} \sum_{\sigma,k_y \in [-\pi, \pi)} \sum_{\zeta \in {\rm occ}} e^{-ik_y a_0} \Psi^{*,\zeta}_{a,\sigma}(2n+1,k_y) \Psi^\zeta_{b,\sigma}(2n,k_y) \\
		\eta^4_{ba}({\bm r}_i^a) &\!=\! \frac{1}{N_y} \sum_{k_y} \sum_{\sigma} \langle e^{+ik_y a_0} \hat C^\dagger_{b,\sigma}(x_i^a + a_0,k_y) \hat C_{a,\sigma}(x_i^a,k_y) \rangle   \!=\! \frac{1}{N_y} \sum_{k_y}\sum_{\sigma,k_y\in[-\pi,\pi)} \sum_{\zeta \in {\rm occ}}e^{+ik_y a_0} \Psi^{*,\zeta}_{b,\sigma}(2n,k_y) \Psi^\zeta_{a,\sigma}(2n-1,k_y) \\
	\eta^4_{ab}({\bm r}_i^b) 	&\!=\! \frac{1}{N_y} \sum_{k_y} \sum_\sigma \langle e^{+ik_y a_0} \hat C^\dagger_{a,\sigma}(x_i^b + a_0, k_y) \hat C_{b,\sigma}(x_i^b, k_y) \rangle \!=\! \frac{1}{N_y} \sum_{k_y} \sum_{\sigma,k_y \in [-\pi,\pi)}  \sum_{\zeta \in {\rm occ}}e^{+ik_y a_0} \Psi^{*,\zeta}_{a,\sigma}(2n+1,k_y) \Psi^\zeta_{b,\sigma}(2n,k_y).
	\end{aligned}
\end{equation}

The central approach in this work is a self-consistent mean-field ansatz for a CSL system with both positive and negative chirality domains. For such a spatially inhomogeneous ansatz, the iteration method is particularly efficient compared to the energy minimization approach and hence will be our main tool in the analysis below. We attempt a self-consistent mean-field description of a multi-domain CSL. For simplicity, we consider a two-domain setup as illustrated in Fig.~\ref{Illustration}(c): the system is finite along $x$ direction and periodic along $y$ direction, with a negative (positive) chirality $-\chi_0$ ($+\chi_0$) CSL domain existing on the left (right). While the chirality is nearly uniform far away from the interface, a profile develops near the interface, leading to a finite domain wall whose width will be determined self-consistently in the following procedure: (1) choose one of the ansatze of the homogeneous system with negative (positive) chirality obtained above, $\eta_{{\bm i} {\bm j}}^{-,a}$ ($\eta_{{\bm i} {\bm j}}^{+,b}$), and assign it to the region $x<x_1$ on the far left ($x>x_2$ on the far right) in Fig.~\ref{Illustration}(c); (2) choose a set of mean-field bonds arbitrarily and assign it to the middle region $x_1<x<x_2$; (3) repeat the iteration process only for the bonds between $x_1$ and $x_2$, with the far left $x<x_1$ and far right $x_2<x$ bonds held fixed; (4) a solution is obtained when the bonds converge. If multiple solutions exist, then the one with the lowest energy at half-filling will be chosen.

\section{Amplitude fluctuation}\label{Fluctuations}
Now we want to introduce the dynamical field:
\begin{equation}
	H_0 \sim \sum_{\bm k} \psi_{\bm k}^\dagger [(u_x^0 + \hat u_x) k_xa_0 \sigma_x + (u_y^0 + \hat u_y)k_y a_0 \sigma_y + (u_z^0 + \hat m_z)\sigma_z] \psi_{\bm k},
\end{equation}
which means we assume that the flux pattern in the chiral spin liquid phase we considered is fixed (i.e. no phase fluctuation for the gauge fields), while the amplitude for the mean-field bonds (i.e., the value of $\rho_x,\rho_y,\lambda_\pm$) can fluctuate. The fluctuations of the mean-field bonds are captured by the newly introduced classical field $\hat u_x$, $\hat u_y$ and $\hat u_z$, which are related to the fluctuations of the original bond amplitude by:
\begin{equation}
	\hat u_x = 2J_1^x \hat \rho_x, \quad \hat u_y =-2J_1^y  \hat \rho_y, \quad \hat m_z =2 (J_2^+ \hat \lambda_+ + J_2^- \hat \lambda_-).
\end{equation}
The total action then reads (with $\partial_\tau$ the derivative with respect to imaginary time):
\begin{equation}
	\begin{aligned}
		S_s = \int d\tau d^2 x \bar \psi_s \big{(} \partial_\tau + u_x^0 \tau^x k_x a_0 + u^0_y \tau^y k_y a_0+ u_z^0 \tau^z + \hat u_x k_xa_0 \tau^x + \hat u_y k_ya_0 \tau^y + \hat m_z \tau^z\big{)} \psi_s.
	\end{aligned}
\end{equation}
Upon doing the transformation, 
\begin{equation}
	\psi_s \rightarrow e^{- i \frac{\pi}{4} \tau^z} \psi, \quad \bar \psi_s \rightarrow i \bar \psi e^{-i \frac{\pi}{4} \tau^z}.
\end{equation}
We shall then arrive at:
\begin{equation}
\begin{aligned}
	S_s &= \int d\tau d^2x \bar \psi (\tau^z \partial_\tau + u_x^0 a_0\partial_x \tau^x + u_y^0 a_0\partial_y \tau^y + u_z^0 \tau_0 + {\color{blue}\hat u_x k_xa_0 i \tau^x + \hat u_y k_ya_0 i\tau^y + \hat m_z \tau_0}) \psi \\
\end{aligned}
\end{equation}


We can treat the blue part as a perturbation. Then we shall have the form:
\begin{equation}
	\begin{aligned}
		S_{\rm eff} &= -\ln {\rm det} [\tau^z \partial_\tau + u_x^0 a_0\partial_x \tau^x + u_y^0 a_0\partial_y \tau^y + u_z^0 \tau_0 +{\color{blue} \hat u_x k_xa_0 i \tau^x + \hat u_y k_ya_0 i\tau^y + \hat m_z \tau_0}] \\
		&= -{\rm Tr} \ln [\tau^z \partial_\tau + u_x^0 a_0 \partial_x \tau^x + u_y^0 a_0\partial_y \tau^y + u_z^0 \tau_0 + {\color{blue}\hat u_x k_x a_0 i \tau^x + \hat u_y k_ya_0 i\tau^y + \hat m_z \tau_0}] \\
		&= -{\rm Tr} \ln [G^{-1} + {\color{blue} {\mathcal V}} ] = -{\rm Tr}\ln G^{-1} - {\rm Tr}\ln [1 +G{\color{blue} {\mathcal V}}] \\
		&= {\rm const} -{\rm Tr}[G{\color{blue} {\mathcal V}}] + \frac{1}{2} {\rm Tr} [G{\color{blue} {\mathcal V}}G{\color{blue} {\mathcal V}})] + \cdots{}
	\end{aligned}
\end{equation}
$(u_x^0, u_y^0 , u_z^0) = (2J_1 \bar \rho_x , 2J_1 \bar \rho_y , 4J_2 \bar \lambda)$. As we have set $\bar \rho_x = \bar \rho_y$, from which we shall have $u_x^0 = u_y^0$, and can further define: $m_z^0 = u_z^0/ u_x^0 = 2J_1 \bar \lambda/J_1 \bar \rho_x $, $k_0 = \hbar \omega/u_x^0 a_0$. Thus the propagator is defined as:
\begin{equation}
    G(k) = \frac{iu_x^0 k_x a_0\tau_x +i u_y^0 k_ya_0 \tau_y + i\hbar \omega \tau_z + u_z^0 \tau_0}{(\hbar \omega)^2 + (u_x^0 k_x a_0)^2 + (u_y^0 k_y a_0)^2 + (u_z^0)^2}  = \frac{1}{u_x^0} \frac{i\tilde k_x \tau_x + i\tilde k_y\tau_y + i\tilde k_0 \tau_z + m_z^0 \tau_0}{\tilde k_x^2 + \tilde k_y^2 + \tilde k_0^2 + (m_z^0)^2},
\end{equation}
where the dimensionless momentum are defined as $(\tilde k_x, \tilde k_y, \tilde k_0) = (k_x a_0, k_y a_0, k_0a_0)$. Similarly, the vertex is defined as:
\begin{equation}
	{\mathcal V} = {\mathcal V}_x + {\mathcal V}_y + {\mathcal V}_z = \hat u_x \tilde k_x i \tau^x + \hat u_y \tilde k_y i\tau^y + \hat m_z \tau_0, \quad  {\mathcal V}_x= \hat u_x \tilde k_x i \tau^x, ~ {\mathcal V}_y = \hat u_y \tilde k_y i\tau^y, ~ {\mathcal V}_z=  \hat m_z \tau_0.
\end{equation}
The leading order term vanishes, and we would love to see the dynamical field from:
\begin{equation}
	\begin{aligned}
		\frac{1}{2 }{\rm Tr} [G{\mathcal V}G{\mathcal V}] = \frac{1}{2} {\rm Tr}\ln [G(\tilde p_x, \tilde p_y,m_z^0) (\hat u_x \tilde p_x i \tau^x + \hat u_y \tilde p_y i\tau^y + \hat m_z \tau_0)G(\tilde k_x, \tilde k_y,m_z^0) (\hat u_x \tilde k_x i \tau^x + \hat u_y \tilde k_y i\tau^y + \hat m_z \tau_0) ].
	\end{aligned}
\end{equation}
Below we will use the dimensionless momentum and drop the tilde on momentum.

\section{The effective action}\label{EFFT}
\begin{figure}[!h]
\centering 
\includegraphics[width=1\columnwidth]{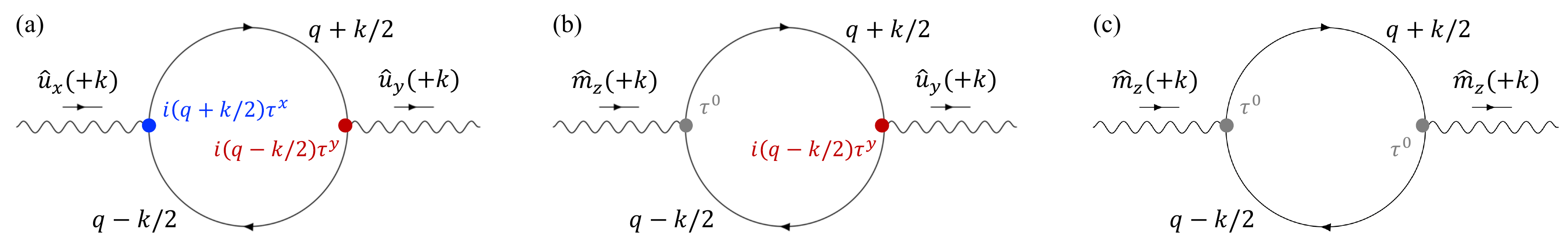}
\caption{\label{FeynmanDiagram} Feynman diagrams for evaluation of $S_{\rm eff}$ given in Eq.~\eqref{Action}, with (a) for a term in $S_{\rm eff}^A$, (b) for a term in $S_{\rm eff}^B$, and (c) for a term in $S_{\rm eff}^C$. }
\end{figure}
One can integrate out the fermions to get the effective field theory for $\hat u_x$, $\hat u_y$, and $\hat m_z$ (or equivalently, the $\hat \rho_x$, $\hat \rho_y$ and $\hat \lambda$). After the lengthy calculation in Section~[\ref{Appendix}], we arrive at the following effective action:
\begin{equation}
\begin{aligned}\label{Action}
	S_{\rm eff} &= S_{\rm eff}^A + S_{\rm eff}^B  + S_{\rm eff}^C
\end{aligned}	
\end{equation}
with the term associated with two velocity vertices:
\begin{equation}
	S_{\rm eff}^A \equiv \sum_{\mu,\nu = x,y}\frac{1}{2} \int \frac{d^3k}{(2\pi)^3} \Pi_{\mu \nu}^A(k) \hat u_{\mu}(+k) \hat u_{-\nu}(k),
\end{equation}
and two mass vertices:
\begin{equation}
	S^B_{\rm eff} \equiv \frac{1}{2} \int \frac{d^3k}{(2\pi)^3} \Pi^B_{zz}(k) \hat m_z(+k) \hat m_z(-k),
\end{equation}
and one velocity vertex and one mass vertex:
\begin{equation}
	S_{\rm eff}^C \equiv \sum_{\mu,\nu = x,y}\frac{1}{2} \int \frac{d^3k}{(2\pi^3)} [\Pi^C_{z\nu} (k)\hat m_z(+k) \hat u_\nu(-k)  + \Pi^C_{\mu z} (k)\hat u_\mu(+k) \hat m_z(-k)].
\end{equation}
We could write the effective action in a simpler way:
\begin{equation}
	S_{\rm eff} \simeq  \frac{\hbar}{2(u_x^0)^2} \int \frac{d^3k}{(2\pi)^3}\begin{pmatrix}
		 \hat u_x(+k) & \hat u_y(+k) & \hat m_z(+k)
	\end{pmatrix}
	{\mathcal S} \begin{pmatrix}
		\hat u_x(-k) \\
		\hat u_y(-k) \\
		\hat m_z(-k)
	\end{pmatrix}
\end{equation}
The matrix element of ${\mathcal S}$ reads (note that we have defined $m = |m_z^0|$):
\begin{equation}
	\begin{aligned}
		{\mathcal S}_{xx}  &= \frac{2m^3}{3\pi} + \frac{m}{12\pi}(k^2-k_x^2), \quad {\mathcal S}_{xy} =  \frac{m^3}{6 \pi} + \frac{m}{24 \pi}(k^2-k_x^2-k_y^2), \quad {\mathcal S}_{xz} = \frac{m^3}{2\pi} + \frac{m}{24 \pi }(k^2-k_x^2),\\
		{\mathcal S}_{yx} &=  \frac{m^3}{6 \pi} + \frac{m}{24\pi}(k^2-k_x^2-k_y^2), \quad {\mathcal S}_{yy}  = \frac{2m^3}{3\pi} + \frac{m}{12 \pi}(k^2-k_y^2), \quad {\mathcal S}_{yz} = \frac{m^2}{2\pi} + \frac{m}{24 \pi }(k^2-k_y^2), \\
		{\mathcal S}_{zx} &= \frac{m^3}{2\pi} + \frac{m}{24 \pi }(k^2-k_x^2), \quad 
		{\mathcal S}_{zy} = \frac{m^3}{2\pi} + \frac{m}{24 \pi }(k^2-k_y^2), \quad {\mathcal S}_{zz}  =  \frac{7m^3}{8\pi}  + \frac{3k^2m}{32 \pi}. 
	\end{aligned}
\end{equation}
such that we have:
\begin{equation}
	{\mathcal S} = {\mathcal M}_0(m_z^0) + {\mathcal T}_k(m_z^0),
\end{equation}
with mass ${\mathcal M}_0(m_z^0)$ and stiffness ${\mathcal T}_k(m_z^0)$ reads:
\begin{equation}
    {\mathcal M}_0 (m_z^0)= \frac{m^3}{24\pi}\begin{pmatrix}
        16& 4& 12\\
        4 & 16 & 12 \\
       12& 12 & 21
    \end{pmatrix},\quad     {\mathcal T}_k (m_z^0)= \frac{m}{24\pi}\begin{pmatrix}
            2(k^2-k_x^2) & (k^2-k_x^2-k_y^2) & (k^2-k_x^2) \\
            (k^2-k_x^2-k_y^2)  & 2(k^2-k_y^2) & (k^2-k_y^2) \\
            (k^2-k_x^2) & (k^2-k_y^2) & \frac{9}{4}k^2
        \end{pmatrix}.
\end{equation}
From the effective action, we can get the free energy for one valley $(0,0)$:
\begin{equation}
	F(m_z^0) = \frac{1}{2u_x^0} \int \frac{d^2k}{(2\pi)^2}\begin{pmatrix}
		 \hat u_x(+k) & \hat u_y(+k) & \hat m_z(+k)
	\end{pmatrix}
	{\mathcal F}(m_z^0) \begin{pmatrix}
		\hat u_x(-k) \\
		\hat u_y(-k) \\
		\hat m_z(-k)
	\end{pmatrix}.
\end{equation}
where:
\begin{equation}
	{\mathcal F}(m_z^0) = {\mathcal M}_0 (m_z^0)+ {\mathcal T}_k (m_z^0).
\end{equation}
This gives a Ginzburg--Landau theory for the three-component vector field. By using the fact that $k^2 = k_x^2 + k_y^2$, we have:
\begin{equation}
	{\mathcal T}_k(m_z^0)= \frac{m}{24\pi}\begin{pmatrix}
            2 k_y^2 & 0 & k_y^2 \\
            0 & 2 k_x^2 & k_x^2 \\
           k_y^2 &  k_x^2 & \frac{9}{4}(k_x^2+k_y^2)
        \end{pmatrix}
\end{equation}
Another valley  $(0, \pi)$ corresponds to ${\mathcal F}(-m_z^0)$, such that we have:
\begin{equation}
	\tilde {\mathcal F}(m_z^0) = {\mathcal F}(m_z^0) + {\mathcal F}(-m_z^0) = {\mathcal M}_0(m_z^0) + {\mathcal T}_k(m_z^0) + {\mathcal M}(-m_z^0) + {\mathcal T}_k(-m_z^0) = 2 {\mathcal M}_0(m) +2 {\mathcal T}_k(m) = 2{\mathcal F}(m_z^0).
\end{equation}

\subsection{Domain wall structure}

Suppose we have the domain structure (breaking the translation invariance) along the $x$ direction, such that $k_y = 0$, and:
\begin{equation}
	{\mathcal T}_k = \frac{m}{24\pi} \begin{pmatrix}
		0 & 0 & 0 \\
		0 & 2k_x^2 &  k_x^2 \\
		0 & k_x^2 & \frac{9}{4} k_x^2
	\end{pmatrix}
\end{equation}
We further integrate over $y$ direction, and then we shall have the total free energy:
\begin{equation}
	\begin{aligned}
		F_T = F(m_z^0) + F(-m_z^0) &\simeq \frac{m}{u_x^0 }\int \frac{dk}{2\pi} [2u_y(+k)u_y(-k) + u_y(+k)m_z(-k)+m_z(+k)u_y(-k) + \frac{9}{4}m_z(+k)m_z(-k)] \\
		&+  \frac{m}{u_x^0 }\int \frac{dk}{2\pi} m^2[16 u_x(+k)u_x(-k) + 4 u_x(+k) u_y(-k) + 12 u_x(+k) m_z(-k)] \\
		&+  \frac{m}{u_x^0 }\int \frac{dk}{2\pi} m^2[4 u_y(+k) u_x(-k) + 16 u_y(+k) u_y(-k) + 12 u_y(+k) m_z(-k)] \\
		&+  \frac{m}{u_x^0 }\int \frac{dk}{2\pi} m^2[12 m_z(+k) u_x(-k) + 12 m_z(+k) u_y(-k) + 21 m_z(+k) m_z(-k)],
	\end{aligned}
\end{equation}
from which, we shall further have in the real space:
\begin{equation}\label{EnergyDomain}
	\begin{aligned}
		F_T = \int [d^2x] {\mathcal F}_T &\simeq 
		 \frac{m}{u_x^0 a_0}\int [dx] [2(\partial_x u_y)^2 + (\partial_x u_y)(\partial_x m_z) + (\partial_x m_z) (\partial_x u_y) + \frac{9}{4} (\partial_x m_z)^2] \\
		&+ \frac{m}{u_x^0 a_0}\int [dx] m^2[16 u_y^2 + 12 u_y m_z + 12 m_z u_y + 21 m_z^2] \\
		&+ \frac{m}{u_x^0 a_0}\int [dx] m^2[16 u_x^2 + 4u_x u_y + 12 u_x m_z + 4 u_y u_x + 12 m_z u_x].
	\end{aligned}
\end{equation}
There should be an interaction term which can be determined from the position of the local minimal, we will show this in the next section. 

\section{Ginzburg Landau theory}
\subsection{Free energy for fluctuations and full Ginzburg Landau theory}\label{GLFullandFluctuation}
Consider the following free energy functional in $d$-dimensions, where the $\phi({\bm x})$ is the full field for certain order parameter: 
\begin{equation}\label{GLInitial}
	F[\phi] = \int [d^dx] \bigg{[} K [\nabla \phi({\bm x})]^2 + \frac{{\mathcal M}}{2} \phi^2({\bm x}) + \frac{{U}}{4} \phi^4({\bm x}) \bigg{]},
\end{equation}
we shall then get the equation of motion which reads:
\begin{equation}
	0 = \frac{\delta F}{\delta \phi} = - K \nabla^2 \phi + {\mathcal M}\phi + U \phi^3,
\end{equation}
which gives a solution for ${\mathcal M} <0$ and $U >0$, 
\begin{equation}
	\phi_0({\bm x}) \equiv \pm \sqrt{-\frac{\mathcal M}{U}} = \pm \phi_0,
\end{equation}
as the local minimal for effective potential:
\begin{equation}
	V_{\rm eff} = \frac{\mathcal M}{2} \phi^2({\bm x}) + \frac{U}{4} \phi^4({\bm x}).
\end{equation}
Note that, if we already know the position, we could also uniquely determine the interaction strength:
\begin{equation}
	U = -\frac{\mathcal M}{\phi_0^2}.
\end{equation}
Now we introduce the fluctuations $\delta \phi$ close to $\pm \phi_0$
\begin{equation}
	\phi({\bm x}) = \pm \phi_0 + \delta \phi,
\end{equation}
then we can write down
\begin{equation}
	\begin{aligned}
		F[\phi] &\!=\! \int [d^dx] \bigg{[} K [\nabla \delta \phi({\bm x})]^2 + \frac{\mathcal M}{2}[\pm \phi_0 + \delta\phi({\bm x})]^2 + \frac{U}{4}[\pm \phi_0 + \delta\phi({\bm x})]^4 \bigg{]} \\
		&\!\approx\! \int [d^dx] \bigg{[} K [\nabla \delta \phi({\bm x})]^2 + \frac{\mathcal M}{2}[\phi_0^2 \pm 2\phi_0 \delta\phi({\bm x}) + (\delta \phi({\bm x}))^2] \\
		&\quad \quad \quad ~+ \frac{U}{4} [\phi^4_0 \pm 4\phi_0^3 \delta \phi({\bf x})+6 \phi_0^2 (\delta\phi({\bm x}))^2 \pm 4 \phi_0 (\delta \phi_0({\bm x}))^3 + (\delta \phi({\bm x}))^4]\bigg{]} \\
		&\!=\! \int [d^dx] \bigg{[} K [\nabla \delta \phi({\bm x})]^2+\frac{\mathcal M}{2} \phi_0^2 + \frac{U}{4}\phi_0^4 + \bigg{[} \frac{\mathcal M}{2} + \frac{6U}{4}\phi_0^2 \bigg{]}(\delta \phi({\bm x}) )^2 \pm U \phi_0 (\delta\phi_0({\bm x}))^3 + \frac{U}{4}(\delta \phi({\bm x}))^4\bigg{]}, \\
	\end{aligned} 
\end{equation}
by dropping the constant term and the odd term, and introducing 
\begin{equation}
	{\mathcal M}_{\rm eff} = \frac{\mathcal M}{2} + \frac{6U}{4} \phi_0^2 = \frac{\mathcal M}{2} - \frac{3U}{2}\frac{\mathcal M}{u} = -{\mathcal M},
\end{equation}
we have the effective field theory for fluctuations solely:
\begin{equation}\label{GLFluctuations}
	\begin{aligned}
		F[\delta\phi] = \int [d^dx] \bigg{[} K [\nabla \delta \phi({\bm x})]^2 + {\mathcal M}_{\rm eff} [\delta \phi({\bm x})]^2 + \frac{U}{4} [\delta \phi({\bm x})]^4 \bigg{]}, \quad {\mathcal M}_{\rm eff} = -{\mathcal M}, \quad U = -\frac{\mathcal M}{\phi_0^2}.
	\end{aligned}
\end{equation}
this is the energy for fluctuations just like the one we derived from Eq.~\eqref{EnergyDomain}, and we can also get the full field theory from the coefficient we derived from the effective field theory. 

\subsection{Generic domain wall}\label{GinzburgLandauDomainWall}
Consider the following free energy functional:
\begin{equation}
	F[\phi] = \int [d^dx] [K (\nabla \phi({\bm x}))^2 + \frac{\mathcal M}{2}\phi^2({\bm x}) + \frac{U}{4} \phi^4({\bm x})].
\end{equation}
We shall then get the equation of motion which reads:
\begin{equation}
	0 = \frac{\delta F}{\delta \phi} = - K \nabla^2 \phi + {\mathcal M} \phi + U \phi^3.
\end{equation}
Here $K$ is the stiffness with respect to the spatial variation of $\phi$, and one can define a length scale $\xi =\sqrt{K/|{\mathcal M}|}$ as the coherence length. Rescaling the free energy with respect to $\phi({\bm x}) \equiv \sqrt{|{\mathcal M}|/U} \Phi$, and one can obtain:
\begin{equation}
	F[\Phi] = \int [d^dx] \bigg{[}K \frac{|{\mathcal M}|}{U} (\nabla \Phi)^2 + \frac{\mathcal M}{2} \frac{|{\mathcal M}|}{U} \Phi^2 + \frac{U}{4}\frac{{\mathcal M}^2}{U^2}\Phi^4 \bigg{]} = \int [d^dx] \frac{{\mathcal M}^2}{U} \bigg{[}\frac{K}{|{\mathcal M}|}(\nabla \Phi)^2 + \frac{{\rm sgn}({\mathcal M})}{2} \Phi^2 + \frac{1}{4} \Phi^4 \bigg{]}.
\end{equation}
Then we get the new EOM:
\begin{equation}
	-\xi^2 \nabla^2 \Phi + {\rm sgn}({\mathcal M}) \Phi + \Phi^3 = 0.
\end{equation}
Consider the case of a domain ${\mathcal M} <0$, we assume that all spatial variation occurs in the $x$-direction, and we set $\Phi(x= x_0) =0$ and $\Phi(x= \infty) = 1$,  we then arrive at:
\begin{equation}
	-\xi^2 \Phi^{\prime\prime}(x) - \Phi + \Phi^3 = 0, 
\end{equation}
which may be written as:
\begin{equation}
	\xi^2 \frac{d^2 \Phi}{dx^2} = \frac{\partial}{\partial \Phi} \bigg{[} \frac{1}{4}(1-\Phi^2)^2 \bigg{]}.
\end{equation}
Multiplying the above equation by $\Phi^\prime(x)$ and integrating once, we have:
\begin{equation}
	\xi^2 \bigg{(} \frac{d\Phi}{dx} \bigg{)}^2 = \frac{1}{2}(1-\Phi^2)^2 + C,
\end{equation}
where $C$ is a constant, which is fixed by setting $\Phi (x\rightarrow \infty) = +1$, which says $\Phi^\prime(\infty) =0$, hence $C= 0$. Integrating once more, we shall have:
\begin{equation}
\Phi(x) =\tanh \bigg{(} \frac{x-x_0}{\sqrt{2}\xi} \bigg{)}	,
\end{equation}
and $x_0$ is determined by the zero points of the system. The width of the domain wall is $\xi$. In terms of the original field, that is:
\begin{equation}
	\phi({\bm x}) = \sqrt{\frac{|{\mathcal M}|}{K}} \Phi = \sqrt{-\frac{\mathcal M}{K}} \tanh \bigg{(} \frac{x-x_0}{\sqrt{2}\xi}\bigg{)}, \quad \xi = \sqrt{\frac{K}{|{\mathcal M}|}} = \sqrt{-\frac{K}{\mathcal M}}.
\end{equation}
The term $K =0$ corresponds to the sharp domain wall.

\subsection{Analysis of the domain wall}\label{SecGLDomain}
\begin{figure}[!h]
\centering 
\includegraphics[width=1\columnwidth]{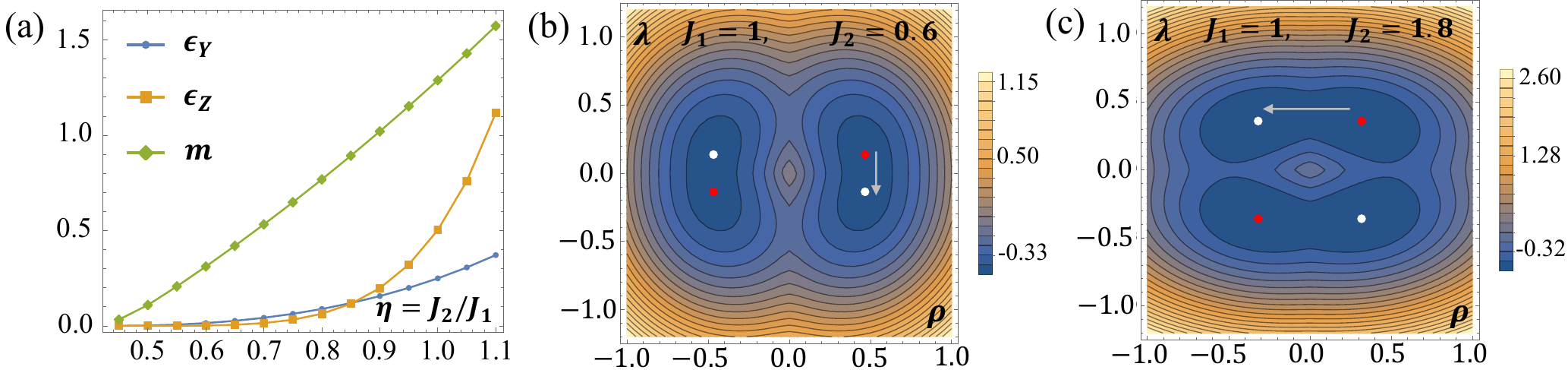}
\caption{\label{ContourPlot} (a) The relations for ${\mathcal E}_Y(\eta)$ (blue), ${\mathcal E}_Z(\eta)$ (yellow), and $m(\eta)$, with $\eta = J_2/J_1$ tell the ratio between the nearest neighbor and second nearest neighbor coupling. The domain wall energy is calculated with respect to homogenous limit (without domain wall) (b) Contour plot of free energy $F(\rho_x^0 = \rho_y^0,\lambda^0)$ for small $\eta = J_2/J_1 = 0.6$, in which we can see the domain wall has the tendency to go from the red point with positive chirality to the white point with opposite $\lambda$ along the arrow. (c) Contour plot of free energy $F(\rho_x^0= \rho_y^0,\lambda^0)$ for large $\eta = J_2/J_1 = 1.8$ (which may not be realistic), which it tends to form the velocity domain wall.}
\end{figure}

The $F_T[u_x,u_y,m_z]$ is positive definite, means all kinds for the fluctuations should increase the energy compared with the homogeneous solution. For a given value of $J_1,J_2$, the mean-field ansatz corresponds to eight points in Hilbert space, each four belonging to one chirality, and we would like to pick the path that connecting two points with opposite chirality with minimal energy cost. 

As these are all fluctuations, we can have by forcing via the boundary condition 
\begin{enumerate}
	\item $u_x$ cannot change sign by these dynamics 
	\item $u_y$ can change sign via nonzero $(\partial_x u_y)^2$, we can force $m_z  = u_x = 0$, this leads to effective energy for fluctuation fields, i.e., Eq.~\eqref{GLFluctuations}:
	\begin{equation}
	\begin{aligned}
		F_T &\simeq \frac{m}{24\pi u_x^0 a_0} \int [dx] {\mathcal F}_Y(x) =  \frac{m}{24\pi u_x^0 a_0}\int [dx] [2a_0^2(\partial_x u_y)^2 - 16m^2  u_y^2 + \frac{U_y}{4}  u_y^4],
	\end{aligned}
	\end{equation}
	by using the relation mentioned in Sec.~[\ref{GLFullandFluctuation}], and defining the full field $\tilde u_y = u_y^0 + u_y$, we arrive at the full Ginzburg--Landau theory, i.e., Eq.~\eqref{GLInitial}
	\begin{equation}
	\begin{aligned}
		F_T &\simeq \frac{m}{24\pi u_x^0 a_0} \int [dx] {\mathcal F}_Y(x) =  \frac{m}{24\pi u_x^0 a_0}\int [dx] [2a_0^2(\partial_x \tilde u_y)^2 - 16m^2 \tilde u_y^2 + \frac{U_y}{4} \tilde u_y^4],
	\end{aligned}
	\end{equation}
	comparing with Eq.~\eqref{GLInitial}, one can read off the coefficients as:
	\begin{equation}
		K = 2a_0^2, \quad \frac{{\mathcal M}}{2} = -16m^2, \quad  U_y = -\frac{\mathcal M}{\phi_0^2}  = \frac{32m^2}{(u_y^0)^2}.
	\end{equation}
	From Sec.~[\ref{GinzburgLandauDomainWall}], we know that such a theory can hold a domain wall if one can impose the twisted boundary condition for $\tilde u_y$, with the correlation length $\xi$ 
	\begin{equation}
		\xi_b = \sqrt{-\frac{K}{\mathcal M}} = \sqrt{\frac{a_0^2}{16m^2}} = \frac{a_0}{4|m|}.	
	\end{equation}
	This leads to the configuration:
	\begin{equation}
	\tilde u_y = u_y^0 \tanh[(x-x_0)/\sqrt{2}\xi_b], \quad u_y^0 = 1.
	\end{equation}
	With the domain wall energy reads:
	\begin{equation}\label{EnergyVelocity}
		\begin{aligned}
			{\mathcal E}_Y &\simeq \frac{m}{24\pi u_x^0 a_0} \int [dx] [2[\partial_x \tilde u_y(x)]^2 - 16 m^2 [\tilde u_y(x)]^2 + \frac{2}{4}\tilde u_y(x)^4 -{\mathcal F}_{Y}(\infty)] \propto \frac{1}{\pi} \int [dx] \{ m^3{\rm sech}^4[2\sqrt{2}m x] \}
		\end{aligned}
	\end{equation}
	As the fluctuations $u_x = m_z = 0$, such that the full field $\tilde u_x$ and $ \tilde m_z$ is spatial independent, we can further derive the configuration for chirality $\chi(x)$:
	\begin{equation}
		\chi(x) = \tilde u_x(x) \tilde u_y(x) \tilde m_z(x) = u_x^0 u_y^0 m_z^0 \tanh[(x-x_0)/\sqrt{2}\xi_b] = \chi_0\tanh[(x-x_0)/\sqrt{2}\xi_b],
	\end{equation} 
	where $\chi_0$ is the chirality for homogeneous case. By slightly modifying Sec.~[\ref{SecJackiw}] (expanding Bloch Hamiltonian around $(0,\pi/2)$), we have the edge dispersion reads:
	\begin{equation}
		E_{\rm edge}({k_y}) = \pm v_{\rm edge} k_y, \quad v_{\rm edge} = 4J_2 \lambda a_0/\hbar.
	\end{equation}
	\item $m_z$ can change sign via the dynamic of non-zero $(\partial_x m_z)^2$, we can force $u_x = u_y = 0$, this leads to the effective energy for fluctuation fields, i.e., Eq.~\eqref{GLFluctuations}
	\begin{equation}
	\begin{aligned}
		F_T \simeq \frac{m}{24 \pi u_x^0 a_0} \int[dx] {\mathcal F}_Z(x) =\frac{m}{24\pi u_x^0 a_0}\int [dx] [\frac{9}{4} a_0^2(\partial_x  m_z)^2 - 21 m^2 m_z^2 + \frac{U_z}{4}  m_z^4], 
	\end{aligned}
	\end{equation}
	by using the relation mentioned in Sec.~[\ref{GLFullandFluctuation}], and defining the full field $\tilde m_z  = m_z^0 + m_z$, we arrive at the full Ginzburg Landau theory, i.e. Eq.~\eqref{GLInitial}:
	\begin{equation}
	\begin{aligned}
		F_T \simeq \frac{m}{24 \pi u_x^0 a_0} \int[d^2x] {\mathcal F}_Z(x) =\frac{m}{24\pi u_x^0 a_0}\int [dx] [\frac{9}{4} a_0^2 (\partial_x \tilde m_z)^2 - 21 m^2\tilde m_z^2 + \frac{U_z}{4} \tilde m_z^4], 
	\end{aligned}
	\end{equation}
	comparing with Eq.~\eqref{GLInitial}, one can read the coefficient as:
	\begin{equation}
		K= \frac{9}{4} a_0^2, ~\frac{\mathcal M}{2} = -21m^2, ~ U_z = -\frac{\mathcal M}{\phi_0^2} = \frac{42m^2}{(m_z^0)^2},
	\end{equation}
	From Sec.~[\ref{GinzburgLandauDomainWall}], we know that such a theory can hold a domain wall if one can impose the twisted boundary condition for $\tilde m_z$, with the correlation length $\xi_z$ 
	\begin{equation}\label{correlationlength}
		\xi_z = \sqrt{-\frac{K}{\mathcal M}} = \frac{3a_0}{2|m|\sqrt{42}}, \quad
		\tilde m_z = m_z^0 \tanh[(x-x_0)/\sqrt{2}\xi_z], \quad m_z^0 = \pm m.
	\end{equation}
	With the domain wall energy reads:
	\begin{equation}\label{EnergyLambda}
		\begin{aligned}
			{\mathcal E}_Z &\simeq \frac{m}{24\pi u_x^0 a_0} \int [dx] [\frac{9}{4} a_0^2 (\partial_x \tilde m_z)^2 - 21 m^2\tilde m_z^2 + \frac{U_z}{4} \tilde m_z^4- {\mathcal F}_Z(\infty)] \propto \frac{1}{\pi u_x^0 a_0} \int [dx] \bigg{[}\frac{21m^5}{16} {\rm sech}^4 \bigg{(}2\sqrt{\frac{7}{3}}mx \bigg{)} \bigg{]},
		\end{aligned}
	\end{equation}
	and the chirality also have the profile 
	\begin{equation}
		\chi(x) = \tilde u_x(x) \tilde u_y(x) \tilde m_z(x) = u_x^0 u_y^0 m_z^0 \tanh[(x-x_0)/\sqrt{2}\xi_z] = \chi_0\tanh[(x-x_0)/\sqrt{2}\xi_z].
	\end{equation} 
	According to Jackiw's theory Eq.~\eqref{JackiwVelocity}, we have the edge dispersion for this type of domain wall as:
	\begin{equation}
		E_{\rm edge}(k_y) = \pm v_{\rm edge} k_y, \quad v_{\rm edge} = 2J_1 \bar \rho_y a_0/\hbar.
	\end{equation}
\end{enumerate}

Now we would like to compare the free energy for velocity domain wall Eq.~\eqref{EnergyVelocity} and lambda domain wall Eq.~[\ref{EnergyLambda}], as shown in Fig.[\ref{ContourPlot}.(a)]. For small $\eta = J_2/J_1$ (which is the physical case as the next-nearest neighbor coupling is weaker than the nearest neighbor coupling), the system will favor the lambda domain wall. This is consistent with the energy plot of Fig.~\ref{ContourPlot}.(b,c), where the transition will happen along the smoother approach, though the phase transition point may be different based on numerics.

\subsection{Domain wall tension for the Ginzburg Landau $\phi^4$ theory}\label{SecTension}
The domain wall energy can be written in terms of dimensionless local gap $\tilde m(x) = m(x)/J_1$ as
\begin{equation}
	F = \int [dx] {\mathcal F}[x] \sim \frac{m_0J_1}{24\pi a_0} \int [dx] \bigg{[} K (\partial_x m)^2 - \frac{r}{2} m^2 + \frac{U}{4}m^4 \bigg{]},
\end{equation}
where $m_0 = 2J_2 \bar \lambda/J_1 \bar \rho_0$ is the dimensionless mass deep in the bulk, which is proportional to the bulk gap in the homogeneous limit. For $J_2 < J_1$, in general $0<m_0 <1$.

According to the effective field theory mentioned in our previous note, $K = \tilde K_0 l_0^2$, with $\tilde K_0$ is a pure number, while $r \propto m_0^2$, and $Um_0^2 = r$. We have the following solution for equation of motions for the boundary condition mentioned in Fig.~[\ref{Illustration}.(b)]:
\begin{equation}
	\tilde m(x) = m_0 \tanh [(x- x_0)/(\sqrt{2} \xi)], \quad \xi = \sqrt{\frac{K}{r}} = \frac{l_0}{m_0},  
\end{equation}
where $l_0$ is a length scale close to the lattice constant. 

The domain wall tension (domain wall energy measured from the homogenous results) reads:
	\begin{equation}\label{EnergyLambda}
		\begin{aligned}
			{\mathcal E} &= \frac{m_0J_1}{24\pi} \int [dx] [K (\partial_x  m)^2 - \frac{r}{2} m + \frac{U}{4} m^4- {\mathcal F}(\infty)] = \frac{m_0^5J_1 \tilde K_0}{96\pi} \int [dx] \bigg{[}  {\rm sech}^4 \bigg{(} \frac{x-x_0}{\sqrt{2} \xi} \bigg{)} \bigg{]} \propto m_0^5J_1 \ll J_1 \sim J_2,
		\end{aligned}
	\end{equation}
where we have used fact $0<m_0<1$ such that $m_0^5 \ll 1$, while $J_1$ and $J_2$ are to the same order.

\section{Singular Ginzburg-Landau $|\phi|^3$ theory}\label{SecPhi3}
Now let's consider the most generic Ginzburg-Landau theory:
\begin{equation}\label{GenericGL}
	F \simeq \frac{m}{u_x^0 a_0} \int [dx] \bigg{[} K (\partial_x m)^2 + V(m) \bigg{]},
\end{equation}
where $V(m)$ is w-shaped with two local minimal, and does not have to be a traditional $\phi^4$ theory.

\subsection{Energy for a homogenous slab and approximation based on Dirac-fermion's spectrum}

Since we know the domain wall solution is $\rho(x) = \bar \rho_x$, $\rho(y) = \bar \rho_y$ and $\lambda (x) = \bar \lambda \tanh x/(\sqrt{2}\xi$), we could take the approximation that the spectrum is that of a massive Dirac fermion, such that we can formulate the energy in terms of the local mass $m$ as (we would like to see the scaling so we drop all the coefficient with dimension such as $v_F$):
\begin{equation}\label{SlabEnergy}
\begin{aligned}
	V(m) &\sim m^2 - \int_{\rm BZ} [d^2k] \sqrt{k^2 + m^2}  \sim  m^2 - \int_0^{k_0} dk k\sqrt{k^2 +m^2} \sim m^2 + \frac{1}{3} |m|^3 - \frac{1}{3} (k^2_0 + m^2)^{3/2},
\end{aligned}
\end{equation}
where $k_0$ is the cutoff from the BZ. Note the presence of a $|m|^3$ term, which is the unique signature from the spectrum of Dirac fermions in 2D, and would be singular at the gapless point $m = 0$.

\begin{figure}[!h]
\centering 
\includegraphics[width=0.7\columnwidth]{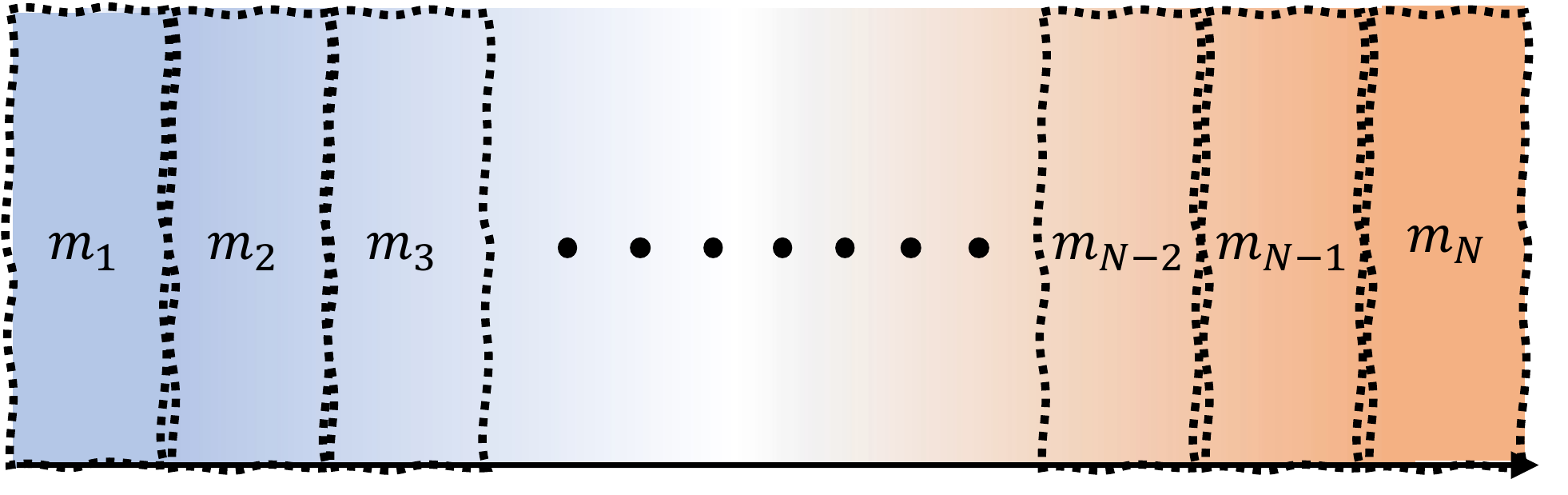}
\caption{\label{Partition} Partition of chiral spin liquid with spatial varying mass. We assume that the mass is uniform in each block $m_i$, with $m_1 \equiv - m_0 = -4J_2\bar \lambda$ and $m_N = +m_0 = + 4J_2 \bar \lambda$.}
\end{figure}

\subsection{Singular Ginzburg-Landau theory for domain walls}
We know the domain wall can be captured by the chirality profile $\chi(x) = \rho_x(x) \rho_y(x) \lambda(x) = \bar \rho_x \bar \rho_y \lambda(x)$. The local mass $m(x)$ is proportional to $\lambda(x)$ as $m(x) = 4J_2\lambda(x)$. So long as the domain varies slow enough, we can divide it into individual slabs, as shown in Fig.~[\ref{Partition}]. One can view the mass as homogenous for $i$-th slab as $m_i$, and the energy for $i$-th slab can be captured by Eq.~\eqref{SlabEnergy}. Thus the potential energy for the domain wall is:
\begin{equation}\label{PotentialContinuous}
	V(m) = \sum_i V(m_i)  \sim  \sum_i \bigg{[}m^2_i + \frac{1}{3} |m_i|^3 - \frac{1}{3} (k^2_0 + m^2_i)^{3/2} \bigg{]} \sim \int [dx]\bigg{[} - \frac{r}{2} m^2 + \frac{S}{3} |m|^3 \bigg{]},
\end{equation}
add back the stiffness term, we have the singular Ginzburg Landau theory:
\begin{equation}
	F \simeq \frac{m}{u_x^0 a_0}  \int dx \bigg{[} K (\partial_x m)^2 + V(m) \bigg{]} = \frac{m}{u_x^0 a_0}  \int dx \bigg{[} K (\partial_x m)^2 - \frac{r}{2} m^2 + \frac{S}{3}|m|^3 \bigg{]}.
\end{equation}

\begin{figure}[!h]
\centering 
\includegraphics[width=1\columnwidth]{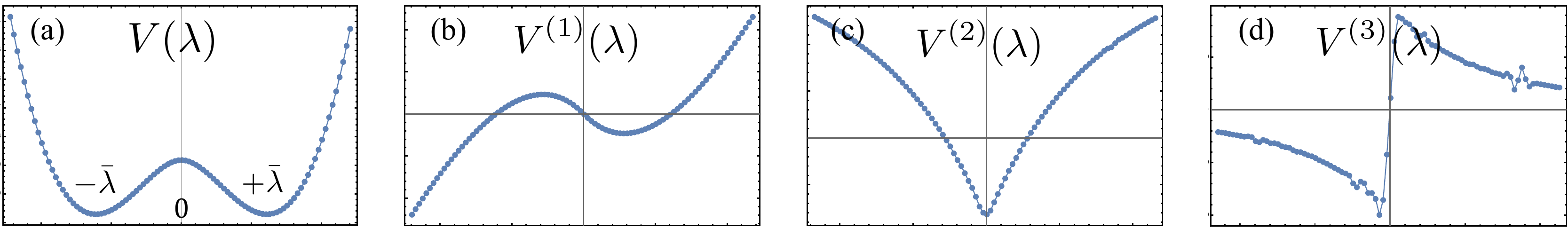}
\caption{\label{Differentiate} (a) $V(\lambda) = V(\bar \rho_x, \bar \rho_y, \lambda)$ of Eq.~\eqref{Energy} with $\rho_x$ and $\rho_y$ fixed to their homogenous value $\bar \rho_x$ and $\bar \rho_y$ respectively, while $\lambda$ is varying. The horizontal axis is $\lambda$ for (a-d). Figure (b-d) are the first, second, and third order differentiate of $V(\lambda)$. Clearly, the discontinuity in (d) for the $\frac{d^3V(\lambda)}{d\lambda^3}$ at $\lambda =0$ implies the existence of $|\lambda^3|$.}
\end{figure}

One can plot Eq.~\eqref{Energy} with $\rho_x$ and $\rho_y$ fixed to the homogenous mean field value $\bar \rho_x$ and $\rho_y$, as the $V(\lambda)$ shown in Fig.~[\ref{Differentiate}.(a)]. There are two local minimals in $V(\lambda)$, and the domain wall captures the tunneling from one local minimal to the other minimal with opposite chirality. The $V(\lambda)$ here may be viewed as the potential function $V(m)$ in the Ginzburg-Landau theory for $m$ in Eq.~\eqref{GenericGL}. The discontinuity of the third order derivative for $V(m)$ shows the existence of $|\lambda|^3$ $(|m|^3)$ in the expansion of $V(\lambda)$ ($V(m)$).

\subsection{Modification to domain wall profile}
Consider the following generic Ginzburg-Landau theory for order parameter $\phi$:
\begin{equation}
	F[\phi] \simeq \frac{m}{u_x^0 a_0}  \int dx \bigg{[} K (\partial_x \phi)^2 + V(\phi) \bigg{]},
\end{equation}
where $V(\phi)$ is an even function with two local minima with a shape of $w$ like Fig.~[\ref{Differentiate}.(a)] (but not necessary to be a $\phi^4$ theory). The equation of motion then reads:
\begin{equation}
	-K \partial_x^2 \phi + \frac{d V(\phi)}{d\phi}= 0.
\end{equation}
Since:
\begin{equation}
	\phi^{\prime\prime}=\frac{d \phi^\prime}{d x}  =  \frac{d \phi^\prime}{d\phi} \frac{d \phi}{d x} = \frac{d\phi^\prime}{d \phi} \phi^\prime = \frac{1}{2}\frac{d[(\phi^\prime)^2]}{d\phi} ,
\end{equation}
then we shall have:
\begin{equation}
	\begin{aligned}
		\frac{K}{2} \frac{d (\phi^\prime)^2}{d \phi} &= \frac{dV(\phi)}{d\phi}
	\end{aligned}
\end{equation} 
integrating from both sides we have:
\begin{equation}
    \frac{K}{2} (\phi^\prime)^2 = V[\phi] + {\rm Const}.
\end{equation}
We assume that when $x \rightarrow \infty$, $\phi(x) \rightarrow \phi_m$ (the local minimal for $V(\phi)$ deep in the bulk) and  $\phi^\prime(x) = 0$, thus we have
\begin{equation}
   \frac{d\phi}{dx}=  \pm \sqrt{\frac{2}{K}[V(\phi) - V(\phi_m)]}
\end{equation}
from which we can solve:
\begin{equation}\label{StructureIntegral}
    x - x_0= \pm \int_0^{\Phi} \frac{d\phi}{\sqrt{\frac{2}{K}[V(\phi) - V(\phi_m)]}}
\end{equation}
the right hand side counts for the time of a classical particle moving in some strange potential field. The profile may no longer have a closed-form analytic solution like tanh as in the $\phi^4$ theory, and we have the decay half-width:
\begin{equation}
    d = \int_0^{\phi_m/2} \frac{d\phi}{\sqrt{\frac{2}{K}[V(\phi) - V(\phi_m)]}} = \sqrt{\frac{K}{r}} \int_0^{\phi_m/2} \frac{d\phi}{\sqrt{2[V(\phi) - V(\phi_m)]/r}} = \eta \xi,
\end{equation}
where $\phi_m$ is the value of order parameter that deep in the bulk (i.e., $x \rightarrow \pm \infty$). Then we shall have $d_1 = \eta_1 \xi$ and $d_2 = \eta_2 \xi$ denotes the width of the domain wall. They have the same order $\xi$ (inverse proportional to bulk gap), but the exact value of $\eta_1$ and $\eta_2$ may be different, as we will find in the following case.

\begin{figure}[!h]
\centering 
\includegraphics[width=1\columnwidth]{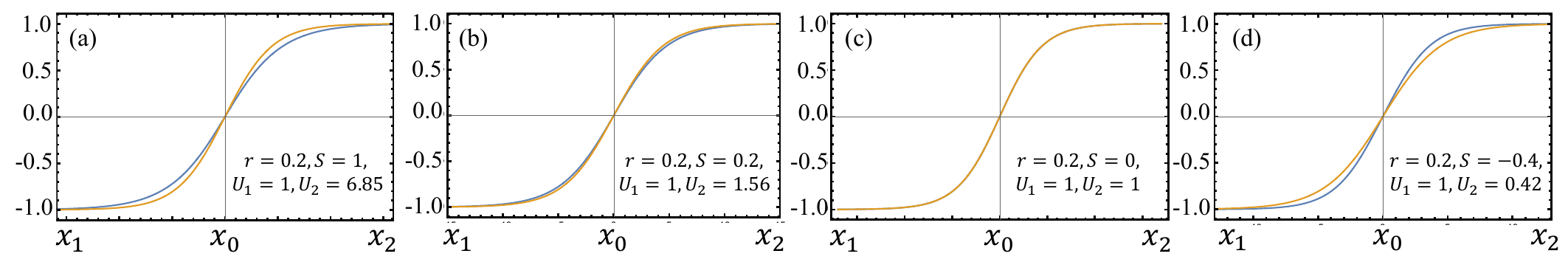}
\caption{\label{Fit01} Profile of $\phi$ from integrating Eq.~[\ref{StructureIntegral}] for two potentials given in Eq.~[\ref{TwoPotential}]. With the data listed in the bottom right of each plot, $V_1(\phi)$ and $V_2(\phi)$ would have same value of local minimal $\phi_m$ as long as the constraint Eq.~[\ref{SameMinimalRequirement}] is satisfied. The blue curve is the solution from integraing $V_1(\phi)$, and the orange curve is from integrating $V_2(\phi)$. For $S=0$, $U_1=U_2$, we have two profile coincides with each other, as shown in (c). For $S>0$, we have wider domain wall as the $|\phi^3|$ is introduced, see in (b). In general, the large $|S|$ is, the greater the modification to the domain wall width will be. When $S<0$, we have the domain wall with $|\phi^3|$ become sharper compared with the $\phi^4$ theory, as shown in (d).}
\end{figure}

Consider two different way to formulate the potential $V(\phi)$, where mass $r$ for them are from integrate out the spinons in our old note which should be identical:
\begin{equation}\label{TwoPotential}
	\begin{aligned}
		V_1(\phi) &= - \frac{r}{2} \phi^2 + \frac{S}{3} |\phi|^3 + \frac{U_1 \phi^4}{4}, \\
		V_2(\phi) &= - \frac{r}{2} \phi^2 + \frac{U_2}{4} \phi^4 .
 	\end{aligned}
\end{equation}
The local minimal for $V_1(\phi)$ and $V_2(\phi)$ are set at same value $\phi_m$, such that:
\begin{equation}
    \begin{aligned}
        \frac{\partial V_1(\phi)}{\partial \phi} \bigg|_{\phi_m} =  \frac{\partial V_2(\phi)}{\partial \phi} \bigg|_{\phi_m}=0, \quad \frac{\partial^2 V_1(\phi)}{\partial \phi^2} \bigg|_{\phi_m} > 0, \quad \frac{\partial^2 V_2(\phi)}{\partial \phi^2} \bigg|_{\phi_m} > 0,
    \end{aligned}
\end{equation}
solving above leads to the following requirement $S,U_1,U_2$:
\begin{equation}\label{SameMinimalRequirement}
    U_2 = \frac{S^2+2rU_1+S\sqrt{S^2 + 4rU_1}}{2r}.
\end{equation}
With Eq.~[\ref{SameMinimalRequirement}], we will have $V_1(\phi)$ and $V_2(\phi)$ has the same value of local minimal. Now we substitute Eq.~\eqref{TwoPotential} into Eq.~\eqref{StructureIntegral}, and conduct the integral numerically, as shown in Fig.~\eqref{Fit01}. From which we find that a positive $|\phi^3|$ will make the domain wall wider than the $\phi^4$ theory.

\section{Effective Chern-Simons theory for mean field chiral spin liquids}
\subsection{Low energy theory}
We can expand the Hamiltonian around the point $(k_x,k_y)\sim(0,0)$, then we shall have:
\begin{equation}\label{LowCSL}
	\begin{aligned}
		H_0 \sim \sum_{\bm k} \tilde \psi^\dagger_{\bm k} [2J_1^x \bar \rho^0_x a_0 k_x \sigma_x + 2(J_2^+ \bar \lambda^0_+  + J_2^- \bar \lambda^0_-) \sigma_y -2J_1^y \bar \rho^0_y a_0 k_y \sigma_z]  \tilde \psi_{\bm k}
	\end{aligned}
\end{equation}
one can further set $\tilde \psi_{\bm k} = \sigma_x \psi_{\bm k}$ to rotate the above to a more convenient way
\begin{equation}
	H_0 \sim \sum_{\bm k} \psi_{\bm k}^\dagger[\hbar v_x k_x \sigma_x + \hbar v_y k_y \sigma_y + \tilde m\sigma_z] \psi_{\bm k},
\end{equation}
with:
\begin{equation}
    \hbar v_x = 2J_1^x \bar \rho^0_x a_0, \quad \hbar v_y = -2J_1^y \bar \rho^0_y a_0, \quad \tilde m = 2(J_2^+ \bar \lambda_+^0 + J_2^- \bar \lambda_-^0).
\end{equation}

\subsection{Edge velocity from Jackiw-Rebbi mechanism and QAHE}\label{SecJackiw}
The Jackiw-Rebbi mechanism~\cite{Jackiw1976,bernevig2013} for QAHE tells that, for a low energy effective theory:
\begin{equation}
	H = \int dxdk_y \{\psi^\dagger_{k_y}(x)[\hbar v_x k_x \sigma_x + \hbar v_y k_y \sigma_y + \tilde m(x) \sigma_z] \psi_{k_y}(x)\},
\end{equation}
where $\tilde m(x)$ change sign at $x=0$, say $\tilde m(x) = \tilde m_0\tanh[(x-x_0)/(\sqrt{2}\xi)]$ for instance, then the edge states on $x =x_0$ will have the following dispersion:
\begin{equation}
	E_{\rm edge}(k_y)= - v_{\rm edge} k_y, \quad v_{\rm edge} = |v_y|,
\end{equation}  
for the low energy theory of chiral spin liquid Eq.~\eqref{LowCSL}, we shall have:
\begin{equation}\label{JackiwVelocity}
	v_{\rm edge} = |2J_1^y \bar \rho^0_y a_0|.
\end{equation}

\subsection{Edge spinon confinement}
We follow Refs.~\cite{Lu2012B,Lai2013} to get the effective field theory from our mean field chiral spin liquids. The two flavors of fermions, $\{ f_\pm\}$, fill the lowest bands with Chern number $C_\pm = 1$. Conserved fermion currents $J^\mu_m$ can be expressed in terms of dynamical U(1) gauge field $a^m_\mu$ as $J^\mu_m = \epsilon^{\mu \nu \lambda} \partial_\nu a^m_\lambda/(2\pi)$, where the summation over repeated $\mu,\nu,\lambda$ is assumed. The fermion band structure in this CSL is described by the following two copies of U(1) Chern Simons theory:
\begin{equation}
    \begin{aligned}
        {\mathcal L}_f &= \frac{\epsilon^{\mu \nu \lambda}}{4\pi} \bigg{(}\sum_{m = \pm} C_m a_\mu^m \partial_\nu a^m_\lambda \bigg{)} + \frac{\epsilon^{\mu \nu \lambda}}{2\pi} A_\mu^{S^z} \partial_\nu \bigg{[}\frac{1}{2}(a_\lambda^+ - a_\lambda^-) \bigg{]} \\
        &= \frac{\epsilon^{\mu \nu \lambda}}{4\pi} C_{IJ} a_{I\mu} \partial_\nu a_{J\lambda} + \frac{\epsilon^{\mu \nu \lambda}}{2\pi} t_I A_\mu^{S^z} \partial_\nu a_{I\lambda},
    \end{aligned}
\end{equation}
where $A^{S^z}_\mu$ is the gauge potential that couples to the $S^z$ spin density and current, $I,J = 1,2$ and:
\begin{equation}
    C =\begin{pmatrix}
        1 & 0 \\
        0 & 1
    \end{pmatrix},
    \quad 
    t = \begin{pmatrix}
        +1/2 \\
        -1/2
    \end{pmatrix}.
\end{equation}

The local no-double occupancy constraint:
\begin{equation}\label{NDConstrain}
    \langle f^\dagger_{i\alpha} f_{i\alpha} \rangle = 1,
\end{equation}
can be written in a covariant form:
\begin{equation}
    \frac{\epsilon^{\mu \nu \lambda}}{2\pi} \sum_m \partial a_\lambda^m = \sum_{m = \pm} J^\mu_m = \bar J^\mu \equiv \frac{\epsilon^{\mu \nu \lambda}}{2\pi}\partial_\nu \bar a_\lambda,
\end{equation}
where $\bar a_\mu$ is a nondynamical (constant) background field, whose density $\bar J^0 = (\partial_x \bar a_y - \partial_y \bar a_x )/2\pi$ is the right hand side of Eq.~\eqref{NDConstrain}. The constriant can be implemented by introducing an extra U(1) gauge field $b_\mu$ as a Lagrangian multiplier:
\begin{equation}
    {\mathcal L}_{\rm con} = \frac{\epsilon^{\mu \nu \lambda}}{2\pi} b_\mu \partial_\nu \big{(} a_\lambda^+ + a_\lambda^- - \bar a_\lambda \big{)}=\frac{\epsilon^{\mu \nu \lambda}}{2\pi} b_\mu \partial_\nu \big{(} \sum_I a_{I\lambda} - \bar a_\lambda \big{)}.
\end{equation}
After integrating out the gauge field $b^\mu$ and $a_{2\mu}$ (or $a^-_\mu$ in the original language), we can obtain the low-energy theory of the CSL:
\begin{equation}
    {\mathcal L}_{\rm CS} = {\mathcal L}_f + {\mathcal L}_{\rm con} = \frac{\epsilon^{\mu \nu \lambda}}{4\pi} K a_\mu \partial_\nu a_\lambda + \frac{\epsilon^{\mu \nu \lambda}}{2\pi} q A_\mu^{S^z} \partial_\nu a_\lambda,
\end{equation}
with $a_\mu$ for $a_\mu^+$ or $a_{1\mu}$. The $1\times 1$ $K$ matrix and $q$-vector are:
\begin{equation}\label{KMatrix}
    K = 2, q = 1.
\end{equation}
The CSL has spin quantum Hall conductance in unit of $1/2\pi$:
\begin{equation}
    \sigma_{xy}^S = q^T K^{-1} q = 1/2,
\end{equation}
in accordance with the bulk spinon transport. 

The corresponding edge theory can be derived from bulk-edge correspondence. The effective edge theory can be written as:
\begin{equation}
    {\mathcal L}_{\rm edge} = \frac{1}{4\pi} K \partial_t \phi \partial_x \phi - V (\partial_x \phi)^2 + \frac{1}{2\pi}q A_0^{S^z}(\partial_x \phi - \partial_t \phi).
\end{equation}
The $K$-matrix and $q$ vector are defined in Eq.~\eqref{KMatrix}. The $V$ is the edge velocity, as $K$ is positive, the edge state is chiral and stable. The $S^z$ density on the edge is given by the defined bosons as:
\begin{equation}
    S^z(x) \simeq q \frac{\partial_x \phi(x)}{2\pi},
\end{equation}
as $q = 1$ it indeed carries $S^z$ quantum number $+1$. The edge boson fields satisfy the Kac-Moody algebra:
\begin{equation}
    [\phi(x), \partial_x \phi(y)] = i 2\pi (K^{-1}) \delta(x-y) = i \pi \delta(x-y),
\end{equation}
or $[\phi(x),\phi(y)]=i(\pi/2){\rm sign}(x-y)$, which is indeed a U(1)${}_2$ theory.

\section{Detailed field theory calculations}\label{Appendix}
\subsection{Useful Dictionary}
Before conducting the calculations for the loop diagram, we acknowledge the following useful formulas for traces:
\begin{equation}
	{\rm Tr} \tau^a \tau^b = 2\delta^{ab}, \quad {\rm Tr} \tau^a \tau^b \tau^c =2 i\epsilon^{abc}, \quad {\rm Tr}(\tau^a \tau^b \tau^c \tau^d) = 2(\delta^{ab} \delta^{cd}  + \delta^{ad} \delta^{bc} - \delta^{ac} \delta^{bd}).
\end{equation}
and for Feynman integrals:
\begin{equation}
	\begin{aligned}
		&\int \frac{d^{2\omega} l }{(2\pi)^{2\omega}} \frac{1}{(l^2 + M^2 + 2 l \cdot{} p)^A} = \frac{\Gamma(A-\omega)}{(4\pi)^\omega \Gamma(A)} \frac{1}{(M^2 - p^2)^{A-\omega}}, \\
		&\int \frac{d^{2\omega} l}{(2\pi)^{2\omega}} \frac{l_\mu}{(l^2 + M^2 + 2 l \cdot{} p)^A} = -\frac{\Gamma(A-\omega)}{(4\pi)^\omega \Gamma(A)} \frac{p_\mu}{(M^2 - p^2)^{A-\omega}}, \\
		&\int \frac{d^{2\omega} l}{(2\pi)^{2\omega}} \frac{l_\mu l_\nu}{(l^2 + M^2 + 2 l \cdot{} p)^A} = \frac{1}{(4\pi)^\omega \Gamma(A)} \times \bigg{[} p_\mu p_\nu \frac{\Gamma(A-\omega)}{(M^2 - p^2)^{A-\omega}} + \frac{1}{2} \delta_{\mu \nu} \frac{\Gamma(A-1-\omega)}{(M^2 - p^2)^{A-1-\omega}} \bigg{]} ,\\
		&\int \frac{d^{2\omega}l}{(2\pi)^{2\omega}} \frac{l_\mu l_\nu l_\rho l_\sigma}{(l^2 + M^2 + 2l \cdot{} p)^A} = \frac{1}{(4\pi)^\omega\Gamma[A]} \bigg{[} p_\mu p_\nu p_\rho p_\sigma \frac{\Gamma(A-\omega)}{(M^2 -p^2)^{A-\omega}}\\
		& + \frac{1}{2}[\delta_{\mu \nu}p_\rho p_\sigma + \delta_{\nu \sigma} p_\mu p_\rho + \delta_{\rho \sigma}p_\mu p_\nu + \delta_{\mu \rho} p_\nu p_\sigma + \delta_{\nu \rho}p_\mu p_\sigma + \delta_{\mu \sigma} p_\rho p_\nu] \frac{\Gamma(A-1-\omega)}{(M^2 -p^2)^{A-1-\omega}}\\
		&+ \frac{1}{4} [\delta_{\mu \nu} \delta_{\rho \sigma} + \delta_{\nu \rho}\delta_{\mu \sigma} = \delta_{\mu \rho} \delta_{\nu \sigma}] \frac{\Gamma(A-2-\omega)}{(M^2 -p^2)^{A-2-\omega}}  \bigg{]} .
	\end{aligned}
\end{equation}

\subsection{Two $\hat u$ vertices (Diagrams like Fig.~[\ref{FeynmanDiagram}.(a)])}
The effective action reads:
\begin{equation}
	S_{\rm eff}^A \equiv \frac{1}{2} \int \frac{d^3k}{(2\pi)^3} \Pi_{\mu \nu}^A(k) \hat u_{\mu}(+k) \hat u_{-\nu}(k).
\end{equation}
With the corresponding polarization function $\Pi^1_{\mu \nu}(k)$, and two $\hat u$ vertices reads: $(q_\mu + k_\mu/2)\tau^\mu$, and $(q_\nu - k_\nu/2)\tau^\nu$ (note that the repeated indices $\mu,\nu$ does not stands for summation, it's just means that the Pauli matrix is bond to the corresponding momentum, say $k_x\tau_x$ and $k_y\tau_y$)
\begin{equation}
    \begin{aligned}
       \Pi^A_{\mu \nu}(k) &=- \int \frac{d^3q}{(2\pi)^3}{\rm Tr} \bigg{[} \frac{m_z^0 + i(q_a +k_a/2) \tau^a}{(m_z^0)^2 + (q+k/2)^2} \tau^\mu\frac{m_z^0+i(q_b - k_b/2)\tau^b}{(m_z^0)^2 + (q-k/2)^2}\tau^\nu\bigg{]} (q_\mu + k_\mu/2) (q_\nu -k_\nu/2) \\ 
       &= -\int \frac{d^3q}{(2\pi)^3} {\rm Tr} \bigg{[} \frac{[m_z^0 + i(q_a + k_a/2)\tau^a]\tau^\mu [m_z^0 + i(q_b - k_b/2)\tau^b ]\tau^\nu}{[(m_z^0)^2 + (q+k/2)^2 ][(m_z^0)^2 + (q-k/2)^2]} \bigg{]} (q_\mu + k_\mu/2) (q_\nu - k_\nu/2) \\
       &= -\int \frac{d^3q}{(2\pi)^3} \int_0^1 [dx] \frac{\Gamma(1+1)}{\Gamma(1)\Gamma(1)} {\rm Tr} \bigg{[} \frac{[m_z^0 + i(q_a + k_a/2)\tau^a]\tau^\mu [m_z^0 + i(q_b-k_b/2)\tau^b]\tau^\nu}{\{ [(q+k/2)^2 + (m_z^0)^2](1-x) + [(q-k/2)^2 +(m_z^0)^2]x \}^2}\bigg{]}(q_\mu + k_\mu/2) (q_\nu - k_\nu/2) \\
       &= -\int \frac{d^3q}{(2\pi)^3} \int_0^1 [dx] {\rm Tr} \bigg{[} \frac{[m_z^0 + i(q_a + k_a/2)\tau^a]\tau^\mu [m_z^0 + i(q_b-k_b/2)\tau^b]\tau^\nu}{\{ q^2 + 2q \cdot{} [1/2-x]k + (m_z^0)^2 +k^2/4 \}^2} \bigg{]} (q_\mu + k_\mu/2) (q_\nu - k_\nu/2) \\
       &= - \int \frac{d^3q}{(2\pi)^3} \int_0^1 [dx] {\rm Tr} \bigg{[}  \frac{[m_z^0 + i(q_a + k_a/2)\tau^a]\tau^\mu [m_z^0 + i(q_b-k_b/2)\tau^b]\tau^\nu}{[(q+(1/2-x)k)^2 + x(1-x)k^2 + (m_z^0)^2]^2}\bigg{]}(q_\mu + k_\mu/2) (q_\nu - k_\nu/2) 
    \end{aligned}
\end{equation}
We can further define:
\begin{equation}
	\tilde q = q + (1/2-x)k, \quad \tilde q + xk = q + k/2, \quad \tilde q - k +x k =q - k/2.
\end{equation}
Then we arrive at:
\begin{equation}
	\begin{aligned}
		\Pi^A_{\mu \nu}(k) &= - \int_0^1 [dx] \int \frac{d^3 \tilde q}{(2\pi)^3} {\rm Tr} \bigg{[} \frac{[m_z^0 + i(\tilde q_a + xk_a)\tau^a]\tau^\mu[m_z^0 + i(\tilde q_b - k_b + xk_b)\tau^b]\tau^\nu}{[\tilde q^2 + x(1-x)k^2 + (m_z^0)^2]^2} \bigg{]} (\tilde q_\mu +x k_\mu)(\tilde q_\nu -k_\nu + xk_\nu) \\
		&= -\int_0^1 [dx] \int \frac{d^3  q}{(2\pi)^3} {\rm Tr} \bigg{[} \frac{[m_z^0 + i( q_a + xk_a)\tau^a]\tau^\mu[m_z^0 + i( q_b - k_b + xk_b)\tau^b]\tau^\nu}{[ q^2 + x(1-x)k^2 + (m_z^0)^2]^2} \bigg{]} ( q_\mu +x k_\mu)( q_\nu -k_\nu + xk_\nu)
	\end{aligned}
\end{equation}
Note that 
\begin{equation}
	\begin{aligned}
		&{\rm Tr} \{[ m_z^0 + i(q_a + xk_a)\tau^a]\tau^\mu [m_z^0 + i(q_b -k_b + xk_b)\tau^b]\tau^\nu \} \\
		=& 2[(m_z^0)^2 + q^2 + x(1-x)k^2]\delta^{\mu \nu} + 4[k^\mu k^\nu -k^2\delta^{\mu \nu}](x-x^2) -4q^\mu q^\nu + (2-4x)(k^\mu q^\nu + k^\nu q^\mu) -2m_z^0 \epsilon^{\mu \nu \lambda} k_\lambda,
	\end{aligned}
\end{equation}
with these, we shall have the response function divided into the following parts:
\begin{equation}
	\begin{aligned} 
	    \Pi^A_{\mu \nu}(k) &= \Pi^{A,1}_{\mu \nu}(k) + \Pi^{A,2}_{\mu \nu}(k) + \Pi^{A,3}_{\mu \nu}(k) + \Pi^{A,4}_{\mu \nu}(k) + \Pi^{A,5}_{\mu \nu}(k) \\
		\Pi^{A,1}_{\mu \nu}(k) &= - \int_0^1 [dx] \int \frac{d^3q}{(2\pi)^3} \frac{2\delta^{\mu \nu}(q_\mu + xk_\mu)(q_\nu -k_\nu +x k_\nu)}{[(m_z^0)^2 + q^2 + x(1-x)k^2]} \\
		\Pi^{A,2}_{\mu \nu}(k) &= -\int_0^1 [dx] \int \frac{d^3q}{(2\pi)^3} \frac{-4q^\mu q^\nu(q_\mu + xk_\mu)(q_\nu -k_\nu +x k_\nu)}{[(m_z^0)^2 + q^2 + x(1-x)k^2]^2} \\
		\Pi^{A,3}_{\mu \nu}(k) &= -\int_0^1 [dx] \int \frac{d^3q}{(2\pi)^3} \frac{4[k^\mu k^\nu -k^2\delta^{\mu \nu}](q_\mu + xk_\mu)(q_\nu -k_\nu +x k_\nu)(x-x^2)}{[(m_z^0)^2 + q^2 + x(1-x)k^2]^2} \\
		\Pi^{A,4}_{\mu \nu}(k) &= -\int_0^1 [dx]\int \frac{d^3q}{(2\pi)^3} \frac{(2-4x)(k^\mu q^\nu + k^\nu q^\mu)(q_\mu + xk_\mu)(q_\nu -k_\nu +x k_\nu)}{[(m_z^0)^2 + q^2 + x(1-x)k^2]^2} \\
		\Pi^{A,5}_{\mu \nu}(k) &= -\int_0^1 [dx] \int \frac{d^3q}{(2\pi)^3} \frac{-2m\epsilon^{\mu \nu \lambda} k_\lambda(q_\mu + xk_\mu)(q_\nu -k_\nu +x k_\nu)}{[(m_z^0)^2 + q^2 + x(1-x)k^2]^2}
	\end{aligned}
\end{equation}

The numerators that are odd in $q$ will vanish. So let us look into the numerator term by term: (again, the repeat indices for $\mu$ and $\nu$ does not stands for summation)
\begin{equation}
\begin{aligned}
	N[\Pi^{A,1}_{\mu \nu}(k)]  &= -2\delta^{\mu \nu}(q_\mu +x k_\mu)(q_\nu + xk_\nu -k_\nu) = -2\delta^{\mu \nu}(q_\mu q_\nu + xk_\mu q_\nu + xq_\mu k_\nu + x^2 k_\mu k_\nu -q_\mu k_\nu -xk_\mu k_\nu)\\
    N[\Pi^{A,2}_{\mu \nu}(k)] &= 4q^{\mu} q^\nu (q_\mu +x k_\mu)(q_\nu  + xk_\nu-k_\nu) = 4q^{\mu}q^{\nu}(q_\mu q_\nu + xk_\mu q_\nu + xq_\mu k_\nu + x^2 k_\mu k_\nu -q_\mu k_\nu -xk_\mu k_\nu) \\
    N[\Pi^{A,3}_{\mu \nu}(k)] &= -4[k^\mu k^\nu - k^2\delta^{\mu \nu}](x-x^2)(q_\mu +x k_\mu)(q_\nu  + xk_\nu-k_\nu)  \\
    &= -4(x-x^2)[k^\mu k^\nu - k^2\delta^{\mu \nu}](q_\mu q_\nu + xk_\mu q_\nu + xq_\mu k_\nu + x^2 k_\mu k_\nu -q_\mu k_\nu -xk_\mu k_\nu) \\
    N[\Pi^{A,4}_{\mu \nu}(k)] &= -(2-4x)[k^\mu q^\nu + k^\nu q^\mu](q_\mu q_\nu + xk_\mu q_\nu + xq_\mu k_\nu + x^2 k_\mu k_\nu -q_\mu k_\nu -xk_\mu k_\nu) \\
    &= -(2-4x)[k^\mu q^\nu q_\mu q_\nu + xk^\mu q^\nu k_\mu q_\nu + x k^\mu q^\nu q_\mu k_\nu + x^2 k^\mu q^\nu k_\mu k_\nu -k^\mu q^\nu q_\mu k_\nu -x k^\mu q^\nu k_\mu k_\nu] \\
    &+ -(2-4x) [k^\nu q^\mu q_\mu q_\nu + xk^\nu q^\mu k_\mu q_\nu + xk^\nu q^\mu q_\mu k_\nu + x^2k^\nu q^\mu k_\mu k_\nu -k^\nu q^\mu q_\mu k_\nu -xk^\nu q^\mu k_\mu k_\nu] \\
    N[\Pi^{A,5}_{\mu \nu}(k)] &= 2m_0^z \epsilon^{\mu \nu \lambda} k_\lambda (q_\mu +x k_\mu)(q_\nu  + xk_\nu-k_\nu) \\
    &= 2m_0^z\epsilon^{\mu \nu \lambda} k_\lambda (q_\mu q_\nu + xk_\mu q_\nu + xq_\mu k_\nu + x^2 k_\mu k_\nu -q_\mu k_\nu -xk_\mu k_\nu)
\end{aligned}
\end{equation}

\begin{equation}
	\begin{aligned}
			N_e[\Pi^{A,1}_{\mu \nu}(k)] &= -2\delta^{\mu \nu} [(x^2-x) k_\mu k_\nu + q_\mu q_\nu ] \\
			N_e[\Pi^{A,2}_{\mu \nu}(k)] &= 4[q^\mu q_\mu q^\nu q_\nu +(x^2 -x)q^\mu q^\nu k_\mu k_\nu ] \\
			N_e[\Pi^{A,3}_{\mu \nu}(k)] &= -4(x-x^2)[k^\mu k^\nu -k^2 \delta^{\mu \nu}][q_\mu q_\nu +(x^2-x)k_\mu k_\nu] \\
			N_e[\Pi^{A,4}_{\mu \nu}(k)] &= -(2-4x)[x q^\nu q_\nu k^\mu k_\mu + (x-1) q^\nu q_\mu k^\mu k_\nu + x q^\mu q_\nu k^\nu k_\mu + (x-1) q^\mu q_\mu k^\nu k_\nu] \\
			N_e[\Pi^{A,5}_{\mu \nu}(k)] &= 2m_0^z \epsilon^{\mu \nu \lambda} k_\lambda[q_\mu q_\nu +(x^2-x)k_\mu k_\nu]
 	\end{aligned}
\end{equation}
With the above, we shall have {\color{blue} by defining $m = |m_0^z|$}:
\begin{equation}
	\begin{aligned}
		\Pi^{A,1}_{\mu \nu}(k) &= \int_0^1 [dx] \int \frac{d^3q}{(2\pi)^3} \frac{-2\delta^{\mu \nu}[(x^2-x)k_\mu k_\nu + q_\mu q_\nu ]}{[q^2 + (m_z^0)^2 + x(1-x)k^2]} \\
		&= \int_0^1 [dx] \frac{\Gamma(1-3/2)}{(4\pi)^{3/2}\Gamma(1)} \frac{-2\delta^{\mu \nu}(x^2-x) k_\mu k_\nu}{[(m_z^0)^2 + x(1-x)k^2]^{1-3/2}} + \int_0^1 [dx] \frac{-2\delta^{\mu \nu}}{(4\pi)^{3/2}\Gamma(1)} \frac{\delta_{\mu \nu}}{2} \frac{\Gamma(1-1-3/2)}{[(m_z^0)^2 + x(1-x)k^2]^{-3/2}} \\
		&=\frac{\delta^{\mu \nu}}{2\pi}  \int_0^1 [dx] \frac{(x^2-x)k_\mu k_\nu}{[(m_z^0)^2 + x(1-x)k^2]^{-1/2}} + \frac{(\delta^{\mu \nu})^2}{6\pi} \int_0^1 [dx] \frac{1}{[(m_z^0)^2 + x(1-x)k^2]^{-3/2}} \\
		&= -\frac{\delta^{\mu \nu} k_\mu k_\nu}{2\pi} \bigg{[} \frac{(-6k^3 m -8km^3 + (-3k^4 -8k^2 m^2 + 16m^4){\rm arccot}[2m/k])}{64 k^3} \bigg{]} \\
		&+\frac{(\delta^{\mu \nu})^2}{6\pi} \bigg{[} \frac{(6k^3 m + 40 k m^3 + 3(k^2 + 4m^2)^2 {\rm arccot}[2m/k])}{64 k} \bigg{]} \\
		&\approx -\frac{\delta^{\mu \nu} k_\mu k_\nu}{2\pi}\bigg{[} -\frac{m}{6} - \frac{k^2}{60 m}  + \frac{k^4}{1120m^3}+ {\mathcal O}(k^6) \bigg{]} + \frac{(\delta^{\mu \nu})^2}{6\pi} \bigg{[} m^3 + \frac{mk^2}{4} + \frac{k^4}{80m} + {\mathcal O}(k^6) \bigg{]}
	\end{aligned}
\end{equation}
and:
\begin{equation}
	\begin{aligned}
		\Pi^{A,2}_{\mu \nu}(k) &= \int_0^1[dx] \int \frac{d^3q}{(2\pi)^3} \frac{4[q^\mu q_\mu q^\nu q_\nu +(x^2-x)q^\mu q^\nu k_\mu k_\nu]}{[q^2 + (m_z^0)^2 + x(1-x)k^2]^2} \\
		&= \int_0^1[dx] \frac{4}{(4\pi)^{3/2}\Gamma(2)} \frac{1}{4} [\delta_{\mu \mu} \delta_{\nu \nu} + \delta_{\mu \nu} \delta_{\nu \mu} + \delta_{\mu \nu} \delta_{\mu \nu}] \frac{\Gamma(2-2-3/2)}{[(m_z^0)^2 + x(1-x)k^2]^{2-2-3/2}} \\
		&+ \int_0^1 \frac{4(x^2-x)k_\mu k_\nu}{(4\pi)^{3/2}\Gamma(2)} \frac{\delta_{\mu \nu }}{2} \frac{\Gamma(2-1-3/2)}{[(m_z^0)^2 + x(1-x)k^2]^{2-1-3/2}} \\
		&=\frac{1}{6\pi} [1 + 2(\delta_{\mu \nu})^2] \int_0^1 [dx]  \frac{1}{[(m_z^0)^2 + x(1-x)k^2]^{-3/2}} -\frac{k_\mu k_\nu \delta_{\mu \nu}}{2\pi}\int_0^1 [dx]  \frac{(x^2-x)}{[(m_z^0)^2 + x(1-x)k^2 ]^{-1/2}} \\
		&= \frac{1}{6\pi}[1+2(\delta_{\mu \nu})^2] \bigg{[} \frac{(6k^3 m + 40 k m^3 + 3(k^2 + 4m^2)^2 {\rm arccot}[2m/k])}{64 k} \bigg{]} \\
		&+ \frac{k_\mu k_\nu \delta_{\mu \nu}}{2\pi}\bigg{[} \frac{(-6k^3 m -8km^3 + (-3k^4 -8k^2 m^2 + 16m^4){\rm arccot}[2m/k])}{64 k^3} \bigg{]}  \\
		&\approx \frac{[1+2(\delta_{\mu \nu})^2]}{6\pi} \bigg{[} m^3 + \frac{mk^2}{4} + \frac{k^4}{80m} + {\mathcal O}(k^6) \bigg{]} + \frac{k_\mu k_\nu \delta_{\mu \nu}}{2\pi} \bigg{[} -\frac{m}{6} - \frac{k^2}{60 m}  + \frac{k^4}{1120m^3}+ {\mathcal O}(k^6) \bigg{]}
	\end{aligned}
\end{equation}
and:
\begin{equation}
	\begin{aligned}
		\Pi^{A,3}_{\mu \nu}(k) &= \int_0^1 [dx] \int \frac{d^3 q}{(2\pi)^3} \frac{-4(x-x^2)[k^\mu k^\nu - k^2\delta^{\mu \nu}][q_\mu q_\nu + (x^2-x)k_\mu k_\nu]}{[q^2 + (m_z^0)^2 + x(1-x)k^2]^2} \\
		&= -\int_0^1 [dx] 4(x -x^2)[k^\mu k^\nu - k^2 \delta^{\mu \nu} ]  \frac{1}{(4\pi)^{3/2}\Gamma(2)} \frac{1}{2}\delta_{\mu \nu} \frac{\Gamma(2-1-3/2)}{[(m_z^0)^2 +x(1-x)k^2]^{2-1-3/2}} \\
		&+ \int_0^1 [dx] 4(x^2-x)^2[k^\mu k^\nu -k^2 \delta^{\mu \nu} ] \frac{\Gamma(2-3/2)}{(4\pi)^{3/2}\Gamma(2)} \frac{k_\mu k_\nu}{[(m_z^0)^2 + x(1-x)k^2]^{2-3/2}} \\
		&= -\frac{\delta_{\mu \nu}[k^\mu k^\nu -k^2\delta^{\mu \nu}]}{2\pi} \int_0^1 [dx] \frac{(x^2-x)}{[(m_z^0)^2 + x(1-x)k^2]^{-1/2}}  + \frac{[k^\mu k^\nu - k^2\delta^{\mu \nu}]}{2\pi} \int_0^1 [dx] \frac{(x^2-x)^2}{[(m_z^0)^2 + x(1-x)k^2]^{1/2}} \\
		&= - \frac{\delta_{\mu \nu}[k^\mu k^\nu - k^2 \delta^{\mu \nu}]}{2\pi} \bigg{[} \frac{-6k^3m -8k^2m^3 + (-3k^4 -8k^2m^2 + 16m^4){\rm arccot}[2m/k]}{64 k^3} \bigg{]} \\
		&+ \frac{[k^\mu k^\nu -k^2 \delta^{\mu \nu}]}{2\pi} k_\mu k_\nu\bigg{[} \frac{6km(k^2-4m^2) + (4k^4 - 8k^2 m^2 + 48m^4){\rm acrcot}[2m/k]}{64 k^5} \bigg{]} \\
		&= - \frac{\delta_{\mu \nu}(k^\mu k^\nu -k^2\delta^{\mu \nu})}{2\pi} \bigg{[} -\frac{m}{6} - \frac{k^2}{60 m} + \frac{k^4}{1120m^3} + {\mathcal O}(k^6)\bigg{]} + \frac{[k^\mu k^\nu - k^2 \delta_{\mu\nu}]k_\mu k_\nu}{2\pi} \bigg{[} \frac{1}{30m} - \frac{k^2}{280 m^3} + \frac{k^4}{1680m^5} + {\mathcal O}[k^6]\bigg{]}
	\end{aligned}
\end{equation}
and:
\begin{equation}
	\begin{aligned}
		&\Pi^{A,4}_{\mu \nu}(k)= \int_0^1 [dx] \int \frac{d^3q}{(2\pi)^3} \frac{-(2-4x)[x q^\nu q_\nu k^\mu k_\mu + (x-1) q^\nu q_\mu k^\mu k_\nu + x q^\mu q_\nu k^\nu k_\mu + (x-1) q^\mu q_\mu k^\nu k_\nu]}{[q^2 + (m_z^0)^2 + x(1-x)k^2]^2} \\
		=& \int_0^1 [dx] \frac{(4x-2)xk^\mu k_\mu}{(4\pi)^{3/2}\Gamma(2)} \frac{\delta^{\nu \nu}}{2} \frac{\Gamma(2-1-3/2)}{[(m_z^0)^2 + x(1-x)k^2]^{2-1-3/2}} + \int_0^1 [dx] \frac{(4x-2)(x-1)k^\mu k_\nu}{(4\pi)^{3/2}\Gamma(2)} \frac{1}{2} \delta^{\mu \nu} \frac{\Gamma(2-1-3/2)}{[(m_z^0)^2 + x(1-x)k^2]^{2-1-3/2}} \\
		+&\int_0^1 [dx] \frac{(4x-2)xk^\nu k_\nu}{(4\pi)^{3/2}\Gamma(2)} \frac{\delta^{\mu \mu}}{2} \frac{\Gamma(2-1-3/2)}{[(m_z^0)^2 + x(1-x)k^2]^{2-1-3/2}} + \int_0^1 [dx] \frac{(4x-2)(x-1)k^\nu k_\mu}{(4\pi)^{3/2}\Gamma(2)} \frac{1}{2} \delta^{\nu \mu} \frac{\Gamma(2-1-3/2)}{[(m_z^0)^2 + x(1-x)k^2]^{2-1-3/2}} \\
		=& -\int_0^1 [dx] \frac{(2x-1)x}{4\pi} \frac{k^\mu k_\mu \delta^{\nu \nu} + k^\nu k_\nu \delta^{\mu \mu}}{[(m_z^0)^2 + x(1-x)k^2]^{-1/2}} - \int_0^1 [dx] \frac{(2x-1)(x-1)}{4\pi}\frac{k^\mu k_\nu\delta^{\nu \mu}+k^\nu k_\mu\delta^{\mu \nu} }{[(m_z^0)^2 + x(1-x)k^2]^{-1/2}}\\
		=& -\frac{k^\mu k_\mu \delta^{\nu \nu} + k^\nu k_\nu \delta^{\mu \mu}}{4\pi} \int_0^1 [dx]  \frac{(2x-1)x}{[(m_z^0)^2 + x(1-x)k^2]^{-1/2}} - \frac{k^\mu k_\nu\delta^{\nu \mu}+k^\nu k_\mu\delta^{\mu \nu} }{4\pi} \int_0^1 [dx] \frac{(2x-1)(x-1) }{[(m_z^0)^2 + x(1-x)k^2]^{-1/2}} \\
		=& -\bigg{[}\frac{k^\mu k_\mu \delta^{\nu \nu} + k^\nu k_\nu \delta^{\mu \mu}}{4\pi} +  \frac{k^\mu k_\nu\delta^{\nu \mu}+k^\nu k_\mu\delta^{\mu \nu} }{4\pi}  \bigg{]} \bigg{[} \frac{(2k^3m -8km^3 + (k^2 + 4m^2)^2 {\rm arccot}(2m/k))}{32k^3}\bigg{]} \\
		=& -\bigg{[}\frac{k^\mu k_\mu \delta^{\nu \nu} + k^\nu k_\nu \delta^{\mu \mu}}{4\pi} +  \frac{k^\mu k_\nu\delta^{\nu \mu}+k^\nu k_\mu\delta^{\mu \nu} }{4\pi}  \bigg{]}  \bigg{[} \frac{m}{6} + \frac{k^2}{120m} - \frac{k^4}{3360m^3} + {\mathcal O}[k^6]\bigg{]}
	\end{aligned}
\end{equation}
and:
\begin{equation}
	\begin{aligned}
		\Pi^{A,5}_{\mu \nu}(k) &= \int_0^1 [dx] \int \frac{d^3q}{(2\pi)^2} \frac{2m_z^0 \epsilon^{\mu \nu \lambda}k_\lambda[q_\mu q_\nu + (x^2-x) k_\mu k_\nu]}{[q^2 + (m_z^0)^2 + x(1-x)k^2]^2} \\
		&= \int_0^1 [dx] \frac{2m_z^0 \epsilon^{\mu \nu \lambda} k_\lambda}{(4\pi)^{3/2}\Gamma(2)} \frac{1}{2}\delta_{\mu \nu} \frac{\Gamma(2-1-3/2)}{[(m_z^0)^2 + x(1-x)k^2]^{2-1-3/2}} + \frac{\Gamma(2-3/2)}{(4\pi)^{3/2}\Gamma(2)}\int_0^1 [dx] \frac{(x^2-x)2m_z^0 \epsilon^{\mu \nu \lambda}k_\mu k_\nu k_\lambda}{[(m_z^0)^2 + x(1-x)k^2]^{2-3/2}} \\
		&= -\frac{\delta_{\mu \nu}m_z^0 \epsilon^{\mu \nu \lambda} k_\lambda}{4\pi}\int_0^1 [dx] \frac{1}{[(m_z^0)^2 + x(1-x)k^2]^{-1/2}} + \frac{m_z^0 \epsilon^{\mu \nu \lambda} k_\mu k_\nu k_\lambda}{4\pi} \int_0^1 [dx] \frac{(x^2-x) }{[(m_z^0)^2 + x(1-x)k^2]^{1/2}} \\
		&= -\frac{\delta_{\mu \nu} m_0^z \epsilon^{\mu \nu \lambda} k_\lambda}{4\pi} \bigg{[} \frac{m}{2} + \frac{i(k^2 + 4m^2)(\ln (-ik+2m)-\ln(ik+2m))}{8k} \bigg{]} \\
		&+  \frac{m_z^0 \epsilon^{\mu \nu \lambda} k_\mu k_\nu k_\lambda}{4\pi}  \bigg{[} - \frac{2k m + (k^2 -4m^2) {\rm arctan}[k/2m]}{4k^3} \bigg{]} \\
		&\approx -\frac{\delta_{\mu \nu} m_z^0 \epsilon^{\mu \nu \lambda} k_\lambda}{4\pi}  \bigg{[} m + \frac{k^2}{12m} - \frac{k^4}{240m^3} + {\mathcal O}[k^6] \bigg{]} + \frac{m_z^0 \epsilon^{\mu \nu \lambda} k_\mu k_\nu k_\lambda}{4\pi} \bigg{[} - \frac{1}{6m} + \frac{k^2}{10m^3} - \frac{3k^4}{1120m^5} + {\mathcal O}[k^6] \bigg{]}
	\end{aligned}
\end{equation}

\subsubsection{Second order to $k$}
Up to second order of $k$, we shall have 
\begin{equation}
\begin{aligned}
	\Pi^{A,1}_{\mu \nu}(k) &\approx \frac{m^3}{6\pi} (\delta^{\mu \nu})^2 + {{\frac{m}{12\pi} \delta^{\mu \nu} k_\mu k_\nu}} +{\frac{m}{24 \pi}(\delta^{\mu\nu})^2 k^2}  \\
	\Pi^{A,2}_{\mu \nu}(k) &\approx \frac{m^3}{6\pi}[1+2(\delta_{\mu\nu})^2] + {\frac{m}{24\pi}[1+2(\delta_{\mu \nu})^2]k^2 - \frac{m}{12\pi}\delta_{\mu \nu} k_\mu k_\nu} \\
	\Pi^{A,3}_{\mu \nu}(k) &\approx {\frac{m}{12 \pi}\delta_{\mu \nu} (k^\mu k^\nu -k^2\delta^{\mu \nu})} \\
	\Pi^{A,4}_{\mu \nu}(k) &\approx {{-\frac{m}{24\pi} [k_\mu^2 + k_\nu^2 + 2k_\mu k_\nu \delta^{\mu \nu}]}} \\
	\Pi^{A,5}_{\mu \nu} (k) & \approx 0
\end{aligned}
\end{equation}
which gives:
\begin{equation}
	\begin{aligned}
		\Pi^A_{\mu \nu}(k) &= \sum_{i} \Pi^{A,i}_{\mu\nu}(k) = \frac{m^3}{6\pi} [1+3(\delta_{\mu \nu})^2] + {\frac{m}{24 \pi}[1+(\delta_{\mu\nu})^2]k^2 - \frac{m}{24\pi} (k_\mu^2 +k_\nu^2)}
	\end{aligned}
\end{equation}
and:
\begin{equation}
	\begin{aligned}
		\Pi^A_{xx}(k) &= \frac{2m^3}{3\pi} + \frac{m}{12\pi}(k^2-k_x^2) \\
		\Pi^A_{xy}(k) &= \frac{m^3}{6\pi} +\frac{m}{24\pi}(k^2-k_x^2-k_y^2) \\
		\Pi^A_{yx}(k) &= \frac{m^3}{6\pi} + \frac{m}{24\pi}(k^2-k_y^2 -k_x^2) \\
		\Pi^A_{yy}(k) &= \frac{2m^3}{3\pi} + \frac{m}{12\pi}(k^2-k_x^2)
	\end{aligned}
\end{equation}

\subsection{Two $\hat m_z$ vertices (Diagrams like Fig.~[\ref{FeynmanDiagram}].(b))}
The effective action reads:
\begin{equation}
	S^B_{\rm eff} \equiv \frac{1}{2} \int \frac{d^3k}{(2\pi)^3} \Pi^B_{zz}(k) \hat m_z(+k) \hat m_z(-k),
\end{equation}
with the term:
\begin{equation}
	\begin{aligned}
		\Pi^B_{zz}(k) &= \int \frac{d^3q}{(2\pi)^3}{\rm Tr} \bigg{[}  \frac{m_z^0 + i(q_a + k_a/2)\tau^a}{(m_z^0)^2 + (q+k/2)^2} m_z^0 \tau_0 \frac{m_z^0 + i(q_b-k_b/2)\tau^b}{(m_z^0)^2 + (q-k/2)^2} m_z^0 \tau_0\bigg{]} \\
		&= (m_z^0)^2\int_0^1 [dx] \int \frac{d^3q}{(2\pi)^3}  \frac{\Gamma(1+1)}{\Gamma(1)\Gamma(1)}{\rm Tr} \bigg{[} \frac{[m_z^0 + i(q_a + k_a/2)\tau^a][m_z^0 + i(q_b-k_b/2)\tau^b)]}{\{ [(m_z^0)^2 + (q+k/2)^2](1-x) + [(m_z^0)^2 + (q-k/2)^2]x \}^2}\bigg{]} \\
		&= (m_z^0)^2\int_0^1 [dx] \int \frac{d^3q}{(2\pi)^3} {\rm Tr} \bigg{[} \frac{[m_z^0 + i(q_a + k_a/2)\tau^a][m_z^0 + i(q_b-k_b/2)\tau^b)]}{\{ q^2 + 2q \cdot [1/2-x]k + (m_z^0)^2 + k^2/4 \}^2} \bigg{]}
 	\end{aligned}
\end{equation}
We can further define:
\begin{equation}
	\tilde q = q + (1/2-x)k, \quad \tilde q+xk = q+k/2, \quad \tilde q-k+xk = q-k/2
\end{equation}
Then we arrive at:
\begin{equation}
	\begin{aligned}
		\Pi^B_{zz}(k) &=  (m_z^0)^2\int_0^1[dx] \int \frac{d^3\tilde q}{(2\pi)^3} {\rm Tr} \bigg{[} \frac{[m_z^0 + i(\tilde q_a + xk_a)\tau^a][m_z^0 + i(\tilde q_b -k_b +xk_b)\tau^b]}{[\tilde q^2 +x(1-x)k^2 +(m_z^0)^2]^2} \bigg{]} \\
		&= (m_z^0)^2\int_0^1 [dx] \int \frac{d^3q}{(2\pi)^3} \frac{(m_z^0)^2 - 2(q_a +xk_a)(q_b + xk_b -k_b)\delta^{ab}}{[q^2 + x(1-x)k^2 + (m_z^0)^2]^2} \\
		&= (m_z^0)^2\int_0^1 [dx] \int \frac{d^3q}{(2\pi)^3} \frac{(m_z^0)^2 -2(q+xk)(q+xk-k)}{[q^2 +x(1-x)k^2 + (m_z^0)^2]^2}\\
		&= (m_z^0)^2\int_0^1 [dx] \int \frac{d^3q}{(2\pi)^3} \frac{(m_z^0)^2 -2q^2 -2(xk)^2 -4qxk +2(q+xk)k}{[q^2 +x(1-x)k^2 + (m_z^0)^2]^2} \\
		&= (m_z^0)^2\int_0^1 [dx] \int \frac{d^3q}{(2\pi)^3} \frac{(m_z^0)^2 -2q^2  +2(x-x^2)k^2}{[q^2 +x(1-x)k^2 + (m_z^0)^2]^2} \\
		&= (m_z^0)^2\int_0^1 [dx] \int \frac{d^3q}{(2\pi)^3} \frac{3(m_z^0)^2 +  4(x-x^2)k^2-2(q^2 +x(1-x)k^2 +(m_z^0)^2) }{[q^2 +x(1-x)k^2 + (m_z^0)^2]^2}\\
		&= (m_z^0)^2\int_0^1 [dx] \int \frac{d^3q}{(2\pi)^3} \bigg{[} \frac{3(m_z^0)^2 + 4(x-x^2)k^2}{[q^2 +x(1-x)k^2 + (m_z^0)^2]^2} - \frac{2}{[q^2 + x(1-x)k^2 + (m_z^0)^2]} \bigg{]}
	\end{aligned}
\end{equation}
With each term reads:
\begin{equation}
	\begin{aligned}
		\Pi^{B,1}_{zz}(k) &= (m_z^0)^2\int_0^1 [dx] \int \frac{d^3q}{(2\pi)^3} \frac{3(m_z^0)^2 + 4(x-x^2)k^2}{[q^2 +x(1-x)k^2 + (m_z^0)^2]^2}  \\
		&= (m_z^0)^2\int_0^1 [dx] \frac{\Gamma(2-3/2)}{(4\pi)^{3/2}\Gamma(2)} \frac{3(m_z^0)^2 + 4(x-x^2)k^2}{[(m_z^0)^2 + x(1-x)k^2]^{2-3/2}} \\
		&= \frac{(m_z^0)^2}{8 \pi} \int_0^1[dx] \frac{3(m_z^0)^2 + 4(x-x^2)k^2}{[(m_z^0)^2 + x(1-x)k^2]^{1/2}} \\
		&= \frac{(m_z^0)^2}{8 \pi} \int_0^1 [dx] \bigg{[} 2m + \bigg{(} k + \frac{2m^2}{k}{\rm arctan}[k/2m] \bigg{)} \bigg{]} \\
		&\approx \frac{m^2}{8\pi}\bigg{[}3m + \frac{5k^2}{12m} - \frac{7k^4}{240m^3} +{\mathcal O}[k^6] \bigg{]}
	\end{aligned}
\end{equation}
and:
\begin{equation}
	\begin{aligned}
		\Pi^{B,2}_{zz}(k) &= (m_z^0)^2\int_0^1 \int \frac{d^3q}{(2\pi)^3} \frac{-2}{[q^2 +x(1-x)k^2 + (m_z^0)^2]} \\
		&=(m_z^0)^2 \int_0^1 [dx]  \frac{(-2)\Gamma(1-3/2)}{(4\pi)^{3/2}\Gamma(1)} \frac{1}{[(m_z^0)^2 + x(1-x)k^2]^{1-3/2}} \\
		&= \frac{(m_z^0)^2}{2\pi} \int_0^1 [dx] \frac{1}{[(m_z^0)^2 + x(1-x)k^2]^{-1/2}} \\
		&= \frac{(m_z^0)^2}{2\pi} \bigg{[} \frac{m}{2} + \frac{i(k^2 + 4m^2)}{8k} \ln \bigg{(} \frac{-ik+2m}{+ik+2m}\bigg{)} \bigg{]} \\
		&= \frac{m^2}{2\pi} \bigg{[} m + \frac{k^2}{12m} - \frac{k^4}{240m^3} + {\mathcal O}[k^6] \bigg{]}
	\end{aligned}
\end{equation}
With above, we have:
\begin{equation}
	\Pi^B_{zz}(k) \approx \frac{m^2}{8\pi} \bigg{[}3m + \frac{5k^2}{12m} - \frac{7k^4}{240m^3} +{\mathcal O}[k^6] \bigg{]}+\frac{m^2}{2\pi} \bigg{[} m + \frac{k^2}{12m} - \frac{k^4}{240m^3} + {\mathcal O}[k^6] \bigg{]}
\end{equation}
to the second order of $k$, would be:
\begin{equation}
	\Pi^B_{zz}(k) \approx \frac{7m^3}{8\pi} + \frac{3m}{32 \pi } k^2
\end{equation}

\subsection{One $\hat u$ vertex and one $\hat m_z$ vertex (Diagrams like Fig.~[\ref{FeynmanDiagram}].(c))}
\begin{equation}
	S_{\rm eff}^C \equiv \frac{1}{2} \int \frac{d^3k}{(2\pi^3)} [\Pi^C_{z\nu} (k)\hat m_z(+k) \hat u_\nu(-k)  + \Pi^C_{\mu z} (k)\hat u_\mu(+k) \hat m_z(-k)]
\end{equation}
with the function:
\begin{equation}
	\begin{aligned}
		\Pi^C_{z \nu} (k) &= \int \frac{d^3q}{(2\pi)^3} {\rm Tr} \bigg{[} \frac{m_z^0 + i(q_a + k_a/2)\tau^a}{(m_z^0)^2 + (q+k/2)^2} m_z^0 \tau_0\frac{m_z^0 + i(q_b -k_b/2)\tau^b}{(m_z^0)^2 + (q-k/2)^2} \tau^\nu \bigg{]} i(q_\nu - k_\nu /2) \\ 
		\Pi^C_{\mu z}(k) &= \int \frac{d^3q}{(2\pi)^3} {\rm Tr} \bigg{[}  \frac{m_z^0 + i(q_a + k_a/2)\tau^a}{(m_z^0)^2 + (q+k/2)^2} \tau^\mu\frac{m_z^0 + i(q_b -k_b/2)\tau^b}{(m_z^0)^2 + (q-k/2)^2}m_z^0 \tau_0 \bigg{]} i(q_\mu + k_\mu/2)
	\end{aligned}
\end{equation}

We shall get them one by one:
\begin{equation}
	\begin{aligned}
		\Pi^C_{z\nu}(k) &=m_z^0  \int_0^1 [dx] \int \frac{d^3q}{(2\pi)^3} \frac{\Gamma(1+1)}{\Gamma(1)\Gamma(1)} {\rm Tr} \bigg{[} \frac{[m_z^0 + i(q_a + k_a/2)\tau^a ][m_z^0 + i(q_b - k_b/2)\tau^b]\tau^\nu}{\{ [(m_z^0)^2 + (q+k/2)^2](1-x) + [(m_z^0)^2 + (q-k/2)^2]x \}^2} \bigg{]} i(q_\nu -k_\nu/2) \\
		&= m_z^0 \int_0^1 [dx] \int \frac{d^3q}{(2\pi)^3} {\rm Tr} \bigg{[} \frac{[m_z^0 + i(q_a + k_a/2)\tau^a ][m_z^0 + i(q_b - k_b/2)\tau^b]\tau^\nu}{[(q+(1/2-x)k)^2 +x(1-x) k^2 + (m_z^0)^2]^2} \bigg{]}i(q_\nu -k_\nu/2) \\
		&= m_z^0 \int_0^1 [dx] \int \frac{d^3 \tilde q}{(2\pi)^3} {\rm Tr} \bigg{[} \frac{[m_z^0 + i(\tilde q_a +xk_a)\tau^a][m_z^0 + i(\tilde q_b - k_b + xk_b)\tau^b]\tau^\nu}{[\tilde q^2 + x(1-x)k^2 + (m_z^0)^2]^2} \bigg{]}i(\tilde q_\nu -k_\nu +xk_\nu)
	\end{aligned}
\end{equation}
The trace in the numerator reads:
\begin{equation}
	\begin{aligned}
		&{\rm Tr} \{ [m_z^0 + i(q_a + xk_a)\tau^a] [m_z^0 + i(q_b + xk_b -k_b)\tau^b]\tau^\nu \} \\
		=&{\rm Tr} \{ [(m_z^0)^2 + i m_z^0(q_a +xk_a)\tau^a + im_z^0 (q_b + xk_b -k_b)\tau^b- (q_a + xk_a)(q_b +xk_b -k_b)\tau^a\tau^b]\tau^\nu \} \\
		=& im_z^0(q_a +xk_a)(2\delta^{a\nu}) + im_z^0 (q_b + xk_b -k_b)(2\delta^{b \nu}) -(q_a +xk_a)(q_b +xk_b -k_b)(2i\epsilon^{ab\nu}) \\
		=& im_z^0 (2)(2q_\nu + 2xk_\nu -k_\nu) +q_ak_b (2i\epsilon^{ab\nu})
	\end{aligned}
\end{equation}
and the total numerator reads:
\begin{equation}
	\begin{aligned}
		N[\Pi^C_{z\nu}(k)] &= -(m_z^0)^2(2)(2q_\nu +2xk_\nu - k_\nu)(q_\nu +xk_\nu -k_\nu) -m_z^0 q_a k_b (q_\nu +xk_\nu -k_\nu) (2\epsilon^{ab\nu}) \\
		N_e[\Pi^C_{z\nu}(k)] &= -2(m_z^0)^2[2q_\nu^2 + (2x-1)(x-1)k_\nu^2] -2m_z^0q_a q_\nu  k_b \epsilon^{ab\nu}
	\end{aligned}
\end{equation}
Such that, we shall see (the term we dropped in the second line is $q_xq_y$ will vanish as it's odd to $q_x$ or $q_y$ separately)
\begin{equation}
	\begin{aligned}
		\Pi^C_{z\nu}(k) &=m_z^0 \int_0^1 [dx] \int \frac{d^3q}{(2\pi)^3} \frac{-2m_z^0 [2q_\nu^2 + (2x-1)(x-1)k_\nu^2]-2q_a q_\nu k_b \epsilon^{ab\nu}}{[q^2 + x(1-x)k^2 + (m_z^0)^2]^2} \\
		&= m_z^0\int_0^1 [dx] \int \frac{d^3q}{(2\pi)^3} \frac{-2m_z^0 [2q_\nu^2 + (2x-1)(x-1)k_\nu^2]}{[q^2 +x(1-x)k^2 + (m_z^0)^2]^2}
	\end{aligned}
\end{equation}
We shall have:
\begin{equation}
	\begin{aligned}
		\Pi^{C,1}_{z\nu} (k) &=m_z^0 \int_0^1 [dx] \int \frac{d^3q}{(2\pi)^3}\frac{-4m_z^0q_\nu^2}{[q^2 + x(1-x)k^2 + (m_z^0)^2]^2} \\
		&= m_z^0\int_0^1 [dx] (-4m_z^0) \frac{1}{(4\pi)^{3/2}\Gamma(2)} \frac{1}{2}\delta_{\nu \nu} \frac{\Gamma(2-1-3/2)}{[(m_z^0)^2 + x(1-x)k^2]^{2-1-3/2}} \\
		&=  \frac{(m_z^0)^2\delta_{\nu \nu}}{2\pi}\int_0^1 [dx] \frac{1}{[(m_z^0)^2 + x(1-x)k^2]^{-1/2}} \\
		&= \frac{(m_z^0)^2 \delta_{\nu \nu}}{2\pi} \bigg{[} \frac{m}{2} + \frac{i(k^2 + 4m^2)\ln \bigg{(} \frac{-ik+2m}{+ik+2m} \bigg{)}}{8k} \bigg{]} \\
		&\approx \frac{m^2\delta_{\nu\nu}}{2\pi} \bigg{[} m+ \frac{k^2}{12m}  - \frac{k^4}{240m^3} + {\mathcal O}(k^6)  \bigg{]}
	\end{aligned}
\end{equation}
and:
\begin{equation}
	\begin{aligned}
		\Pi^{C,2}_{z\nu}(k) &= m_z^0\int_0^1 [dx] \int \frac{d^3q}{(2\pi)^3} \frac{-2m_z^0(2x-1)(x-1)k_\nu^2}{[q^2 + x(1-x)k^2 + (m_z^0)^2]^2} \\
		&= (-2m_z^0)m_z^0k_\nu^2 \int_0^1 [dx] (2x-1)(x-1) \frac{\Gamma(2-3/2)}{(4\pi)^{3/2}\Gamma(2)} \frac{1}{[(m_z^0)^2 + x(1-x)k^2]^{2-3/2}} \\
		&= -\frac{k_\nu^2 (m_z^0)^2}{4\pi} \int_0^1 [dx] \frac{(2x-1)(x-1) }{[(m_z^0)^2 + x(1-x)k^2]^{1/2}} \\
		&= -\frac{k_\nu^2 (m_z^0)^2}{4\pi} \bigg{[} \frac{(-2km + (k^2 + 4m^2){\rm arccot}[2m/k])}{2k^3} \bigg{]} \\
		&= -\frac{k_\nu^2 m^2}{4\pi} \bigg{[} \frac{1}{6m} - \frac{k^2}{120 m^4} + \frac{k^4}{1120m^5} + {\mathcal O}[k^6] \bigg{]}
	\end{aligned}
\end{equation}

Thus we have:
\begin{equation}
	\Pi^C_{z\nu}(k) = \frac{m^2}{2\pi} \bigg{[} m+ \frac{k^2}{12m}  - \frac{k^4}{240m^3} + {\mathcal O}(k^6)  \bigg{]} -\frac{k_\nu^2 m^2}{4\pi} \bigg{[} \frac{1}{6m} - \frac{k^2}{120 m^4} + \frac{k^4}{1120m^5} + {\mathcal O}[k^6] \bigg{]},
\end{equation}
and to the second order of $k$:
\begin{equation}
	\Pi^C_{z\nu}(k) \approx \frac{m^3}{2\pi} + \frac{m}{24 \pi } (k^2-k_\nu^2)
\end{equation}

Similarly, we shall have:
\begin{equation}
	\begin{aligned}
		\Pi^C_{\mu z}(k) &= \int \frac{d^3q}{(2\pi)^3} {\rm Tr} \bigg{[} \frac{[m_z^0 + i(q_a +k_a/2)\tau^a]\tau^\mu[m_z^0 + i(q_b -k_b/2)\tau^b]}{[(m_z^0)^2 + (q+k/2)^2][(m_z^0)^2 + (q-k/2)^2]} m_z^0 \tau_0 \bigg{]} i(q_\mu + k_\mu/2) \\
		&=  m_z^0\int \frac{d^3q}{(2\pi)^3} \int_0^1 [dx] \frac{\Gamma(1+1)}{\Gamma(1)\Gamma(1)} {\rm Tr} \bigg{[} \frac{[m_z^0 + i(q_a +k_a/2)\tau^a]\tau^\mu[m_z^0 + i(q_b -k_b/2)\tau^b]}{\{ [(m_z^0)^2 + (q+k/2)^2](1-x) + [(m_z^0)^2 + (q-k/2)^2]x \}^2} \bigg{]} i(q_\mu +k_\mu/2) \\
		&= m_z^0\int_0^1 [dx] \int \frac{d^3q}{(2\pi)^3} {\rm Tr} \bigg{[} \frac{[m_z^0 + i(q_a +k_a/2)\tau^a]\tau^\mu[m_z^0 + i(q_b -k_b/2)\tau^b]}{[(q+(1/2-x)k)^2 + x(1-x)k^2 + (m_z^0)^2]^2} \bigg{]}i(q_\mu + k_\mu/2) \\
		&= m_z^0\int_0^1 [dx] \int \frac{d^3q}{(2\pi)^3}{\rm Tr} \bigg{[} \frac{[m_z^0 + i(\tilde q_a + xk_a)\tau^a]\tau^\mu[m_z^0 + i(\tilde q_b -k_b + xk_b)]}{[\tilde q^2 +x(1-x)k^2 + (m_z^0)^2]^2} \bigg{]}i(\tilde q_\mu + xk_\mu)
	\end{aligned}
\end{equation}
The trace in the numerator reads:
\begin{equation}
	\begin{aligned}
		&{\rm Tr} \{ [m_z^0 + i(q_a + xk_a)\tau^a]\tau^\mu[m_z^0 + i(q_b +xk_b - k_b)\tau^b]  \} \\
		=& {\rm Tr} \{ [m_z^0 \tau^\mu + i(q_a + xk_a)\tau^a \tau^\mu][m_z^0 + i(q_b +xk_b -k_b)\tau^b]\} \\
		=&{\rm Tr} [(m_z^0)^2 \tau^\mu +im_z^0 (q_a + xk_a)\tau^a \tau^\mu +im_z^0 (q_b +xk_b -k_b) \tau^\mu\tau^b- (q_a +xk_a)(q_b +xk_b -k_b)\tau^a \tau^\mu \tau^b] \\
		=& im_z^0(q_a +xk_a) (2\delta^{a\mu}) + im_z^0 (q_b +xk_b-k_b)(2\delta^{\mu b}) -(q_a +xk_a)(q_b +xk_b -k_b)(2i\epsilon^{a\mu b}) \\
		=& im_z^0(2)(2q_\mu + 2xk_\mu -k_\mu) +q_a k_b (2i\epsilon^{a\mu b}),
	\end{aligned}
\end{equation}
and the total numerator reads:
\begin{equation}
	\begin{aligned}
		N[\Pi^C_{\mu z}(k)] &= -2(m_z^0)^2(2q_\mu + 2xk_\mu -k_\mu)(q_\mu +xk_\mu) - m_z^0q_a k_b (q_\mu +xk_\mu)(2\epsilon^{a\mu b}) \\
		N_e[\Pi^C_{\mu z}(k)] &= -2(m_z^0)^2 [2q_\mu^2 +(2x-1)xk_\mu^2 ]-2m_z^0q_aq_\mu k_b \epsilon^{a\mu b}
	\end{aligned}
\end{equation}

We obtain:
\begin{equation}
	\begin{aligned}
		\Pi^C_{\mu z}(k) &=m_z^0 \int_0^1 [dx] \int \frac{d^3q}{(2\pi)^3} \frac{-2m_z^0 [2q_\mu^2 + (2x-1)xk_\mu^2]-2q_a q_\mu k_b \epsilon^{a\mu b}}{[q^2 + x(1-x)k^2 + (m_z^0)^2]^2} \\
		&= m_z^0\int_0^1 [dx] \int \frac{d^3q}{(2\pi)^3} \frac{-2m_z^0 [2q_\mu^2 + (2x-1)xk_\mu^2]}{[q^2 +x(1-x)k^2 + (m_z^0)^2]^2}
	\end{aligned}
\end{equation}
We then have:
\begin{equation}
	\begin{aligned}
		\Pi^{C,1}_{\mu z} (k) &=m_z^0 \int_0^1 [dx] \int \frac{d^3q}{(2\pi)^3}\frac{-4m_z^0q_\mu^2}{[q^2 + x(1-x)k^2 + (m_z^0)^2]^2} \\
		&=m_z^0 \int_0^1 [dx] (-4m_z^0) \frac{1}{(4\pi)^{3/2}\Gamma(2)} \frac{1}{2}\delta_{\mu \mu} \frac{\Gamma(2-1-3/2)}{[(m_z^0)^2 + x(1-x)k^2]^{2-1-3/2}} \\
		&=  \frac{(m_z^0)^2\delta_{\mu \mu}}{2\pi}\int_0^1 [dx] \frac{1}{[(m_z^0)^2 + x(1-x)k^2]^{-1/2}} \\
		&= \frac{(m_z^0)^2 \delta_{\mu \mu}}{2\pi} \bigg{[} \frac{m}{2} + \frac{i(k^2 + 4m^2)\ln \bigg{(} \frac{-ik+2m}{+ik+2m} \bigg{)}}{8k} \bigg{]} \\
		&\approx \frac{m^2\delta_{\mu\mu}}{2\pi} \bigg{[} m+ \frac{k^2}{12m}  - \frac{k^4}{240m^3} + {\mathcal O}(k^6)  \bigg{]}
	\end{aligned}
\end{equation}
and:
\begin{equation}
	\begin{aligned}
		\Pi^{C,2}_{\mu z}(k) &= m_z^0\int_0^1 [dx] \int \frac{d^3q}{(2\pi)^3} \frac{-2m_z^0(2x-1)(x-1)k_\mu^2}{[q^2 + x(1-x)k^2 + (m_z^0)^2]^2} \\
		&= m_z^0(-2m_z^0)k_\mu^2 \int_0^1 [dx] (2x-1)(x-1) \frac{\Gamma(2-3/2)}{(4\pi)^{3/2}\Gamma(2)} \frac{1}{[(m_z^0)^2 + x(1-x)k^2]^{2-3/2}} \\
		&= -\frac{k_\mu^2 (m_z^0)^2}{4\pi} \int_0^1 [dx] \frac{(2x-1)(x-1) }{[(m_z^0)^2 + x(1-x)k^2]^{1/2}} \\
		&= -\frac{k_\mu^2 (m_z^0)^2}{4\pi} \bigg{[} \frac{(-2km + (k^2 + 4m^2){\rm arccot}[2m/k])}{2k^3} \bigg{]} \\
		&\approx -\frac{k_\mu^2 m^2}{4\pi} \bigg{[} \frac{1}{6m} - \frac{k^2}{120 m^4} + \frac{k^4}{1120m^5} + {\mathcal O}[k^6] \bigg{]}
	\end{aligned}
\end{equation}
We shall have:
\begin{equation}
	\Pi^C_{\mu z}(k) = \frac{m^2}{2\pi} \bigg{[} m + \frac{k^2}{12m} - \frac{k^4}{240 m^3} +{\mathcal O}[k^6]\bigg{]}  -\frac{k_\mu^2m^2}{4\pi} \bigg{[} \frac{1}{6m} - \frac{k^2}{120 m^4} + \frac{k^4}{1120m^5} + {\mathcal O}[k^6] \bigg{]}
\end{equation}
and to the second order of $k$, that is:
\begin{equation}
	\Pi^C_{\mu z}(k) = \frac{m^2}{2\pi} + \frac{m}{24 \pi }(k^2-k_\mu^2)
\end{equation}

\end{document}